\documentclass[twocolumn]{aastex631}
\usepackage{booktabs}
\usepackage{longtable}
\usepackage{array}
\usepackage{amsmath}
\usepackage{cleveref}

\definecolor{newred}{rgb}{0.1,0.6,0.75}
\definecolor{grey}{rgb}{0.2,0.27,0.57}
\definecolor{freeblue}{rgb}{0.2,0.25,0.45}
\definecolor{freeblue2}{rgb}{0.2,0.25,0.4}
\definecolor{blue}{rgb}{0.45,0.41,0.98}
\definecolor{myblue}{rgb}{0.1,0.5,0.5}
\definecolor{guillaume}{rgb}{0.,0.5,0.65}
\definecolor{louloublue}{rgb}{0.2,0.2,0.65}
\definecolor{loulougreen}{rgb}{0.1,0.35,0.75}
\definecolor{violet}{rgb}{0.5,0.,0.5}

\makeatletter
\usepackage{etoolbox}
\patchcmd\H@refstepcounter{\protected@edef}{\protected@xdef}{}{}
\makeatother

\newcolumntype{i}{>{\scriptsize}r}

\shorttitle{CSOs and TDEs}
\shortauthors{Readhead et al.}

\graphicspath{{./}{figs/}}

\crefname{equation}{Eq.}{Eqs.}
\Crefname{equation}{Equation}{Equations}
\crefname{figure}{Fig.}{Figs.}
\Crefname{figure}{Figure}{Figures}
\crefname{table}{Table}{Tables}
\Crefname{table}{Table}{Tables}
\crefname{section}{Section}{Sections}
\Crefname{section}{Section}{Sections}

\usepackage[nolist,nohyperlinks]{acronym}
\begin{acronym}
    \acro{ads}[ADS]{Astrophysics Data System}
    \acro{agn}[AGN]{Active Galactic Nucleus}
    \acroplural{agn}[AGN]{Active Galactic Nuclei}
    \acro{bologna}[Bologna]{Bologna Complete Sample of Nearby Radio Sources}
    \acro{cd}[CD]{Compact Double}
    \acro{cjf}[CJF]{Caltech Jodrell Bank flat-spectrum}
    \acro{class}[CLASS]{Cosmic Lens All-Sky Survey}
    \acro{cso}[CSO]{Compact Symmetric Object}
    \acro{css}[CSS]{Compact Steep Spectrum}
    \acro{ddrg}[DDRG]{Double-Double Radio Galaxy}
    \acroplural{ddrg}[DDRGs]{Double-Double Radio Galaxies}
    \acro{emerlin}[eMERLIN]{Enhanced Multi Element Remotely Linked Interferometer Network}
    \acro{fsrq}[FSRQ]{Flat Spectrum Radio Quasar}
    \acro{fwhm}[FWHM]{Full Width at Half Maximum}
    \acro{ps}[PS]{Peaked Spectrum}
    \acro{hfp}[HFP]{High-Frequency Peaker}
    \acro{mso}[MSO]{Medium Symmetric Object}
    \acro{ned}[NED]{NASA/IPAC Extragalactic Database}
    \acro{ovro}[OVRO]{Owens Valley Radio Observatory}
    \acro{pr}[PR]{Pearson Readhead}
    \acro{ps}[PS]{Peaked-Spectrum}
    \acro{prcj1}[PR+CJ1]{Pearson Readhead and First Caltech Jodrell Bank}
    \acro{rfc}[RFC]{Radio Fundamental Catalog}
    \acro{vips}[VIPS]{VLBA Imaging and Polarimetry Survey}
    \acro{vla}[VLA]{Very Large Array}
    \acro{vlba}[VLBA]{Very Long Baseline Array}
    \acro{vlbi}[VLBI]{Very Long Baseline Interferometry}
\end{acronym}

\begin{document}

\title{Compact Symmetric Objects - III \\Evolution  of the  High-Luminosity Branch  and
a Possible Connection with Tidal Disruption Events
}

\correspondingauthor{Anthony Readhead}
\email{acr@caltech.edu}

\author{A. C. S Readhead}
\affiliation{Owens Valley Radio Observatory, California Institute of Technology,  Pasadena, CA 91125, USA}
\author{V. Ravi}
\affiliation{Owens Valley Radio Observatory, California Institute of Technology,  Pasadena, CA 91125, USA} 
\author{R. D. Blandford}
\affiliation{Kavli Institute for Particle Astrophysics and Cosmology, Department of Physics,
Stanford University, Stanford, CA 94305, USA}
\author{A. G. Sullivan}
\affiliation{Kavli Institute for Particle Astrophysics and Cosmology, Department of Physics,
Stanford University, Stanford, CA 94305, USA}
\author{J. Somalwar}
\affiliation{Owens Valley Radio Observatory, California Institute of Technology,  Pasadena, CA 91125, USA} 
\author{M. C.  Begelman}
\affiliation{JILA, University of Colorado and National Institute of Standards and Technology, 440 UCB, Boulder, CO 80309-0440, USA} 
\affiliation{Department of Astrophysical and Planetary Sciences, 391 UCB, Boulder, CO 80309-0391, USA}
\author{M.  Birkinshaw}
\affiliation{School of Physics, H.H. Wills Physics Laboratory, University of Bristol, Tyndall Avenue, Bristol BS8 1TL, UK} 
\author{I. Liodakis}
\affiliation{Finnish Center for Astronomy with ESO, University of Turku, Vesilinnantie 5, FI-20014, Finland}
\author{M. L. Lister}
\affiliation{Department of Physics and Astronomy, Purdue University, 525 Northwestern Avenue, West Lafayette, IN 47907, USA}
\author{T. J. Pearson}
\affiliation{Owens Valley Radio Observatory, California Institute of Technology,  Pasadena, CA 91125, USA}
\author{G. B. Taylor}
\affiliation{Department of Physics and Astronomy, University of New Mexico, Albuquerque, NM 87131, USA}
\author{P. N. Wilkinson}
\affiliation{Jodrell Bank Centre for Astrophysics, University of Manchester, Oxford Road, Manchester M13 9PL, UK} 
\author{N. Globus}
\affiliation{Department of Astronomy and Astrophysics, University of California, Santa Cruz, CA 95064, USA}
\author{S. Kiehlmann}
\affiliation{Institute of Astrophysics, Foundation for Research and Technology-Hellas, GR-70013 Heraklion, Greece}
\author{C. R. Lawrence}
\affiliation{Jet Propulsion Laboratory, California Institute of Technology, 4800 Oak Grove Drive, Pasadena, CA 91109, USA}
\author{D. Murphy}
\affiliation{Jet Propulsion Laboratory, California Institute of Technology, 4800 Oak Grove Drive, Pasadena, CA 91109, USA}
\author{S. O'Neill}
\affiliation{Owens Valley Radio Observatory, California Institute of Technology,  Pasadena, CA 91125, USA}
\author{V. Pavlidou} 
\affiliation{Institute of Astrophysics, Foundation for Research and Technology-Hellas, GR-70013 Heraklion, Greece}
\affiliation{Department of Physics and Institute of Theoretical and Computational Physics, University of Crete, 70013 Heraklion, Greece}
\author{E. Sheldahl}
\affiliation{Department of Physics and Astronomy, University of New Mexico, Albuquerque, NM 87131, USA}
\author{A. Siemiginowska}
\affiliation{Center for Astrophysics|Harvard and Smithsonian, 60 Garden St., Cambridge, MA 02138, USA}
\author{K. Tassis} 
\affiliation{Institute of Astrophysics, Foundation for Research and Technology-Hellas, GR-70013 Heraklion, Greece}
\affiliation{Department of Physics and Institute of Theoretical and Computational Physics, University of Crete, 70013 Heraklion, Greece}

\begin{abstract}
 We use a sample of 54  Compact Symmetric Objects (CSOs) to confirm that there are two unrelated CSO classes: an edge-dimmed, low-luminosity class (CSO~1), and an edge-brightened, high-luminosity class (CSO~2). Using blind tests, we show that CSO~2s consist of   three sub-classes:  CSO 2.0, having prominent hot-spots at the leading edges of narrow jets and/or narrow lobes; CSO~2.2, without prominent hot-spots, and with broad jets and/or lobes;  and CSO~2.1, which exhibit mixed properties.   Most  CSO 2s  do not evolve into larger jetted-AGN, but spend their whole life-cycle as CSOs of size $\lesssim$500 pc and age $\lesssim$5000 yr. The minimum energies needed to produce the radio luminosity and structure in CSO~2s range from $\sim~10^{-4}\,M_\odot{c}^2$ to $\sim7\,M_\odot{c}^2$.   We show that the transient nature of  most CSO~2s, and their birthrate, can be explained through ignition in the tidal disruption events of giant stars. We also consider possibilities  of tapping the spin energy of the supermassive black hole, and tapping the energy of the accretion disk. Our results demonstrate that  CSOs constitute a large family of AGN in which we have thus far studied  only the brightest.   More comprehensive  CSO studies, with higher sensitivity, resolution, and dynamic range, will revolutionize our understanding of AGN and the central engines that power them.
\end{abstract}


\section{Introduction}
\label{sec:intro}

\begin{figure*}
    \centering
    \includegraphics[width=1.0\linewidth]{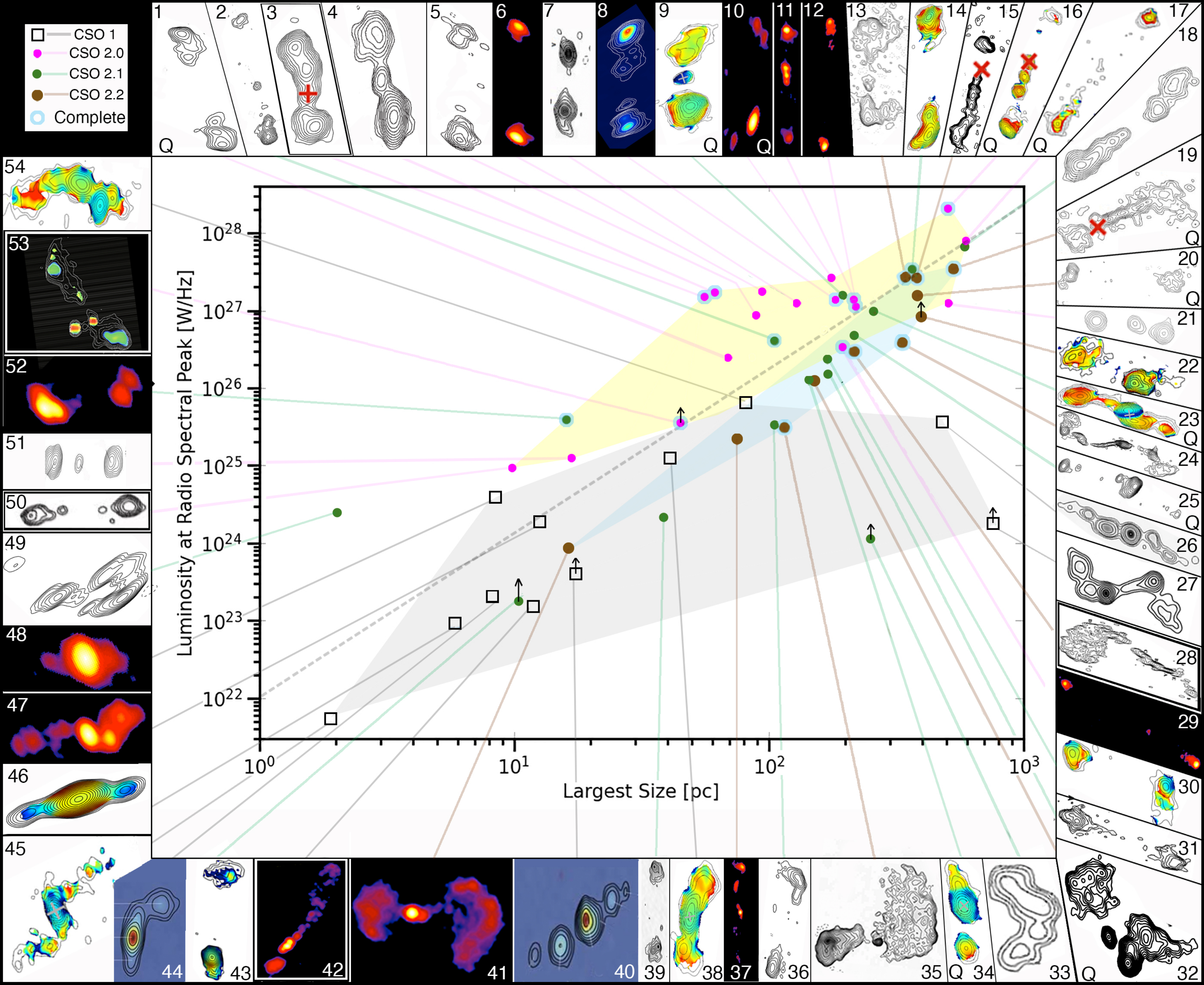}
    \caption{The radio astronomer's  ``Hertzsprung--Russell diagram''  \citep{1982IAUS...97...21B} for CSOs: The distribution in the Luminosity--Size $(P,D)$ plane of the 54~bona fide CSOs for which we have spectroscopic redshifts. Open square symbols: CSO~1 objects. Pink, green and brown filled circles:  CSO~2.0, 2.1 and 2.2 objects, respectively. Up arrows on symbols indicate CSOs with rising spectra below 1 GHz, for which we have only a lower limit on the peak luminosity. The light blue annuli around the symbols indicate objects in the complete PR+CJ1+PW samples. The diagonal dashed line, dividing the $(P,D)$ plane in two, has been drawn by eye to maximise the ratio of CSO~2.0 objects above the line to those below the line, and to maximise the ratio of CSO~2.2 objects below the line to those above the line.     The references for the 54 radio emission maps in this Figure are indicated by the dagger symbol$^\dagger$ in the individual source notes in  Appendix~B. Many of the maps have their true sky orientations rotated for arrangement purposes.    CSOs for which there is evidence of more than one epoch of activity are indicated by double borders - i.e., the CSO~1: \# 42; the CSO~2.0s: \#3, \#50 and \#53; and the CSO~2.2: \# 28. In these five cases, the size plotted refers to the most recent activity (see text). Quasars are indicated by the letter ``Q''. The other CSOs are radio galaxies.  The faint yellow, blue, and gray regions show the areas of the $(P,D)$ plane occupied by the CSO~2.0, CSO~2.2 and CSO~1 objects, respectively, making it clear that they predominantly occupy different regions of the $(P,D)$ plane (see text). }
    \label{plt:linsizedist}
\end{figure*}

  Compact Symmetric Objects (CSOs) are extragalactic radio sources less than 1\,kpc in extent that show emission on both sides of their center of activity  and in which the observed emission is not dominated by relativistic motion towards the observer  \citep{1994ApJ...432L..87W,1996ApJ...460..612R}. 
   A major development in the study of CSOs was the discovery   that there are two, distinct, classes of CSOs: an edge-dimmed low luminosity class (CSO 1), and an edge-brightened high-luminosity  class (CSO 2) \citep{2016MNRAS.459..820T} -see their \S 4.4 and their Fig. 11.

   Compact radio sources associated with AGN exhibit a wide  variety of morphological and radio spectroscopic types.  This paper cannot do justice to the enormous amount of work that has been done in this field, and which is the foundation upon which it is built, but to place CSOs in the wider context, the reader  is referred to the excellent and exceptionally comprehensive review of the field by \citet{2021AandARv..29....3O} hereafter OS21, which is both up-to-date and also traces it back to its origins. 
  
  From the earliest studies, CSOs have often been designated as ``young'' objects, or as objects showing ``recurrent'' activity \citep[see, e.g.,][]{1980ApJ...236...89P,1980ApJ...241L..73M,1988ApJ...328..114P,1990AA...232...19B,1992ApJ...396...62C,1996ApJ...460..612R,2014MNRAS.438..463O}.  In this study we show that these descriptors, ``young'' and ``recurrent'', are  misleading. We find strong evidence supporting the hypothesis that  most CSO 2s exhibit a full life cycle from birth to death as CSOs, and so they cannot correctly be described as  ``young'', although they  can correctly be described as    ``short-lived'',   as compared to  classes of larger jetted-AGN,  such as FR I and FR II objects \citep{1974MNRAS.167P..31F}.   The hypothesis  that high-luminosity CSOs are short-lived was first put forward by \citet{1993AAS...182.5307R}, hereafter R93.  Throughout the previous CSO literature,  including in R93,   the terms ``young'' and ``short-lived'' have been treated as synonymous,  which they are not.  To avoid confusion with the earlier literature we will use the terms ``early-life'', ``mid-life'' and ``late-life'' to describe the different evolutionary  stages of CSO 2s. 

In this paper  we  present evidence which supports the hypothesis that $\sim 40\%$  of the bona fide CSO~2s we have identified are ``early-life'', $\sim 30\%$   are ``mid-life'', and $\sim 30\%$ are ``late-life'', compared to their overall lifetime,  which ranges from   $\lesssim 100$ yr to $\sim 5000$ yr. We then discuss the energies, ages, birth rates, and possible origins of CSO 2s in the context of single star capture by a supermassive black hole (SMBH), as was first suggested by \citet[hereafter R94]{1994cers.conf...17R}, with its implications for tidal disruption events \citep[TDEs;][]{1988Natur.333..523R} and the late stages of evolution of TDEs with radio jets (jetted-TDEs).  The possibility that CSO 2s are fuelled by the capture of single stars by an SMBH has also been suggested by \citet{2012ApJ...760...77A}.
 We also consider two other possible origins of the energy of CSO 2s, namely, the spin of the SMBH and the accretion disk.

An unfortunate development of the last 20 years in the study of CSOs is that there are now many mis-classified CSOs in the literature.  This is what led to the present study.  

This paper is the third of three papers on the morphological radio properties of CSOs in which we explore the phenomenology of CSOs uncontaminated by misidentified objects.  In the first paper (Kiehlmann et al. 2023a, in press, hereafter Paper~1),  we added two new criteria, based on variability and speed, to the CSO selection criteria and  undertook a detailed survey of the literature with the  sole purpose of filtering out the mis-classified CSOs. This  enabled us to identify 79~bona fide CSOs.  Our 79 bona fide CSOs include  three complete samples\footnote{{ A ``complete sample'' is defined to be a sample that includes all objects down to a given flux density limit over a given area of sky \citep{1968MNRAS.139..515P,1968ApJ...151..393S,1970MNRAS.151...45L}.}} that  are therefore suitable for statistical studies.  In the second paper (Kiehlmann et al. 2023b, in press, hereafter Paper~2), we determined the fractions of CSOs in these complete samples.  All of the CSOs in these complete samples are CSO 2s.  We used the subset of  17 bona fide CSO 2s in these complete samples to show that CSO 2s form a distinct subclass of jetted-AGN that has a sharp upper cutoff in size at $\approx 500$ pc.  

Another  potentially misleading descriptor in CSO studies is the use of the term ``recurrent'' for CSOs in which there is evidence of more than one epoch of activity. Five of our bona-fide CSOs  show evidence of more than one epoch of activity, with a distinct gap, or drop in surface brightness, between the emission regions associated with each episode. In our view it should not be assumed that any previous activity was a manifestation of the same physical process as that which produces CSOs,  and we present evidence showing that these earlier 
 epochs of activity were a factor 30-1000 times more energetic than the CSOs.  For this reason we should use the descriptor ``episodic'',   rather than ``recurrent'', for this class of rare CSOs, which comprises  $\sim$6\%  of the bona fide CSOs we have identified. Four CSO 2 objects and one CSO 1 object, out of our total of 79 bona fide CSOs, were found to be episodic.

Throughout this paper we adopt the convention $S_\nu \propto
\nu^{\alpha}$ for spectral index $\alpha$, and, for consistency with our other papers, we use the cosmological
parameters $\Omega_m = 0.27$, $\Omega_\Lambda = 0.73$ and $H_0 = 71 \;
\mathrm{km\; s^{-1} \;\,Mpc^{-1}}$ \citep{2009ApJS..180..330K}. None of the conclusions would be changed were we to adopt the best model of the Planck Collaboration  \citep{2020A&A...641A...6P}. In this paper we follow the lead of  OS21 in their comprehensive review of peaked spectrum sources, and refer to gigahertz-peaked spectrum (GPS) sources as peaked spectrum (PS) sources.

\begin{deluxetable*}{lcDl@{\hskip5mm}|@{\hskip5mm}lcDl}
\tablecaption{Key to the CSOs in \cref{plt:linsizedist} in numerical order$^*$}.
\decimals
\tablehead{ID \# 	&	Source Name	&	\multicolumn2c{$z$}&	Class	&ID \# &	Source Name	&	\multicolumn2c{$z$}&	Class}
\startdata
1	&	J1111+1955	&	0.299	&	CSO~2.0	& 28	&	J1347+1217$^\dag$	&	0.121	&	CSO~2.2	\\
2	&	2022+6136$^\dag$	&	0.227	&	CSO~2.1	& 29	&	J1944+5448$^\dag$	&	0.263	&	CSO~2.0	\\
3	&	J0111+3906$^\dag$	&	0.668	&	CSO~2.0	& 30	&	J1440+6108	&	0.445365	&	CSO~2.1	\\
4	&	J1735+5049$^\dag$	&	0.835	&	CSO~2.0	& 31	&	J1816+3457	&	0.245	&	CSO~2.1	\\
5	&	J2203+1007	&	1.005	&	CSO~2.0	& 32	&	J1158+2450	&	0.203	&	CSO~2.2	\\
6	&	J1734+0926	&	0.1813	&	CSO~2.0& 33	&	J1508+3423	&	0.045565	&	CSO~2.1	\\
7	&	J1939$-$6342	&	0.735	&	CSO~2.0		& 34	&	J1234+4753	&	0.373082	&	CSO~2.1	\\
8	&	J0029+3456$^\dag$	&	0.772137	&	CSO~2.1& 35	&	J0119+3210$^\dag$	&	0.0602	&	CSO~2.2	\\
9	&	J0741+2706	&	1.601115	&	CSO~2.0		& 36	&	J1414+4554	&	0.186	&	CSO~2.1	\\
10	&	J0943+1702	&	0.518	&	CSO~2.0	& 37	&	J1945+7055	&	0.101	&	CSO~2.2	\\
11	&	J0713+4349$^\dag$	&	0.46	&	CSO~2.0	& 38	&	J1311+1658	&	0.081408	&	CSO~1	\\
12	&	J1035+5628$^\dag$	&	0.517	&	CSO~2.0	& 39	&	J0855+5751	&	0.025998	&	CSO~2.1	\\
13	&	J1326+3154$^\dag$	&	0.37	&	CSO~2.2	& 40	&	J1025+1022	&	0.045805	&	CSO~1	\\
14	&	J1609+2641$^\dag$	&	0.473	&	CSO~2.1	& 41	&	J0131+5545	&	0.03649	&	CSO~2.2	\\
15	&	J1400+6210$^\dag$	&	0.431	&	CSO~2.2	& 42	&	J1148+5924	&	0.01075	&	CSO~1	\\
16	&	J1227+3635$^\dag$	&	1.975	&	CSO~2.0& 43	&	J1205+2031	&	0.024037	&	CSO~2.1	\\
17	&	J1159+5820	&	1.27997	&	CSO~2.0		& 44	&	J0909+1928	&	0.027843	&	CSO~1	\\
18	&	J0825+3919	&	1.21	&	CSO~2.1	& 45	&	J1220+2916	&	0.002	&	CSO~1	\\
19	&	J1244+4048$^\dag$	&	0.813586	&	CSO~2.2& 46	&	J0906+4124	&	0.027	&	CSO~1	\\
20	&	J1313+5458	&	0.613	&	CSO~2.2	& 47	&	J1559+5924	&	0.0602	&	CSO~1	\\
21	&	J1120+1420	&	0.362	&	CSO~2.0	& 48	&	J1254+1856	&	0.1145	&	CSO~1	\\
22	&	J1434+4236	&	0.452	&	CSO~2.2		& 49	&	J1723$-$6500	&	0.01443	&	CSO~2.1	\\
23	&	J1644+2536	&	0.588	&	CSO~2.1& 50	&	J1247+6723	&	0.107219	&	CSO~2.0	\\
24	&	J2355+4950$^\dag$	&	0.238	&	CSO~2.2	& 51	&	J1511+0518	&	0.084	&	CSO~2.0	\\
25	&	J1915+6548	&	0.486	&	CSO~2.1	& 52	&	J1407+2827$^\dag$	&	0.077	&	CSO~2.1	\\
26	&	J1602+5243	&	0.105689	&	CSO~1	& 53	&	J0405+3803$^\dag$	&	0.05505	&	CSO~2.0	\\
27	&	J2327+0846	&	0.02892	&	CSO~1&	54	&	J0832+1832	&	0.154	&	CSO~1	\\
\enddata
\tablecomments{ The ID~\# is the reference number  in \cref{plt:linsizedist}.   $^*$A version of this table in Right Ascension order may be found in \cref{tab:bonafides2} in Appendix~A. Other common names of these objects are given in Paper 1, which also gives references for the redshifts.   $^\dag$ indicates objects in the complete PR+CJ1+PW sample.}
\label{tab:bonafides}
\end{deluxetable*}

\section{ The Bona Fide CSO Sample and the Scope of this Paper}\label{sec:bonafide}

In this paper we focus on the 54  bona fide CSOs with spectroscopic redshifts that we identified in Paper 1.  This sample is listed in \cref{tab:bonafides}.  Representative images of these 54 objects are shown in \cref{plt:linsizedist}. As first pointed out by  \citet{1982IAUS...97...21B}, ``This is the radioastronomer's H-R diagram. To make it easy to associate the images with the source data,  the order in \cref{tab:bonafides} is the number of each object in \cref{plt:linsizedist}.    To facilitate looking up individual sources by name, a version of \cref{tab:bonafides} with the objects listed in order of right ascension is given in \cref{tab:bonafides2} in Appendix~A.  A  brief  description of each source is given in Appendix B.

The distribution of our 54~bona fide CSOs between the CSO~1 and CSO~2 classes in the luminosity ($P$) -- size ($D$) plane is shown in \cref{plt:linsizedist}, where  CSO~1s and CSO~2s are represented by empty squares and filled circles, respectively.  The $(P,D)$ values are taken from Paper 1, which lists the relevant references.  We see that CSO~1s tend to have lower luminosities than CSO~2s, and hence tend to occupy a different region of the $(P,D)$ plane. The luminosities refer to that at the peak of each CSO's radio spectrum, and the size is measured from the radio contour maps, as described in Paper 1.  In six cases the peak occurs below the lowest frequency at which the source has been observed.  These are indicated by the up arrows in Fig.~\ref{plt:linsizedist}.

\subsection{The Emerging Picture of the CSO Family}\label{sec:family}

\begin{figure}[!t]
 \centering
 \includegraphics[width=0.7\columnwidth]{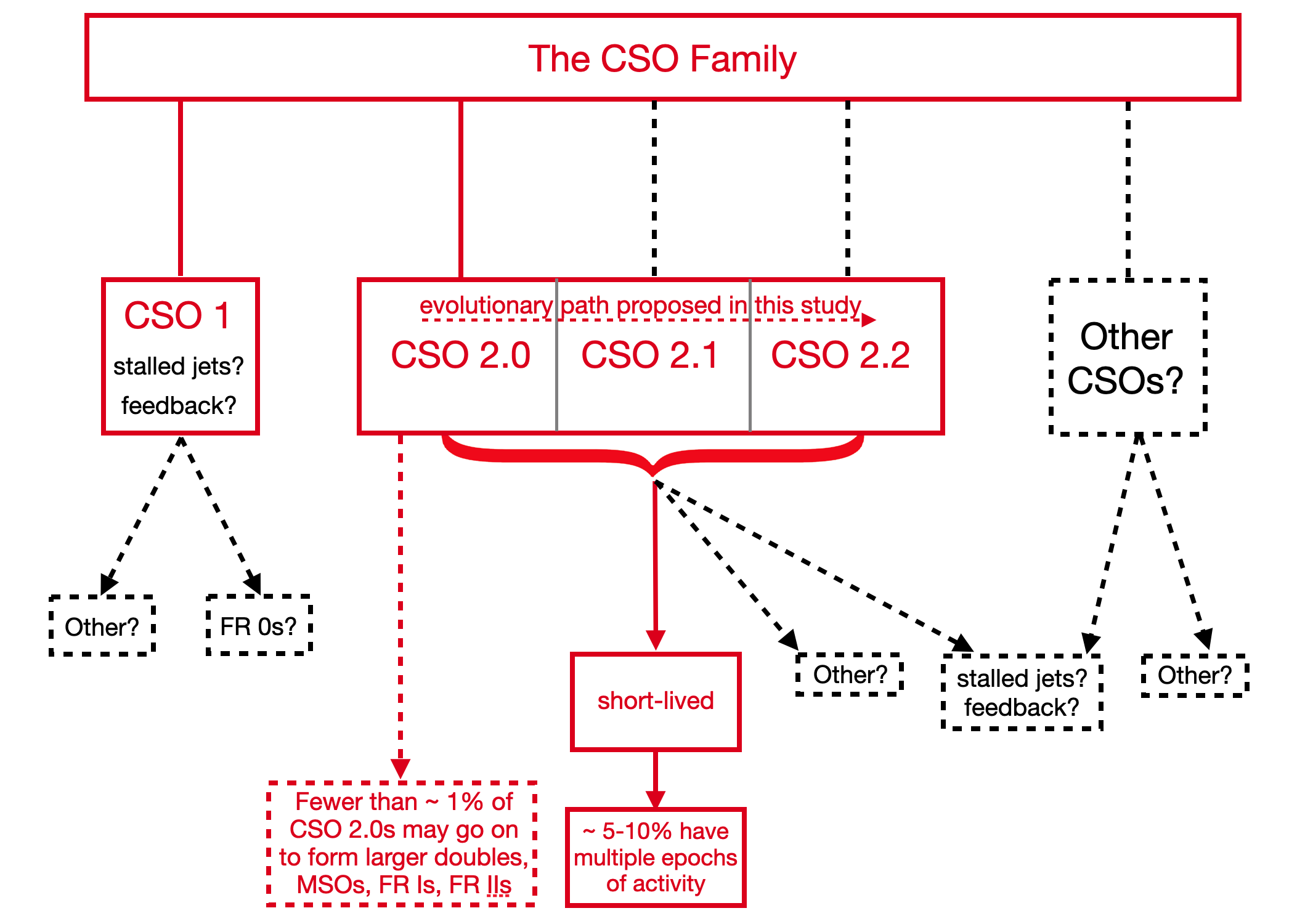}
 \caption{ The complex emerging picture of the CSO family thus far. The solid boxes, lines, and arrows indicate distinctions that in our view have been firmly established in these three papers.  The dotted lines indicate hypotheses that remain to be proven. The red items are all discussed in varying amounts of detail in Papers 1, 2 and 3.  The black items are other possibilities that have been discussed in the literature, and/or that are likely to be verified or discovered as we observe fainter CSOs and dig deeper into the CSO luminosity function.   For  details on FROs, FR Is, FR IIs, stalled jets, feedback, and multiple epochs of activity the reader is referred to OS21.} 
 \label{plt:Family}
\end{figure}

Since Medium Symmetric Objects (MSOs) \citep{1995AandA...302..317F,1996ApJ...460..612R}, FR Is and FR IIs \citep{1974MNRAS.167P..31F} are all symmetric double radio sources that are larger than CSOs, it is highly likely that they began as CSO 2s before reaching 1 kpc in size.  Since the timescales are too long to confirm this in individual cases, we take as a working hypothesis that all of these larger scale doubles did in fact pass through an early CSO 2 phase. In Paper 2 we saw that fewer than $\sim$1\% of CSO 2s can evolve into larger scale radio doubles, such as FR Is and FR IIs,   and only $\sim 5\%$ can evolve into MSOs and CSS sources. Thus, the whole family of CSOs probably embraces at least the following three classes of CSOs, which we present as a schematic diagram in Fig. \ref{plt:Family}:

1. CSO 1s, which, since they are edge-dimmed, might be important for feedback and star formation, and/or may be related to FR 0s \citep{2015A&A...576A..38B, 2023arXiv230708379B}. 

2. The 99\% of CSO 2s that that do not go on to form MSOs, FR Is, or FR IIs, and are the  main focus of this paper.

3. The $\lesssim 1 \%$ of CSO 2s that we hypothesize do go on to form MSOs, FR Is, and/or FR IIs.

This subject remains in its infancy, since so far we have studied mostly high-luminosity CSOs.  We may well discover further distinct classes of CSOs, or other interesting connections, as we delve deeper into the CSO luminosity function. We have tried to leave room for this in the schematic of  Fig. \ref{plt:Family}.  We emphasize that CSOs may well be formed by a variety of mechanisms.  

Thus, this paper is not intended to provide a complete description of CSOs, which is, in any case, impossible at this early stage in their study, but rather to draw attention to the peculiar properties of $\gtrsim 99\%$ of CSO 2s. Thus, we focus primarily on \#2 above and  the possible connection with TDEs. We are limited in our complete samples to the brightest, and hence most luminous, CSOs - i.e. to CSO 2s, of which, as shown in Paper 2,  the majority are short lived and exhibit a sharp cutoff in size at $\sim$500 pc. In addition, we particularly wish to draw attention to the potential of CSO 2s for addressing many interesting questions of AGN and relativistic jet physics through the combination of  time domain astronomy and high resolution imaging.

The follow-up of these high-luminosity CSO 2 observations with observations that include both lower-luminosity CSO 1s and CSO 2s is   clearly  of paramount importance in understanding AGN   (OS21), the creation of relativistic jets   (OS21), and feedback   (OS21).  For these reasons, we have embarked upon a program to observe the steep-spectrum counterpart of the incomplete VLBA Imaging and Polarimetry Survey (VIPS) \citep{2007ApJ...658..203H} flat spectrum sample, which will extend the CSO luminosity function almost an  order of magnitude below the level considered in this study.  

In view of our focus on high luminosity objects, in this paper we do not discuss in detail the properties of CSO 1s.  CSO 1s are sufficiently different, morphologically, to CSO 2s, that we think they merit intensive study on their own. In view of the similar edge-dimmed morphologies, it is possible that CSO 1s are related to FR 0s, and may be of great interest in the study of feedback \citep{2021AN....342.1087D}. 

Another important consideration is that of potentially ``frustrated'' CSOs \citep{1984Natur.308..619W,1984AJ.....89....5V,1991ApJ...380...66O,1998PASP..110..493O}, objects in which the small size is due to confinement by the interstellar medium of the host galaxy.    We are aware of these factors that complicate the study of CSOs. Again, a  study of this interesting class of objects is beyond the scope of the present paper, but we include this possibility in Fig. \ref{plt:Family}.

Similarly, while in this paper we advance the hypothesis that many CSO 2s are formed by single star capture, we are not suggesting that {\it all\/} CSO 2s are formed in this way, and we cannot yet even be certain that {\it any\/} CSO 2s are formed in this way.   In \S \ref{sec:conuncon} we consider at length alternative scenarios for the formation of CSO 2s. 

It should therefore be clear that it is not our intent in this paper to attempt to describe the whole phenomenology of CSOs. Any such description will require a concerted effort, pushing to lower luminosities, and this will take many years, possibly several decades,  to accomplish.

\section{CSO~1 and CSO~2 Classes}\label{sec:csoclass}

The sole criterion we apply to distinguish between CSO~1s and CSO~2s is the ``edge-dimmed'' {\it vs.}~``edge-brightened'' morphological criterion. We do not apply any luminosity constraints on either class. By ``edge-dimmed'' we mean that moving out from the center of activity the surface brightness drops.  By ``edge-brightened'' we mean that moving out from the center of activity the surface brightness first increases, but that it may drop towards the outer extremities of the source.

As can be seen from Fig.~\ref{plt:linsizedist}, there is a wide variety of CSO~2s morphologies, but we have managed to classify these using three sub-classes, which, as we show later, likely corresponds to an evolutionary sequence.

\vskip 6pt
\noindent
CSO~2.0: These are CSOs in which there are two distinct outer lobes, making them edge-brightened, with hot spots  at their outer edges opposite the nucleus, narrow jets (if visible at all), and  lobes not much wider than the hot spots. These are indicated by pink symbols in \cref{plt:linsizedist}, with pink lines connecting their positions  in the $(P,D)$ plane to their corresponding images.

\vskip 6pt
\noindent
CSO~2.2:   These are CSOs in which there are two distinct outer lobes but the hot spots are not dominant or are invisible, or the lobes are well-resolved normal to the jet axis. The hotspot, or hot spots, do not need to be located at the end of the lobe opposite the core. These are indicated by brown points in \cref{plt:linsizedist}, with brown lines connecting the points in the $(P,D)$ plane to their corresponding images. 

\vskip 6pt
\noindent
CSO~2.1: These are ``indeterminate'', or ``intermediate'',  CSO 2s, which have some of the properties of CSO~2.0s and some of the properties of CSO~2.2s.  For example, they might look like CSO~2.0s on one side of the nucleus and CSO~2.2s on the other side, or they might have hot spots that are not located at the extremities of their envelopes.  These are indicated by green points in \cref{plt:linsizedist}, with green lines connecting the points in the $(P,D)$ plane to their corresponding images. Two objects in this ``indeterminate'' class (\#33 and \#49 in \cref{plt:linsizedist}) have very peculiar morphologies that seem unrelated in any way to other CSO~2s, so they are ``indeterminate'', but the remainder are all mixtures of CSO~2.0 and CSO~2.2 morphologies, and may be designated as ``intermediate''. 

\begin{deluxetable*}{c@{\hskip 8mm}ccccccc}
\tablecaption{Binomial Tests of the Numbers of Successes in the Blind Test Classifications}
\tablehead{Number of&Teams in&Expected&Expected&Observed&Observed&Binomial\\
CSOs&Agreement&Fraction&Number&Number&Fraction&Probability}
\startdata
54&4&0.0156&1 & 26&  0.47&$1.3\times 10^{-32}$\\
54&3 or 4&0.203&10& 46&  0.85&$2.4\times 10^{-24}$\\
54&2&0.703&38& 8& 0.15&$3.4\times 10^{-17}$\\
54&0&0.0937&5&0 &0&$4.9\times 10^{-3}$\\
\enddata
\tablecomments{Each row shows the result for the number of teams in agreement given in the second column.  We see that all four teams agreed on the classification for 26 of the 54~CSOs, whereas only 1 was expected in the case of purely random class assignments.  We also see that three or four of the teams were in agreement on 46 of the CSOs, whereas only 10 were expected.  As shown here, were the distribution random, we would expect 5 cases where none of the four teams were in agreement, whereas in fact there were no such cases.}
\label{tab:binomial}
\end{deluxetable*}

Our reasons for assigning each of the 54~CSOs to the four classes, CSO~1, CSO~2.0, CSO~2.1, or CSO~2.2, are given for each object individually in Appendix~B. To ensure that we were not biasing the distributions of and conclusions about the different CSO~2 classes in the $(P,D)$ plane, and to assess the reliability and reproducibility of our CSO classification scheme, we carried out blind tests to classify the 54~bona fide CSOs for which we have spectroscopic redshifts.    These blind tests  and their results  are described in Appendix~C.

 In the blind tests we used four teams, and their classifications as well as the final classification we adopted are given in Table \ref{tab:blindtests} in Appendix C. The significance of these classifications can be tested against the hypothesis that the CSOs are randomly distributed. Results are given in \cref{tab:binomial}.  Since there are four classes and four teams, there are $4^4=256$ possible outcomes of the blind test on each of the 54~CSOs being classified.  The details of the calculations of the expected fractions and binomial probabilities shown in Table \ref{tab:binomial} are given in Appendix C. 

The  binomial probabilities, shown in the seventh column of \cref{tab:binomial} leave no room for doubt that the classification system is one that is astrophysically meaningful, and therefore provides a useful description of CSOs.  Thus, in spite of the variety of morphologies exhibited by  CSOs, they can reliably be assigned to one of the four morphological classes defined above.

\begin{deluxetable*}{c@{\hskip 8mm}cccc}
\tablecaption{The Numbers and Percentages of CSOs in Different Classes}
\tablehead{Class&Number&Percentage&Percentage\\
&in Class&cf. Total&cf. CSO 2s}
\startdata
CSO 1&11&20\%$\pm 7\%$&-&-\\
CSO 2.0&17&31\%$\pm9\%$&40\%$\pm11\%$\\
CSO 2.1&15&28\%$\pm8\%$&35\%$\pm10\%$\\
CSO 2.2&11&20\%$\pm7\%$&26\%$\pm9\%$\\
\enddata
\tablecomments{The numbers and fractions of CSOs in different CSO classes for the full total of 54 CSOs with spectroscopic redshifts, and for the 43 CSO 2s with spectroscopic redshifts.}
\label{tab:fractions}
\end{deluxetable*}

\begin{figure}
    \centering
    \includegraphics[width=0.8\linewidth]{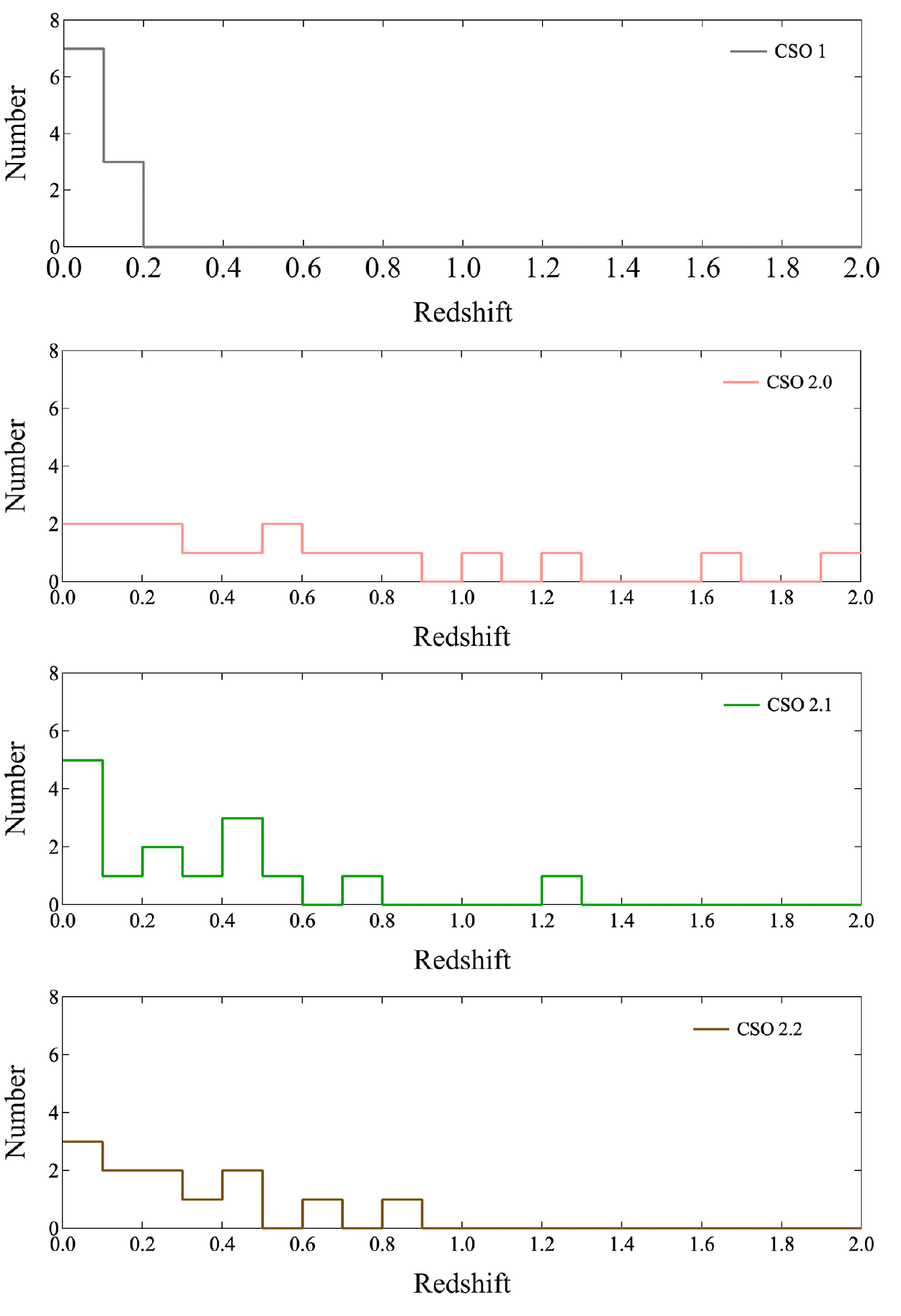}
    \caption{Redshift distributions of the CSO classes.}
    \label{plt:redshift}
\end{figure}

\section{Distributions of CSO\lowercase{s} Amongst the Different Classes}\label{sec:distributions}

In Papers 1 and 2 we discussed a variety of selection effects, that are mainly of relevance to the size distribution of CSO 2s. In this paper we focus on selection effects that are of particular relevance to the distributions of the different classes of CSOs that we have identified and their distributions in the $(P,D)$ plane.

\subsection{Selection Effects  Amongst the Four CSO Classes}\label{sec:selection}

There are many selection biases in this sample, and in particular there is a strong redshift selection bias, so the interpretation of the data on these 54 sources must be treated with care.  But the 54 CSOs populate the $(P,D)$ plane densely enough for us to be able to look for variations in morphology with $P$ and $D$  amongst the high-luminosity CSOs, above 10 pc in size.
It is important to consider any  biases that might affect the distribution of the four classes of CSOs in the $(P,D)$ plane. For this purpose we show  in Fig. \ref{plt:redshift} the redshift distributions of the different classes of CSOs. It is immediately clear that CSO 1s are seen predominantly at lower redshifts than CSO 2s.  This is undoubtedly a selection effect due to the fact that CSO 1s are low luminosity objects and that their symmetric regions, being edge-dimmed, are of low surface brightness.  Both of these effects will push high redshift  CSO 1s below the detection levels of the finding surveys, which have, thus far, been restricted to high flux densities. It is to be expected, therefore, that CSO 1s will be found at higher redshifts as the survey flux density limits are extended downwards.

We will not discuss the redshift distribution of CSO 2.1s since they have mixed CSO 2.0 and CSO 2.2 characteristics.  We therefore compare the redshift distributions of the CSO 2.0s and CSO 2.2s shown in Fig. \ref{plt:redshift}.  Although the two distributions overlap substantially, it appears, by eye, to be the case that the CSO 2.2s are concentrated at somewhat lower redshifts than the CSO 2.0s.  The Kolmogorov-Smirnoff two-sample test gives a KS test statistic of 0.55, and a p-value=0.0026.  So it is rather unlikely that these two distributions are drawn from the same parent population, although the possibility that they are cannot be ruled out definitively. 

 If this is due to selection bias, the implication is that our sample is missing CSO 2.2 objects at high redshifts.  This would not be surprising because, like the CSO 1 sources, CSO 2.2 objects lack bright hot spots and their lobes have lower surface brightnesses than those of CSO 2.0 objects.  Thus this selection effect is in the sense that it biases against  detecting CSO 2.2s in the lower right hand quadrant of the $(P,D)$ plane.  Since the selection effect actually biases against the trend we already see in Fig. \ref{plt:linsizedist}, our  finding  regarding the bifurcation of the CSO 2.0s {\it vs.\/} the CSO 2.2s in the $(P,D)$ plane is robust.

\begin{figure*}
    \centering
    \includegraphics[width=1.0\linewidth]{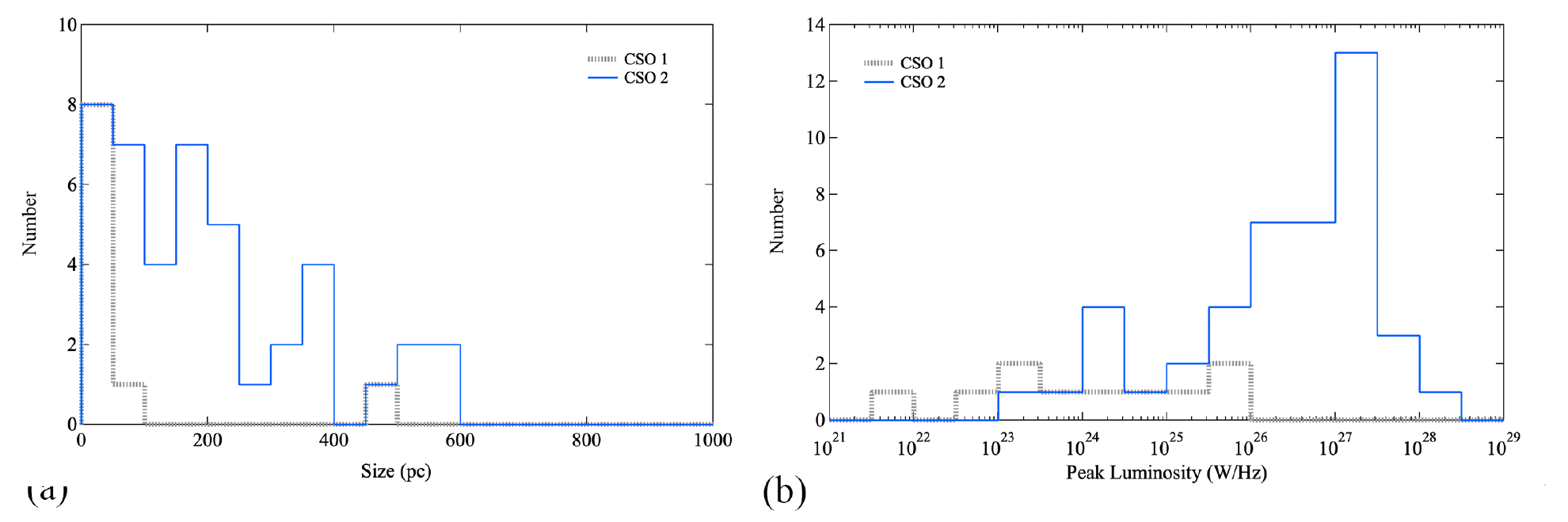}
    \caption{Comparison of CSO~1s and CSO~2s. (a) The distributions of the largest projected linear sizes. (b) The distributions of the peak luminosities.}
    \label{plt:comp1vs2}
\end{figure*}

The distributions in size and luminosity of CSO~1s {\it vs.\/} CSO~2s are shown in \cref{plt:comp1vs2}. The sizes and luminosities are taken from Paper 1.  As in Paper~1, the ``peak luminosities'' are the luminosities at the frequency of the peak of the spectrum for CSOs with peaked spectra, and at the lowest observed frequency for  six 
CSOs  which have monotonically falling spectra. In \cref{plt:comp1vs2}  it can be seen that CSO~1s are on average smaller and less luminous than CSO~2s.     These are clearly two very different classes of objects: the CSO~1s tend to be nearby, low-luminosity objects, whereas the CSO~2s have higher luminosities that lie in the range of FR~I and FR~II objects. 

 While CSO 1s  are clearly a very interesting class of jetted-AGN, there are no immediately apparent morphological trends amongst the bona fide CSO 1s in our sample that can be followed up,  and we will have to wait for larger, deeper surveys to investigate the evolution of CSO 1s.  For this reason we focus for the rest of this paper on the CSO 2s, amongst which we find that there are interesting morphological trends.

\subsection{Comparison of CSO~2.0s and CSO~2.2s}\label{sec:comp2pt0vs2pt2}

\begin{figure*}
    \centering
    \includegraphics[width=1.0\linewidth]{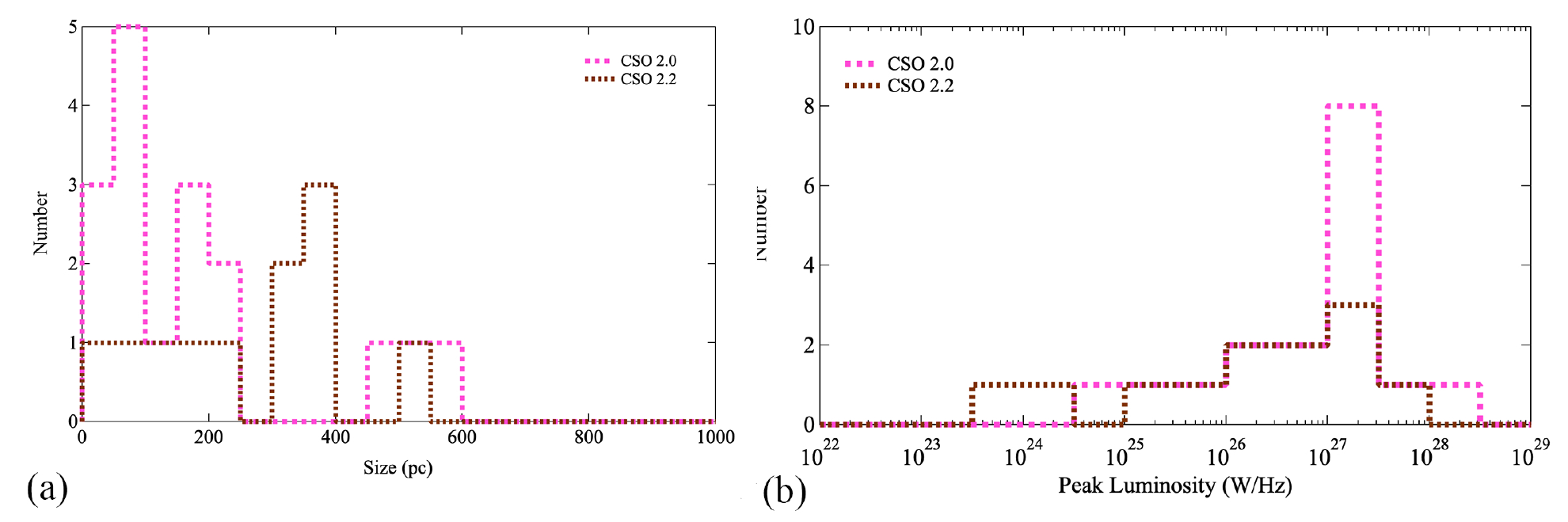}
    \caption{Comparison of the largest projected linear sizes of CSO~2.0s and CSO~2.2s; (b) Comparison of the luminosities of CSO 2.0s and CSO 2.2s; }
    \label{plt:comp2pt0vs2pt2}
\end{figure*}

The distributions in  size and luminosity of CSO~2.0s {\it vs.\/} CSO~2.2s are shown in \cref{plt:comp2pt0vs2pt2}, where it can be seen that on average the CSO~2.0s are smaller than CSO~2.2s, and   on average their luminosities are larger.
  For the whole sample of 54 CSOs with spectroscopic redshifts, the median size of the CSO~2.0s is 129\,pc, and the median size of the CSO~2.2s is 340\,pc, i.e., the ratio of the median sizes is $\sim 2.6:1$. For the PR+CJ1+PW sub-sample these numbers are 190 pc and 340 pc, i.e. a ratio  of the median sizes of $\sim 1.8:1$.
For the whole sample of 54 CSOs with spectroscopic redshifts, the  median luminosities of CSO~2.0s and CSO~2.2s are 12  and $4 \times 10^{26}$ W\,Hz$^{-1}$, respectively,   i.e. the ratio of the median luminosities is $\sim 3:1$. For the PR+CJ1+PW sub-sample these numbers $1.4 \times 10^{27}$ W\,Hz$^{-1}$, and $1 \times 10^{27}$ W\,Hz$^{-1}$, respectively,   i.e., the ratio of the median luminosities is $\sim 1.4:1$. As we have discussed in Papers 1 and 2, the whole sample is heavily affected by sampling effects, so it is the complete PR+CJ1+PW sub-sample that we should use, and for this we see that the CSO 2.0s are significantly smaller than the CSO 2.2s, but their luminosities are comparable.

\vskip 5pt
\noindent
{\it Summary of Classifications in the $(P,D)$ Plane}: The areas of the $(P,D)$ plane occupied by the CSO~1s, 2.0s and 2.2s are shown by the faint gray, yellow and blue highlighted regions in \cref{plt:linsizedist}, respectively.  These make it clear that these different classes occupy different, but overlapping,  portions of the $(P,D)$ plane.

\section{The Evolution of CSO~2\lowercase{s} }\label{sec:evolution}

In this paper we aim only at a very basic approach to the evolution of CSO~2s,  We wish to determine whether or not there is clear evidence for the evolution of CSO~2s in the $(P,D)$ diagram shown in \cref{plt:linsizedist}.  Fundamentally, we wish to determine whether the hypothesis of R93 that most high-luminosity CSOs are ``short-lived'' is correct or not.  We see in the $(P,D)$ plot that CSO~2.0s are all larger than 20\,pc.  If we consider only the region of the $(P,D)$ diagram  above 20\,pc, we see that   there is a tendency  for the
CSO~2.0s to be located toward the upper left hand side   (yellow region) of the
distribution, and the CSO~2.2s toward the lower right hand side (blue region). 

\begin{figure*}
    \centering
    \includegraphics[width=1.0\linewidth]{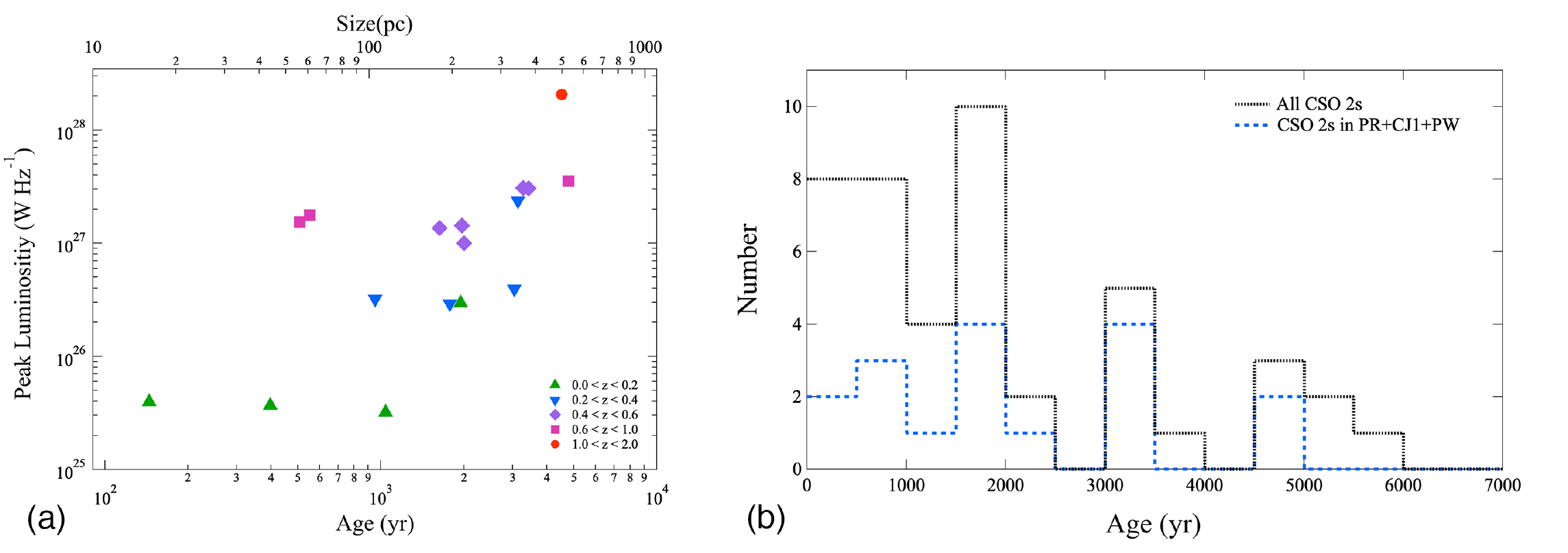}
    \caption{Sizes, ages and luminosities of CSO~2s, assuming a hot-spot separation speed of $v_{\rm s} = 0.36c$ (see text). (a) The observed luminosities,  ages,  and redshifts for the CSO~2s in the complete PR+CJ1+PW sample. The sample is divided into five redshift bins, as indicated in the legend. The  trend of increasing luminosity with size is due to the steep luminosity function (see text). The large jumps in luminosity in individual bins are all associated with jumps in redshift. (b) The distribution of luminosity with age: Black dotted line: the 44 CSO~2 objects with spectroscopic redshifts; blue dashed line: the 17 CSO~2 objects in the PR, CJ1 and PW complete samples. }
    \label{plt:prcj1pwages} 
\end{figure*}

\subsection{Selection Effects in the CSO 2 $(P,D)$ Distribution}\label{sec:effects}

In \cref{plt:linsizedist}, there is an upper envelope of the CSO~2s that rises from $P\sim 10^{25}$ W\,Hz$^{-1}$ at $D\sim 10$\,pc to $P\sim 3 \times  10^{28}$ W\,Hz$^{-1}$ at $D\sim 500$\,pc.  This is likely due to the combined effect of  (i)~the finite resolution of VLBI ($\approx 4$ mas), so that the smallest angular size CSOs recognizable in 8 GHz VLBI surveys, $\approx 30$\,pc at $z=1$; and (ii)~the CSO luminosity function.   In Fig.~\ref{plt:prcj1pwages}(a) we show the $(P,D)$ diagram for the PR+CJ1+PW sample broken up into different redshift bins.  Note that  the upper envelope has the same characteristics as that of the whole sample, shown in \cref{plt:linsizedist}. We see that the luminosity increases with redshift bin.  This tells us that the lower luminosity CSO 2s seen in the low redshift bins do not all evolve into high luminosity objects. Conversely, the higher luminosity objects seen in the higher redshift bins are not seen at low redshifts due to the steepness of the luminosity function.

The lower envelope in \cref{plt:linsizedist} is due to the selection effect of the flux density limits of the complete samples.  We see that the lower envelope of the CSO~2s  is more or less flat at $P\sim 10^{24}$\,W\,Hz$^{-1}$ from $D\sim 10$\,pc to  $D\sim 500$\,pc.  There are several selection effects  responsible for the lack of objects below  $P\sim 10^{24}$\,W\,Hz$^{-1}$, including spectral index, and surface brightness sensitivity limits.

\begin{figure}
\centering
    \includegraphics[width=0.8\columnwidth]{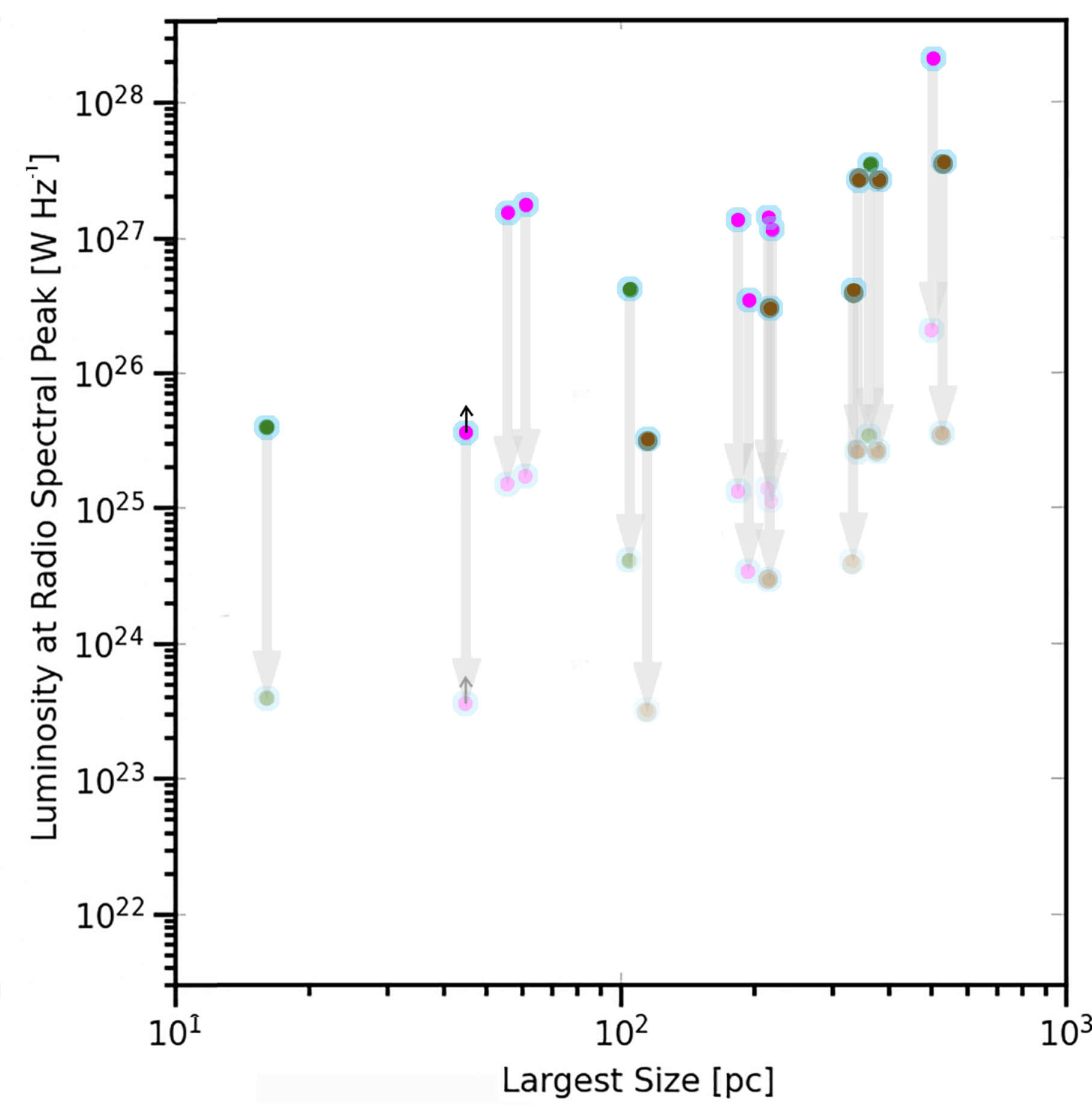}
    \caption{Observing complete samples to lower flux density limits. The 17 CSO 2s in the PR+CJ1+PW samples from Fig. 1 are shown by the dark symbols. An identical sample in which each object is $100\times$ weaker, such as might be discovered in a complete sample of radio sources with a 5 GHz flux density cutoff of 7 mJy,  is shown by the light symbols at the ends of the gray arrows.}
    \label{plt:faintercso2}
\end{figure}

If we imagine reducing the flux densities of each of the objects in the PR+CJ1+PW  sample of Fig. \ref{plt:prcj1pwages}(a), while their other properties are unaffected, they will each follow a track in the $(P,D)$ plane indicated by the gray arrow attached to each object in Fig. \ref{plt:faintercso2}. For example, were we to reduce the flux density of each object, by two orders of magnitude they would appear at the ends of the gray arrows in the positions shown by the corresponding faint symbols.

Note that this reduction in flux densities has populated part of the lower right quadrant of the $(P,D)$ plane that previously was unpopulated. Since CSOs must fade after their fuelling ceases, there must be many low-luminosity CSOs with large sizes - i.e. with sizes in the $\approx$50 pc to $\approx$500 pc range.  Were we to reduce the flux density limit below the lower limit of the PJ+CJ1+PW sample of S$_{\rm 8\, GHz}=700$ mJy, sources could then begin to populate the lower right quadrant of the $(P,D)$ plane.  

\subsection{Evolution in the CSO 2 $(P,D)$ Distributon}\label{sec:evolcso2}

We have shown in Paper~2 that  most ($\gtrsim 99\%$) CSO~2s constitute a distinct population of jetted-AGN that is not related to other classes of jetted-AGN, including CSS, PS, FR~I and FR~II objects, and that the CSO 2 size distribution cuts off well below the 1\,kpc CSO selection criterion.  We now advance the hypothesis that  most CSO~2s pass through all the stages of their evolution, from birth through death,  as CSOs.

 Before discussing CSO 2s \textit{per se}, critical context is provided by  the $(P,D)$ diagram of FR II objects \citep{1997MNRAS.292..723K}, and  the``Baby  Cygnus A's'' model discussed by \citet{1996cyga.book..209B}, who considers the equipartition case in which the particle and magnetic field energy densities are equal, resulting in simple scaling relationships. Suppose the source luminosity $P$ is dominated by an emission region with volume, $V_{\rm em}$,  and pressure, $p_{\rm em}$. Then $P \propto  p_{\rm em}^{7/4} \, V_{\rm em}$. $V_{\rm em}$ could be comparable to the volume of the entire cocoon or some smaller volume representing the lobes, which evolves inside the cocoon as the overall source grows.  If the electrons reside in the emitting region for a time $t_{\rm em}$, which could be the age of the source or shorter, then $p_{\rm em}  \propto L_j \, t_{\rm em}/V_{\rm em}$, where $L_j$ is the jet power.  Thus $P \propto  L_j^{7/4} t_{\rm em}^{7/4} V_{\rm em}^{-3/4}$. For constant velocity of jet advance and constant $L_j$, and if $t_{\rm em} \sim$ age $\propto D$, the scaling is $P \propto D^{7/4}V_{\rm em}^{-3/4}$, i.e. $P \propto D^{-1/2}$ (assuming $V_{\rm em} \propto D^3)$. Thus, in equipartition, and with constant jet power, the luminosity should decrease with increasing source size. The expectation for CSO 2s, therefore, might well be that the luminosity will drop as the size increases, but unfortunately, as discussed in \S \ref{sec:effects}, the $(P,D)$ diagram of CSO 2s is dominated by the luminosity function.

Such a decrease in luminosity with increasing source size is evident among the FR II sources from the complete sample of 3CRR FR II radio sources studied by \citet{1997MNRAS.292..723K}, and becomes even more clear in the sub-sample with $z<1$ studied by \citet{2013ApJ...767...12G,2013ApJ...769..129S}, where allowance is made for the intrinsic spread in  the jet power $L_j $.   

It should be remembered that both FR IIs and CSO 2s were small and weak at birth, and hence must have first appeared at the bottom left of the $(P,D)$ diagram.  This must then have been followed by a period of growth, when the jet power was increasing, in order for them to end up as high-luminosity radio doubles. In the case of the 3CRR FR IIs this occurs at $D< 8$ kpc \citep{1997MNRAS.292..723K}. In the case of CSO 2s, the scale at which the $P$ changes from increasing  to decreasing with $D$ remains to be determined.  It is possible that in CSO 2s this occurs over the observationally accessible size range.  Clearly, if this is the case, then  the $(P,D)$ relationship of CSO 2s  will provide invaluable insight into both their own and  FR II evolution.  To exploit this possibility  we need higher frequency observations of CSO 2s.

\begin{deluxetable*}{c@{\hskip 8mm}ccccccc}
\tablecaption{Distributions of CSO~2.0s, CSO~2.1s and CSO~2.2s in the $(P,D)$ Plane}
\tablehead{Class&Designation&Number&Fraction & Number& Fraction&Ratio of Fractions\\
&&Upper Left&Upper Left&Lower Right&Lower Right&Upper/Lower\\}
\startdata
CSO~2.0 &early-life&15 & 88\%& 2& 12\% &7.5\\
CSO~2.1 &mid-life&5 & 33\%& 10& 67\% &0.50\\
CSO~2.2 &late-life&1 & 8\%& 11&92\% &0.09\\
\enddata
\tablecomments{The dashed line in  \cref{plt:linsizedist} is drawn to maximise the ratio of CSO~2.0s to CSO~2.2s in the upper left and to maximise the ratio of CSO~2.2s to CSO~2.0s in the lower right.  The numbers of CSO~2.0s and 2.2s in the upper left and lower right is shown in this table.  This shows an astonishingly  clear bifurcation of the morphologies of CSO 2s in the previously unexplored $(P,D)$ plane.}
\label{tab:distributions}
\end{deluxetable*}

The bifurcation in the morphologies of CSO~2.0s and CSO~2.2s, shown by the yellow and blue regions of the $(P,D)$ plane in \cref{plt:linsizedist}, prompted us to approximately separate these two classes in the $(P,D)$ plane by drawing the dashed line from bottom left to upper right.  The dashed line in \cref{plt:linsizedist} has been drawn by eye to to maximise the ratio of CSO~2.0s above the line to those below it,  and to maximise the ratio of CSO~2.2s below the line to those above it. 

The numbers and fractions of CSOs 2s in the three sub-classes are shown in \cref{tab:distributions}.  These are striking: we see that the fractions of CSO~2.0s and CSO~2.2s each flip by a factor $\sim 10$ across this line, while the fraction of CSO~2.1s is virtually unchanged.
 In \cref{plt:linsizedist}  the points  in the upper left half of the figure are predominantly pink (CSO~2.0), while below it they are predominantly brown (CSO~2.2). This is clearly not a random effect, and we have shown in \S \ref{sec:effects} that the only selection bias we have been able to conceive of goes in the opposite direction. 
 For these reasons we hypothesize that the observed bifurcation of the CSO 2.0 and CSO 2.2 classes in the $(P,D)$ plane is due to the evolution of CSO 2s,  such that ${\rm CSO \;2.0 } \rightarrow {\rm CSO \;2.2}$, with a possible intermediate class of CSO 2.1, starting from an initial spread of $\sim 2$ orders of magnitude in peak luminosity.

If we now imagine the trajectory of an individual CSO~2 in the $(P,D)$ plane, we see that the stages of growth and decay will match those  of the $(P,D)$ plane that are dominated by the pink and brown points, respectively.  We suggest, therefore that CSO~2.0s (pink points) are objects that are increasing in    $D$, and that CSO~2.2s (brown points) are objects that have slowed or stopped increasing in   $D$. Given this scenario, and given that CSO~2.0s are on average significantly smaller than CSO~2.2s, we would expect the separation speeds, $v_{\rm s}$,  of the outer edges of the lobes straddling the nucleus to be higher in CSO~2.0s than in CSO~2.2s.  We return to this point in the next section.

We note that the spread in CSO~2 luminosities at small $D$ is at least a factor of $10^2$, so it appears that the CSO 2s below the upper envelope in \cref{plt:linsizedist} are a mixture of unevolved ``early-life'', intermediate ``mid-life'', and evolved ``late-life''  objects.  This is seen clearly in the distribution of pink, green and brown points.

Note that the CSO 2.1 sources exhibit features common to both CSO 2.0s and CSO 2.2s. In addition, as shown in Table \ref{tab:distributions} they are much more evenly distributed across the dashed line that demarcates the regions dominated by the CSO 2.0s and 2.2s.  In both respects, they are intermediate in properties between the CSO 2.0s and the CSO 2.2s. It is for these reasons that we regard them as an intermediate step between CSO 2.0s and CSO 2.2s in the evolutionary sequence  that we have suggested, and refer to them as ``mid-life''.  However, it remains possible that CSO 2.1s   are not necessarily a mid-life stage between CSO 2.0s and CSO 2.2s but may represent CSO~2s in asymmetric environments which causes one side to experience turbulence but not the other. Such  asymmetry could be caused by an environment with an average velocity parallel with the CSO jet axis, inducing a high Reynolds number in one jet but not the other. Additionally the ambient density may not be symmetric about the core, leading to the observed differences in the two CSO~2.1 lobes. In any case, CSO~2.1 represent a middle-ground between CSO~2.0s and CSO~2.2s in appearance and physical properties.  It should also be remembered that, for those CSO 2s whose axes are not very close to the plane of the sky, the two sides are seen at significantly different ages, and that this could also lead to one side having apparent CSO 2.0 morphology and the other having apparent CSO 2.2 morphology even if the two sides evolve identically.

\subsection{Other Studies of Short-Lived Compact Jetted-AGN}\label{sec:other}
There have been several studies of short-lived, as opposed to young, compact jetted-AGN, which are often referred to as ``fading'' sources, characterized by steep spectra, indicative of the energy supply having been switched off, and lack of hot spots or  jets \citep{2005A&A...440...93K,2006A&A...450..945K,2005AA...441...89G,2008ASPC..386..176G,2010MNRAS.408.2261K,2010MNRAS.402.1892O,2015ApJ...809..168C,2023MNRAS.522.3877O}. Morphologically, therefore, these faders have the same characteristics as the CSO 2.2 objects we discuss in this paper. Many of the faders are low-luminosity objects, but the sample studied by \citet{2023MNRAS.522.3877O} are high-luminosity objects, and several of them are A-class CSO candidates (see Paper 1), and therefore highly likely to be bona fide CSOs.  We have deep multi-frequency VLBA observations of these objects in hand and are presently analyzing these.

Given the similarities of faders and  CSO 2.2s objects, the question naturally arises as to why the CSO 2s discussed in this paper should not simply be combined with CSS and other PS objects and treated as a whole?  Indeed, this would have been our own inclination had we not come across the unexpected sharp cutoff in CSO 2 sizes presented in Paper 2.  The statistical tests of Paper 2 show that there is only a 1 in $\sim 6000$ probability that this cutoff is just a random result.  What this means is that, while it is indeed possible that CSO 2s are drawn from the same population as the high-luminosity CSS sample, for example,  studied by \citet{2023MNRAS.522.3877O},  this is highly unlikely. In the VIPS flat-spectrum sample the bona fide CSO 2s show the same size distribution as do those from the PR+CJ1+PW sample discussed in Paper 2.  We also find that, in the PR+CJ1+PW CSO 2 sample, the numbers of  steep spectrum objects ($\alpha < -0.5$) is approximately equal to the number of flat spectrum objects ($\alpha \geq -0.5$). Our proposed program of VLBA observations of  332 compact steep spectrum sources will remove the spectral index cutoff of $\alpha \geq -0.5$ that defines the VIPS sample at present.    Taking an extremely conservative estimate, by removing the flat spectrum cutoff from the VIPS sample, we  will at least treble the number of CSO 2s in complete samples, to $\sim$50.  If the steep spectrum VIPS CSO 2 objects show the same size distribution as the flat spectrum VIPS CSO 2 objects that we have in hand,  then the level of significance of the size cutoff will  increase to a p-value that will make an absolutely compelling case for the \textit{distinctive} nature of the vast majority of CSO 2s. In addition, we note that by pushing the flux density limit on complete samples down to S$_{\textrm{8.5 \, }}\textrm{GHz}=85$ mJy, including steep spectrum sources, one accesses the population of CSO 2s with spectral peaks at $\gtrsim 1$ Jy below 100 MHz.

\section{The Ages of CSO~2\lowercase{s}}\label{sec:ages}

\citet{1998AandA...337...69O} discussed the separation speed of the CSO 0710+439 in detail, and found that the hot-spot separation speed is $ v_{\rm s} = (0.354\pm0.041)c$ (assuming $H_0 = 71 \,{\rm km \, s^{-1} \, Mpc^{-1}}$).  From this and the size of the source they derived an age of $1100 \pm 100$ years, and concluded that CSOs are young,  by which, in our interpretation, they mean ``short-lived''. \citet{2005ApJ...622..136G} carried out a detailed study of 20~CSOs for which they could estimate the ages, or lower limits to the ages. They measured kinematic ages that ranged from $20\pm 4$\,yr to $3000 \pm 1490$\,yr.  Combining their kinematic ages with those of \citet{2003PASA...20...69P} they found that 7~out of 13~CSOs had kinematic ages below 500\,yr. They also showed that the ages dropped off steeply above $\sim 2000$\,yr, and concluded that CSOs could not evolve into FR~I or FR~II objects.

\citet{2003PASA...20...69P} considered the separation speed of the hot spots in eight of our bona fide CSO 2s. An unweighted average of their data yields $\overline{v_{\rm s}} = (0.30 \pm 0.04)c$, for $H_0 = 71 \,{\rm km \, s^{-1} \, Mpc^{-1}}$.

\citet{2012ApJ...760...77A} studied a sample of 46 CSOs and CSO candidates. Of these 25 are amongst the bona fide cadidates in our sample selected in Paper 1, and 24 have spectroscopic redshifts. Amongst the 24 bona fide CSOs with spectroscopic redshifts, 14 have measured separation speeds listed by \citet{2012ApJ...760...77A}, with speeds ranging between 0.08c and 0.96c, and an average of $\overline{v_{\rm s}} = (0.42 \pm 0.07)c$, and 5 have upper limits on the separation speed. If we use those upper limits as actual values of $v_{\rm s}$, then we find $\overline{v_{\rm s}} = (0.38 \pm 0.06)c$, whereas if we use $v_{\rm s}=0 c$ for the 5 objects having upper limits then we find that $\overline{v_{\rm s}} = (0.31 \pm 0.07)c$.

The evolutionary hypothesis we are proposing in this paper predicts that CSO~2.0s should, on average, have higher separation speeds than the CSO~2.2s, which have likely slowed down. So we should be able to differentiate between CSO~2.0s and CSO~2.2s when considering CSO 2 separation speeds. Returning to the study of  \citet{2003PASA...20...69P}, we find that 5 of their 10 objects are bona fide CSO~2.0s, and for these their results yield a hot-spot separation speed of $\overline{v_{\rm s}} =(0.36\pm0.04) c$.  In the \citet{2012ApJ...760...77A} study there are 9 CSO 2.0s with measured separation speeds and one with an upper limit.  For the 9 with measured separation speeds we find an average of $\overline{v_{\rm s}} = (0.41 \pm 0.08)c$. If we use the single upper limit as actual value of $v_{\rm s}$, we find $\overline{v_{\rm s}} = (0.39 \pm 0.08)c$, whereas if we use $v_{\rm s}=0 c$ for this object we find that $\overline{v_{\rm s}} = (0.37 \pm 0.09)c$. For the remainder of this paper we  assume that the mean separation speed of CSO 2.0s is $\overline{v_{\rm s}} =0.36 c$,  the average of the measured speeds of bona fide CSO 2.0 objects, and as a simplifying assumption we assume this separation speed for {\it all\/} CSO 2s during ``early-life''. 

The distribution of ages for the 44 CSO~2 objects for which we have spectroscopic redshifts,  based on our measured largest angular sizes and assuming a constant hot-spot separation speed of $0.36 c$, is shown in \cref{plt:prcj1pwages}(b).  In Paper~2 we showed that there are 17 bona fide CSO 2s with spectroscopic redshifts in the complete samples of Pearson and Readhead (PR) \citep{1988ApJ...328..114P}, the first Caltech--Jodrell Bank survey (CJ1) \citep{1995ApJS...98....1P}, and the Peacock and Wall (PW) survey \citep{1982MNRAS.198..843P}.  The distribution of ages for these 17 CSO 2s, assuming a hot-spot separation speed of $0.36c$, is shown in Figs. \ref{plt:prcj1pwages}(a) and (b). We note that the ages in both of these samples extend to $\approx$5000 yr, so we will use 5000 yr as the typical lifetime of the CSO 2s we are discussing. 

In Fig. \ref{plt:linsizedist} the CSO 2s in complete samples are indicated by the light blue annuli around the points.  We see that there are eight CSO~2.0s, three CSO~2.1s, and six CSO~2.2s.  Thus, in complete samples, the number of ``early-life'' CSO 2.0s is about the same as that of  ``late-life'' CSO 2.2s, and these are both significantly larger than the number of CSO 2.1s in the ``mid-life'' phase.  We conclude  that CSO 2s spend roughly half their lifetimes in the ``early-life''  phase, then pass through the ``mid-life'' stage quickly, and spend about half their lifetime in the ``late-life'' phase.   Since, as seen in Fig. \ref{plt:linsizedist}, both CSO 2.1s and CSO 2.2s have lower peak luminosities than CSO 2.0s of comparable size, we assume that the luminosity begins to drop off significantly once CSO 2s pass on to the ``early-life'' and ``mid-life'' stages.

\begin{figure*}
    \centering
    \includegraphics[width=1.0\linewidth]{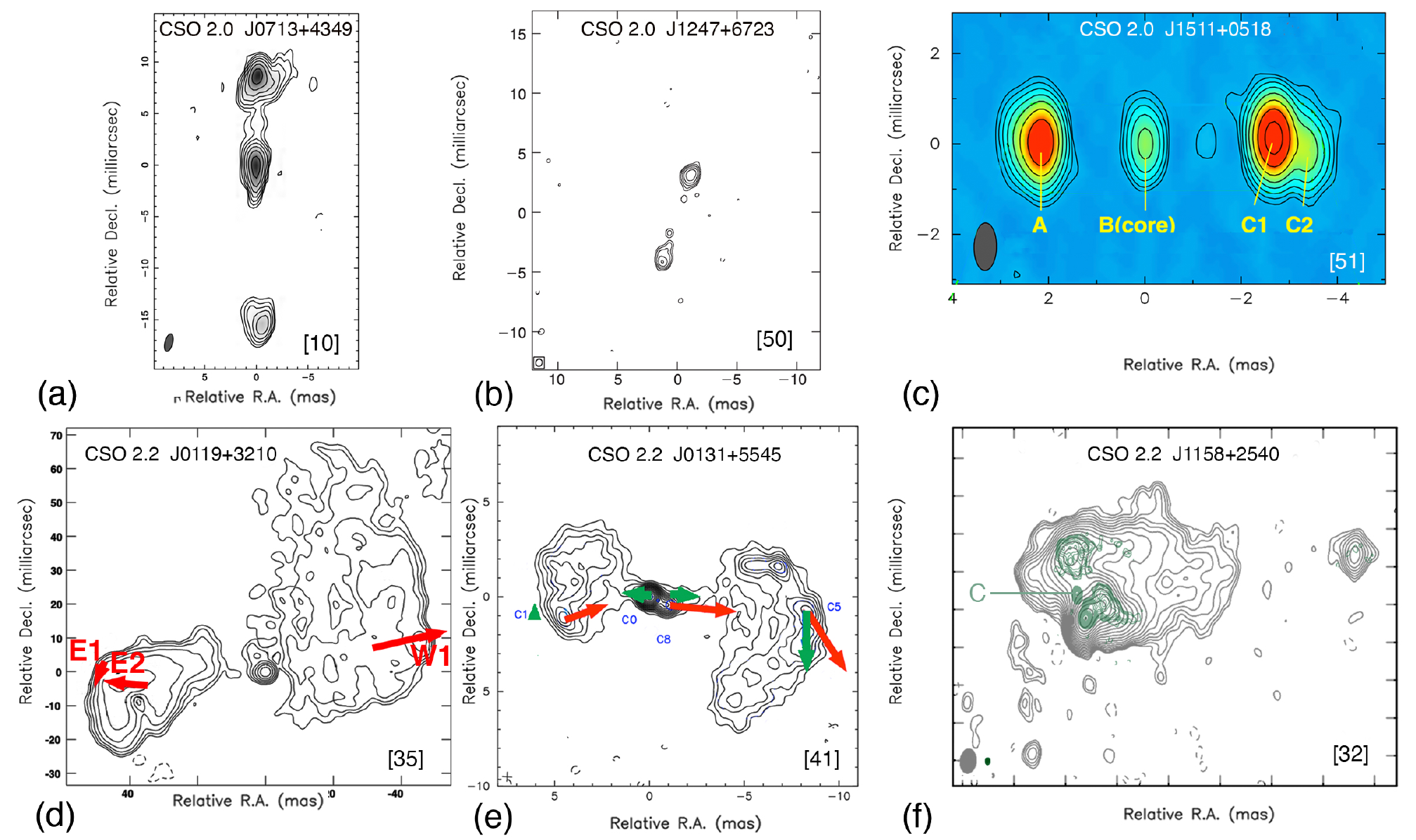}
    \caption{Three illuminating examples of CSO~2.0s and  CSO2.2s, selected to illustrate speed comparisons. CSO~2.0s: (a): J0713+4349 \citep{1998AandA...337...69O}; (b): J1247+6723 \citep{2003PASA...20...16M}; (c): J1511+0518 \citep{2012ApJS..198....5A}; CSO~2.2s: (d): J0119+3210 \citep{2003AandA...399..889G}; (e) J0131+5545 \citep{2020ApJ...899..141L}; (f) J1158+2540 \citep{2008ApJ...684..153T}.  The identification numbers of these six objects in \cref{plt:linsizedist} are indicated in the [square] brackets.  }
    \label{plt:sixmaps}
\end{figure*}

It is difficult to measure the separation speeds in CSO~2.2s because they usually do not have bright hot spots at the outer edges of their outer lobes; however, in cases where they can be measured, they are found to be lower than for CSO~2.0s.  For example,  in the CSO~2.2 object J2355+4950 \citet{2003PASA...20...69P} measured a hot-spot separation speed of $(0.12\pm0.03) c$, and the CSO~2.2 object J1945+7055 has a hot-spot separation speed of $(0.024 \pm 0.006)c$ \citep{2009ApJ...698.1282T}. So the CSO 2.2 object J2359+4950 has a separation speed of about one third that of the average CSO~2.0, whereas in the CSO 2.2 object J1945+7055 the separation speed is about one seventh of the average CSO~2.0 separation speed. 

In \cref{plt:sixmaps} we compare the separation speeds in three CSO~2.0s with three CSO~2.2s, chosen to illustrate and explore the possible differences in morphology and separation speeds in these two classes. 
The separation speeds of the three CSO~2.0s are easy to determine because they have well-defined hot spots at the leading edges of their outer lobes.  These are $(0.52 \pm 0.03)c$, for J0713+4349 \citep{2003PASA...20...69P}; $(0.24 \pm 0.015)c$, for J1247+6723 \citep{2003PASA...20...69P}; and 
$(0.19 \pm 0.03)c$, for J1511+0518 \citep{2012ApJS..198....5A}. The separation speeds of the CSO 2.2s are not easy to determine.
  We therefore discuss these individually below.

\subsection{Separation speed in the CSO~2.2 J0119+3210}\label{sec:J0119}

\citet{2003AandA...399..889G} have carried out a detailed  study of the motion of components in J0119+3210. The speeds they derive for various components are shown in \cref{plt:sixmaps}(d), where we highlight components E1, E2 and W1 with red arrows. \citet{2003AandA...399..889G} derived a significant separation speed between components E2 and W1.  However, we see that component E1 is closer to the leading edge of the eastern lobe, and  is moving almost due south.  It is by no means clear to us, therefore, that the outer edges of J0119+3210 are actually separating, and that the measured relative motions do not apply to components moving along the jets.

\subsection{Separation speed in the CSO~2.2 J0131+5545}\label{sec:J0131}

In Fig.~\ref{plt:sixmaps}(e) we show the stacked 15\,GHz image of \citet{2020ApJ...899..141L} and their measured velocities (red arrows) of the components C1, C8, and C5 relative to component C0, which they assume to be stationary in the rest frame of the central engine.

We propose here a different interpretation of these relative motions based on the morphologies of  the CSO~2.2s J0111+396, J0713+4345, and J2355+4950, which were studied in detail by R96 and \citet{1996ApJ...463...95T}, who showed  that the nuclei are not coincident with bright components.     \citet{2020ApJ...899..141L} have identified the nucleus of J0131+5545 as component C0. We suggest that, as with the above three CSO 2s, the nucleus in J0131+5545 is not coincident with the bright component C0. The jet axis is almost east-west, and, using the velocities measured by \citet{2020ApJ...899..141L}, shown by the red arrows, we see that C1 is moving west with a speed of $(0.166 \pm 0.007)c$ relative to C0,    while C5 is moving west with a speed of $(0.188 \pm 0.013)c$ relative to C0.  Thus, the difference between C1 and C5 in westward motion relative to C0 is $(0.022 \pm 0.015)c$, consistent with zero. We suggest, therefore, that the nucleus lies between components C0 and C8, and  that component C0  is moving eastward relative to both the nucleus and the two outer components, C1 and C5.  Subtracting off the mean westward speed of $(0.177 \pm 0.015)c$ from the speeds derived by \citet{2020ApJ...899..141L}, shown by the red arrows, we derive the speeds shown by the green arrows.  On this interpretation, the outward motion of the lobes has stopped and the material at the outer edges of the lobes at C1 and C5 is moving almost normal to the jet axis and parallel to the interface of the lobe with the interstellar medium.

Note that on this alternative interpretation, the speeds of components C0 and C8, which we assume to straddle the central engine, are oppositely directed and of similar speed.  Thus, while components C0 and C8 might represent a newly reborn CSO 2 and a new phase of activity, as  proposed  by \citet{2020ApJ...899..141L}, in this paper we adopt the alternative explanation that these are simply high surface brightness regions of the jets, such as are commonly seen in CSO 2s. There are now more VLBI data from 2022 which may help to determine which interpretation of the motion is correct, and we are pursuing this analysis in greater detail, but it may well require VLBI astrometry to settle this question.

\subsection{Separation speed in the CSO~2.2 J1158+2540}\label{sec:J1158}
In \cref{plt:sixmaps}(f) we have over-plotted the 5\,GHz map (gray contours) and the 15\,GHz map (green contours) of J1158+2540 observed by \citet{2008ApJ...684..153T}.  The core is seen clearly in the 15\,GHz map, and is indicated by the letter ``C''.  The CSO is aligned north-south, and  \citet{2008ApJ...684..153T} found that the separation of the peaks in these components appears to be decreasing.
This is very likely a result of the``dentist's drill'' effect \citep{1974MNRAS.166..513S,2019MNRAS.482.3718P}, but it does indicate a very slow speed of advance or even that the advance has stopped.  From the morphology of this object, as seen clearly in the multi-frequency maps of \citet{2008ApJ...684..153T}, the advance does appear to have stopped while the source expands normal to the jet axis. 

\vskip 6pt \noindent
{\it Summary and conclusion:} Thus, where we can measure them, the  separation speeds of CSO 2.2s are on average less than those in CSO 2.0s. Our conclusion from this section is therefore that, based on the small number of objects for which measurements are available, the separation speeds of CSO 2s are consistent with the evolutionary hypothesis that CSO~2.0s evolve into CSO~2.2s.  Our hypotheses regarding the negligible separation speeds in J0119+3210 and J0131+5545 could be checked by astrometric VLBI.

\section{The Energies of CSO 2\lowercase{s}}\label{sec:energies}
In this section we estimate the radio emission energy requirements for CSO~2s. We first point out that the highest luminosity  CSO~2s in Fig.~1---i.e., numbers 15, 16, 17, 18, and 19---are all highly asymmetric.  It is  likely that the emission from these objects is slightly relativistically boosted, and we therefore exclude them from our energy estimates. In our selection of CSOs we deliberately specify only that emission should be seen on both sides of the nucleus, without requiring that the emitting regions straddling the nucleus have comparable flux densities.  We do this in order  not to exclude CSOs with emission regions that are mildly beamed towards the observer from our sample.  However, for the purposes of energy calculations, any objects suspected of having flux densities boosted even by factors of a few should clearly be excluded.  From Fig.~1, we see, therefore, that the most luminous CSO~2s suitable for the discussion of the energies of CSO~2s have luminosity $\sim 3 \times 10^{27}$\,W\,Hz$^{-1}$.

R94 pointed out that the energy requirements for CSO~2s are $\sim 1 M_\odot\,c^2$, and suggested that CSO 2s are formed via single star capture by a  SMBH in an elliptical galaxy nucleus. R96 carried out a detailed study of the PR CSO 2s, and estimated that the maximum energy requirement of CSO 2s is $\sim 20 M_\odot c^2$.  We show below that this estimate is a factor $\sim 2$--3 too high.  Note that these energy estimates are based solely on the radio emission regions, and do not include any energy associated with the X-ray and $\gamma$-ray emission, nor do they include any energy expended in expanding into the surrounding medium.

R96 used multifrequency global VLBI observations with continuous 12-hour tracks, so their maps of the CSO 2s in the PR sample are of as high quality as is obtainable today with VLBI.  Using their measured flux densities and angular sizes  for the model-fitted components in J0111+3906 and J2355+4950, we have recalculated the R96 equipartition energies in these objects for the $\Lambda$CDM cosmology.  The total equipartition energy requirements of J0111+3906 and J2355+4950 are given in Table \ref{tab:energies}.

The multifrequency global network VLBI maps of R96 provide the most accurate total equipartition energy estimate of a CSO 2 for J2355+4950.  We therefore develop a scaling relation relative to J2355+4950 to provide a quick method of estimating the approximate equipartition energy for any CSO~2, for which one has the peak luminosity, and the lobe size.

  The total equipartition energy, $U_{_{\rm T,eq}}  \propto L_{\rm bol}^{4/7}\, D_{\rm lobes}^{9/7}$ \citep{1990agn..conf.....B,2011hea..book.....L}, where $L_{\rm bol}$ is the bolometric luminosity and $D_{\rm lobes}$ is the physical size of the radio emitting region. Since $L_{\rm bol}$ scales with the peak luminosity, we see that $U_{_{\rm T,eq}}  \propto L_{\rm peak}^{4/7} D_{\rm lobes}^{9/7}$. The peak luminosities of J0111+3906 and J2355+4950 are $1.54\times 10^{27} \,{\rm W \,Hz^{-1}}$ and $3.95\times 10^{26} \,{\rm W\,Hz^{-1}}$, respectively.  Their mean lobe sizes are $\sim 1.5$ pc and $\sim 47$ pc, respectively.  The equipartition energy values derived from the maps and model fitting of R96, adjusted to the $\Lambda$CDM cosmology, and by the scaling relation, are shown in \cref{tab:energies}.

\begin{deluxetable}{c@{\hskip 8mm}ccc}
\tablecaption{The Energies of CSO~2s Assuming Equipartition }
\tablehead{CSO&$U_{\rm tot,eq}$&$U_{\rm tot,eq}$\\
&from maps&by scaling\\
&[$M_\odot\,c^2$]&[$M_\odot\,c^2$]
}
\startdata
J0111+3906&0.57&0.68\\
J0131+5545&0.051&0.041\\
J1205+2031&$1.5\times 10^{-4}$&$5.1\times 10^{-4}$\\
J2355+4950 &2.06&$2.06^*$\\
Highest Luminosity CSOs&-&7\\
\enddata
\tablecomments{$^*$ by definition, since the others are scaled to J2355+4950 (see text).  The energies determined from maps used the sizes and flux densities of all individual components, whereas the energies derived by scaling used generic lobe sizes, peak luminosities and the relationship $U_{_{\rm T,eq}}  \propto L_{\rm peak}^{4/7} D_{\rm lobes}^{9/7}$.}
\label{tab:energies}
\end{deluxetable}

We next consider the smallest and weakest CSO~2.2 in our sample, J0131+5545, which is shown in Fig.~1 \#41, and in Fig.~\ref{plt:sixmaps}(e). This is is a ``late-life'' CSO 2 in  which the energy supply has likely been switched off, or is rapidly decreasing, so that this object has very likely reached its energy peak.  We use the observed values of  flux densities and spectral indices from \citet{2020ApJ...899..141L}, and an estimated mean lobe size from their maps of 4.5 mas, i.e., 33 pc (see Fig.~\ref{plt:sixmaps} [e]), to obtain an observed total equipartition energy of 0.051$M_\odot \,c^2$.  

Given these observed measures of the total equipartition energies in CSO~2s, and the general agreement with the  scaling relation, we can be confident that the maximum energy observed in CSO~2s may be estimated from the  observed peak luminosity of $3 \times 10^{27}$ W/Hz, and mean lobe size of $\sim 50$ pc. Scaling this to the energy of J2355+4950, we find that the maximum total equipartition energy of CSO~2s is $\sim 7 M_\odot \, c^2$.

The least luminous CSO~2 in our sample is the CSO~2.1 object \#43 in Fig.~1, J1205+2031 (NGC 4093). This CSO 2 has been mapped with the European VLBI Network by \citet{2021MNRAS.506.1609C}. From their 5\,GHz VLBI image we have estimated the angular sizes and flux densities of the two lobes, and, assuming a spectral index of $\alpha=-0.7$ for both lobes, we obtain a total energy of $\sim 1.5 \times 10^{-4} M_\odot \, c^2$ for this object.

We see, therefore, that the energies of CSO 2s observed thus far range from $\sim 1.5 \times 10^{-4} M_\odot\, c^2$ to $\sim 7 M_\odot \, c^2$.  This refers only to the energy associated with the radio emitting regions. The total energies of CSO 2s, including the X-ray and $\gamma$-ray emission and the work done in expanding into the interstellar medium likely increases the total energy by a factor of a few.

\begin{figure*}
    \centering
    \includegraphics[width=0.8\linewidth]{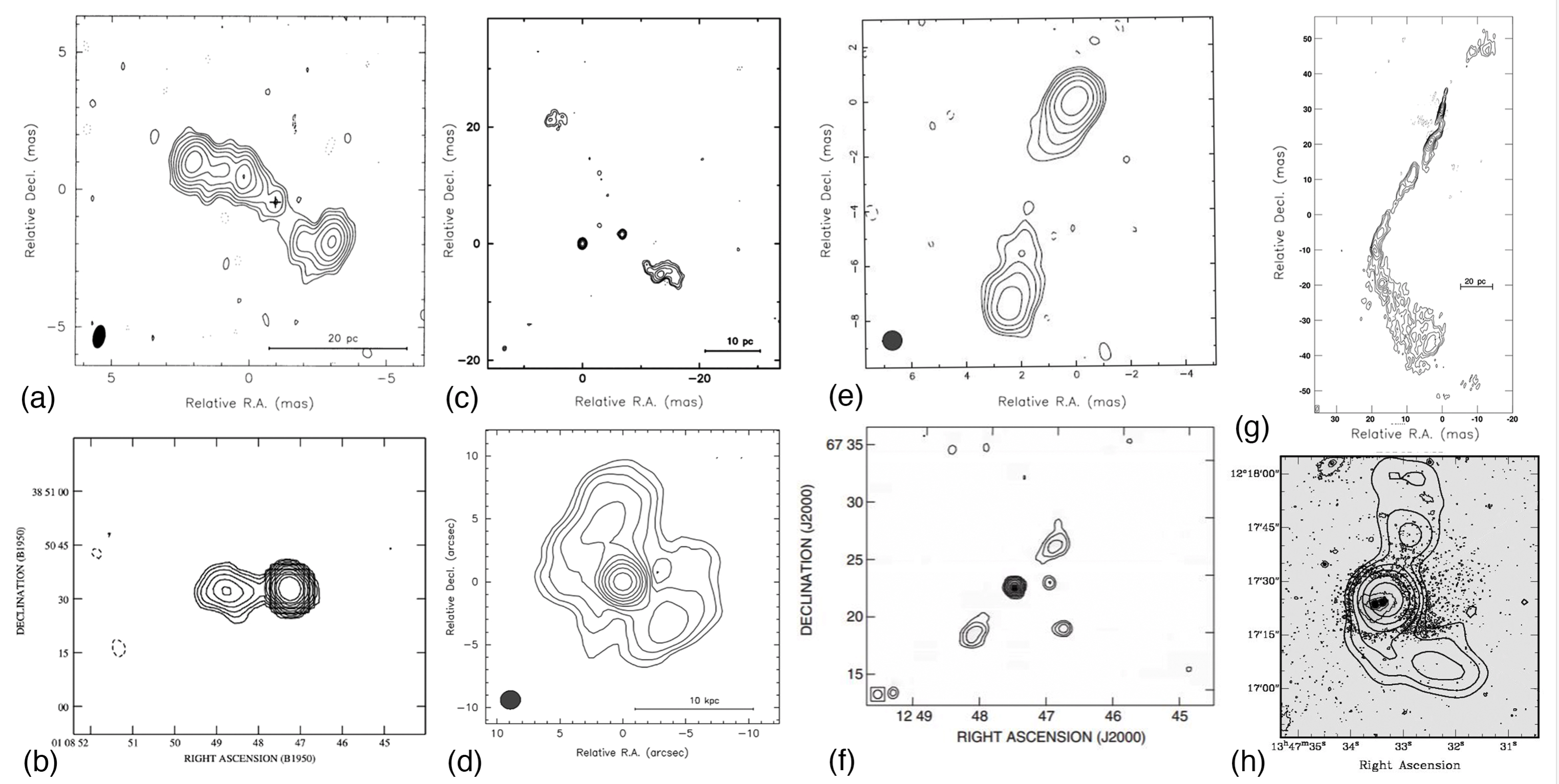}
    \caption{CSO~2s showing evidence of a previous epoch of activity. The top row shows pc-scale radio maps, and the bottom row shows radio maps of the same objects at kpc scales.  0108+388 (J0111+3906): (a) 15 GHz VLBI map \citep{1996ApJ...463...95T}, and  (b) 1.4 GHz VLA map \citep{2003PASA...20..118S}. 
   0402+379 (J0405+3803): (c) 15 GHz VLBA map \citep{2004ApJ...602..123M}, 1.4 GHz VLA map \citep{2004ApJ...602..123M}.  1245+676 (J1247+6723): (e) 15 GHz VLBA map \citep{2009AN....330..149P} and (f) 1.4 GHz VLA map \citep{2003PASA...20...16M}. 1345+12 (J1347+1217): (g) 15 GHz VLBA map  \citep{2003ApJ...584..135L} and (h) 1.4 GHz VLA map \citep{2005AandA...443..891S} }
    \label{plt:twoepochs}
\end{figure*}

\section{Multiple Epochs of Activity in CSO 2\lowercase{s}}\label{sec:multiple}

\citet{1990AA...232...19B} discovered evidence of a previous epoch of activity in the  CSO 2 J0111+3906. Thus,  in the course of the literature search described in Paper~1, we also searched for any CSOs in which there was evidence of a previous epoch of activity.  We identified these objects through gaps in the emission between the small-scale and larger-scale features, or through dramatic reduction in surface brightness between the small-scale and the large-scale structure.  Amongst the 79 bona fide CSOs we identified in Paper 1,  we found 4 bona fide CSO 2s showing evidence of previous epochs of activity, which are shown in Fig.~\ref{plt:twoepochs} and listed in Table \ref{tab:previous}. This suggests that the fraction of CSOs showing evidence of a previous epoch of activity lies in the range $\sim 5\%$ to 10\%.

A possible evolutionary model, in which multiple bursts  of activity in CSOs in particular, and in compact jetted-AGN in general, has been suggested  by \citet{1997ApJ...487L.135R}.  In their model the object shows bursts of activity lasting $10^4$yr spaced $10^5$yr apart.    \citet{2009ApJ...698..840C} considered the source evolution between repetitive outbursts and found that low Eddington accretion rates are required for short duration outbursts ($<10^3$ years).
The maximum size of the radio source in their case was dependent on the initial jet power, duration of the outburst and the properties of the interstellar medium. While these models may well apply in some CSOs, neither of them explains the sharp cutoff in the CSO 2 size distribution at $\sim 500$pc that we found in Paper 2.

\begin{deluxetable*}{llcccccl}
\tablecaption{CSO 2s Showing Evidence of a Previous Epoch of Activity }
\tablehead{B1950&J2000&Redshift&CSO&Large scale&Large scale&$\Delta$PA&References \\
Name&Name&&Size [pc]&Size [kpc]&Energy [$M_\odot c^2$]&&
}
\startdata
0108+388&J0111+3906&0.669&56.0&80&$\sim 6,600$&$\sim46^\circ$&1,2,3\\
0402+379&J0405+3803&0.05505&44.4&16&$\sim 250$&$\sim1^\circ$&4\\
1245+676&J1247+6723&0.10700&23.6&1400&$\sim 2,700$&$\sim23^\circ$&5,6,7,8,9,10,11,12,13\\
1345+12&J1347+1217&0.122&215.4&140&$\sim 320$&$\sim5^\circ$& 14,15,16,17\\
\enddata
\tablecomments{1: \citet{1990AA...232...19B}, 2: \citet{1999AaA...341...44S}, 3: \citet{2002NewAR..46..263C}, 4: \citet{2004ApJ...602..123M} 5: \citet{1992ApJS...79..331W},  6:\citet{2001AaA...370..409L}, 7: \citet{2001AaA...378..826L}, 8: \citet{2003PASA...20...16M}, 9: \citet{2004MNRAS.352..112B}, 10: \citet{2006MNRAS.366.1391S}, 11: \citet{2007MNRAS.375L..31S}, 12: \citet{2009AN....330..149P}, 13: \citet{2012ApJ...760...77A}, 14: \citet{1998AaAS..131..303S}, 15: \citet{2005AandA...443..891S}, 16: \citet{2014MNRAS.438..463O}, 17: \citet{2017MNRAS.468.4992P}}
\label{tab:previous}
\end{deluxetable*}


 We have calculated the energies of the larger components in these CSO 2s, which are listed in Table \ref{tab:previous} using the methods of the previous section.   As can be seen in the table, these energies exceed those of CSO 2s by at least a factor 30. It should be born in mind that these relics of an earlier epoch of activity are no longer being fueled and thus the energies calculated here for the larger components are likely considerably lower than the total energy associated with the earlier active phase.

 We have also measured the position angles of the innermost jetted features of the CSOs and compared these with the large-scale structure.  The differences in position angle, $\Delta$PA, are shown in Table \ref{tab:previous}.  We note that in the case of the CSO 0108+388, the jet bends from an initial $\Delta$PA $\sim 46^\circ$ towards $\Delta$PA $\sim 0^\circ$  in the north-eastern jet, and a similar bend is seen in the south-western jet.  This could be due to pressure gradients in the interstellar medium of the host galaxy.

  Because of their seemingly different origins, we conclude that there is no reason to reject the single star capture hypothesis on the basis of the evidence of previous epochs of activity in some CSO 2s. It is entirely possible that single star capture could have occurred in a galaxy that previously hosted an AGN fueled through a different means. In this scenario, jet launching during the renewed period of activity may indeed be favored by the presence of a fossil, magnetized disk \citep{Kelley2014}. A weak jet was inferred for the nearby TDE ASASSN-14li \citep{pvv18}, and a powerful relativistic jet was inferred for the candidate TDE VT\,J024345.70$-$284040.0 \citep{2023ApJ...945..142S}, both of which exhibited weak AGN activity prior to the tidal disruption events.

\section{CSO host galaxy properties}
\label{sec:hostgalaxy}

\begin{deluxetable}{c@{\hskip 8mm}ccc}
\caption{CSO 2 host galaxy stellar and black hole masses.}
\label{tab:masses}
\tablehead{CSO&$\log M_*/M_\odot$&$\log M_{\rm BH}/M_\odot$
}
\startdata
J0111+3906 & $11.58^{+0.12}_{-0.16}$ & $ 9.35 \pm 0.34 $ \cr
J0713+4349 & $11.92^{+0.11}_{-0.10}$ & $ 9.76 \pm 0.33 $ \cr
J1035+5628 & $11.91^{+0.13}_{-0.12}$ & $ 9.74 \pm 0.34 $ \cr
J1400+6210 & $11.83^{+0.05}_{-0.08}$ & $ 9.65 \pm 0.31 $ \cr
J2022+6136 & $12.18^{+0.01}_{-0.01}$ & $ 10.06 \pm 0.30 $ \cr
J2355+4950 & $11.61^{+0.08}_{-0.20}$ & $ 9.41 \pm 0.39 $ \cr
\enddata
\end{deluxetable}

The physical properties of CSO 2 host galaxies can inform our understanding of their formation channels. If CSO 2s are caused by discrete accretion events, like tidal disruption events as hypothesized by this work, we may expect that their host galaxies would show specific features. For example, TDE properties are likely to  correlate with the SMBH mass. In this section, we determine some of the physical properties of the CSO 2 host galaxies. We focus on the galaxy stellar and black hole masses. We briefly and qualitatively comment on the presence of star formation and AGN activity.

We only consider the six CSO 2s for which we have  high-quality archival optical spectra from \cite{1996ApJS..107..541L}. Our CSO 2s are sufficiently distant that, in some cases, public survey photometry alone is not sufficient to set reliable constraints on the galaxy properties, so we conservatively focus on the subsample with spectra.


We measure host galaxy stellar masses by fitting the galaxy spectral energy distributions (SEDs). We simultaneously fit the galaxy spectrum and public photometry. We adopt the spectra presented by \cite{1996ApJS..107..541L}. We mask out any regions affected by nebular emission, sky lines, or standard star features. We perform optical/IR photometry for each of these objects using images from the Pan-STARRS and WISE surveys, respectively. We use the \texttt{LAMBDAR} code \citep[][]{2016ascl.soft04003W} to obtain psf-matched aperture magnitudes, where the fiducial aperture is defined as $2.5\times$ the Kron magnitude in the Pan-STARRS $r$-band image to define the fiducial aperture size. We then simultaneously fit the photometry and spectra for each galaxy using the \texttt{bagpipes} code \citep[][]{2018MNRAS.480.4379C}. We assume a $\tau$-model star formation history, \cite{2000ApJ...533..682C} extinction, and a fixed metallicity. We do not include any AGN component. We include three additional components to enable the spectrum fit: we float the velocity dispersion, a second order polynomial calibration vector, and a white noise component to account for any underestimated systematic uncertainties. We perform the fit using recommended procedures, and report the resulting stellar masses and $1\sigma$ uncertainties in Table~\ref{tab:masses}. These masses are consistent within a factor of ${\sim}0.5$ dex with results obtained using just the photometry, and using simple mass-to-light ratio scaling relations \citep[e.g.][]{2003ApJS..149..289B}; we are confident that the stellar masses are all $\log M_*/M_\odot \gtrsim 11$.

If the observed light includes a significant contribution from AGN emission, these masses are likely overestimated. We do not expect this to be the case, however, based on the optical spectra. All the galaxy optical spectra are broadly consistent with old elliptical galaxies. The optical spectrum for J0111+3906 looks like a typical quiescent galaxy. No strong [\ion{O}{3}] emission or other emission lines are detected, and stellar absorption features consistent with old stars are visible. The remaining five CSO 2s show AGN-like emission lines (e.g., [\ion{O}{3}], [\ion{N}{2}] emission), similar to Seyfert galaxies. No obvious broad components are detected. The stellar continuum is weak for most of these five, but absorption features consistent with old stars are weakly detected in a few. Given the lack of strong broad emission lines from this subsample or other features suggesting a bright AGN continuum, we do not expect that the stellar masses are hugely overestimated.

We can convert these stellar masses to black hole masses using the bulge mass-black hole mass relation from \cite{2013ARA&A..51..511K}: 
\begin{align}
    \frac{M_{\rm BH}}{10^9\,M_\odot}= 0.49^{+0.06}_{-0.05}  \big(\frac{M_{\rm bulge}}{10^{11}\,M_\odot}\big)^{1.16\pm0.08};\nonumber \\ 
    \,{\rm intrinsic\,scatter}=0.29\,{\rm dex}.
\end{align}
We assume $M_* \approx M_{\rm bulge}$, as is appropriate for elliptical galaxies. The resulting black hole masses are reported in Table~\ref{tab:masses}. For galaxies with $M_* \gtrsim 10^{11}\,M_\odot$, the black hole masses are $M_{\rm BH} \gtrsim 5\times10^8\,M_\odot$. 

CSO host galaxies properties have previously been considered by \cite{2010ApJ...713.1393W}, who used {\it Spitzer} mid-infrared spectra and photometry to study the stellar populations, AGN activity, and black hole masses of an inhomogeneously selected sample of eight nearby CSOs ($z\lesssim 0.1$). Most of their CSOs show evidence for AGN activity and moderate star formation; one did not show detectable AGN/star-formation emission. This is qualitatively similar to the objects discussed in this section, although our data set is very different from that used by \cite{2010ApJ...713.1393W} so we cannot perform a like-to-like comparison. They find black hole masses in the range $10^{8.2-8.8}\,M_\odot$, which are significantly lower than our BH masses which are in the range $10^{9.3-10.1}\,M_\odot$. This trend may be in part caused by the higher redshifts of our objects, which are in the range $z\approx 0.2-0.7$. 

\begin{figure*}[!t]
 \centering
 \includegraphics[width=0.8\linewidth]{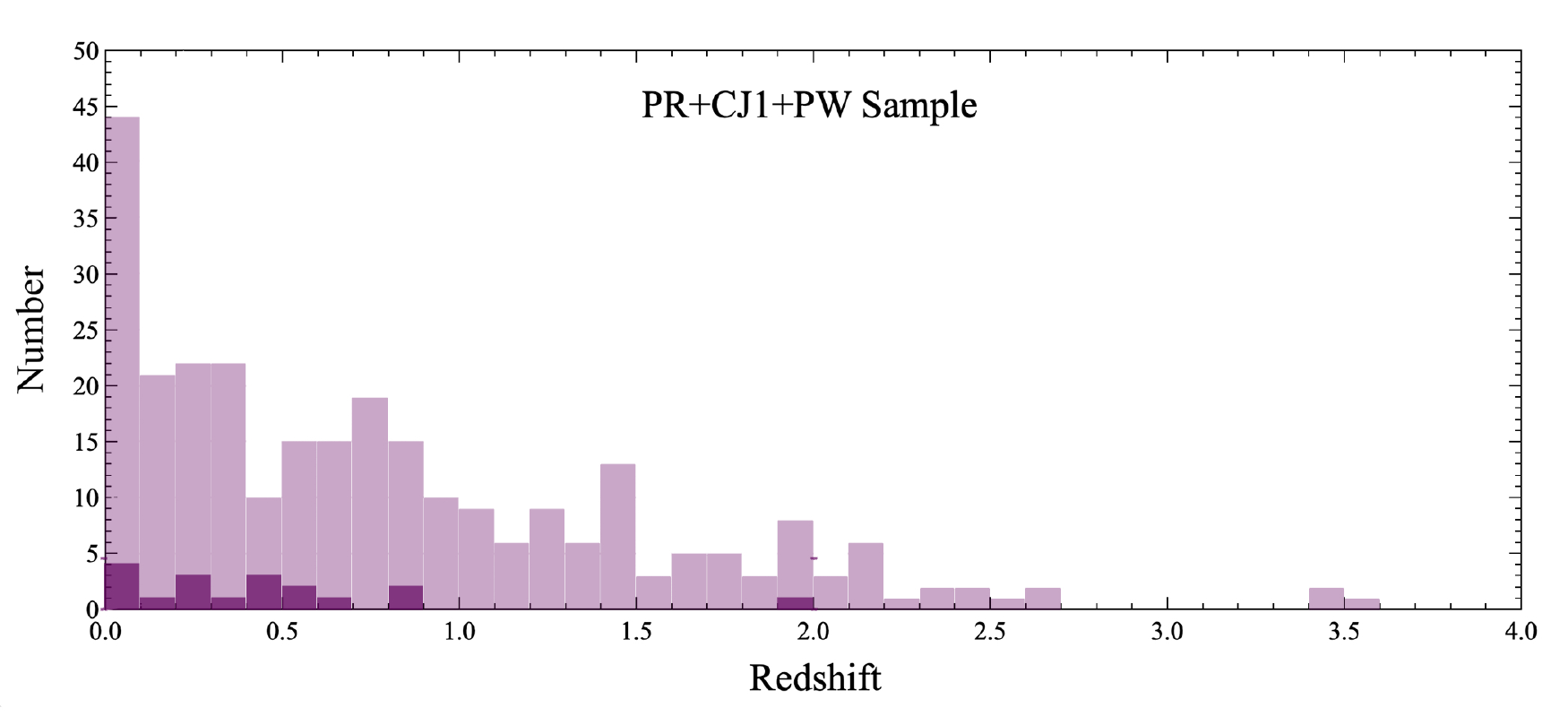}
 \caption{The redshift distribution for the PR+CJ1+PW complete sample.  The light shaded  distributions show the complete sample. The dark shaded regions show the CSO 2s.  Note that these distributions are not stacked vertically, so the values on the ordinate represent the total numbers of sources and the numbers of CSO 2s in each sample.  }
 \label{plt:histogramredshift}
\end{figure*}

More likely, much of the difference may be caused by differences in the methodology used to compute the black hole masses. \cite{2010ApJ...713.1393W} adopt a calibration that measured the relation between $V$-band bulge luminosities and black hole mass for AGN; we adopt the more recent and, arguably, robust \cite{2013ARA&A..51..511K} calibration measured for classical bulges and elliptical galaxies. It is established that the \cite{2013ARA&A..51..511K} is an upward revision of previous black hole mass relations, as discussed in that work. Moreover, as mentioned earlier, we are also using distinct data sets from \cite{2010ApJ...713.1393W}. A like-to-like comparison of our sample to \cite{2010ApJ...713.1393W} would require a detailed and uniform study of the host galaxy properties of all these objects, which is beyond the scope of this work. Instead, we think it conservative to simply state that the CSO 2s from both our sample and that of \cite{2010ApJ...713.1393W} tend to have similarly high black hole masses $\gtrsim 10^8\,M_\odot$.

\begin{figure*}
    \centering
    \includegraphics[width=1.0\linewidth]{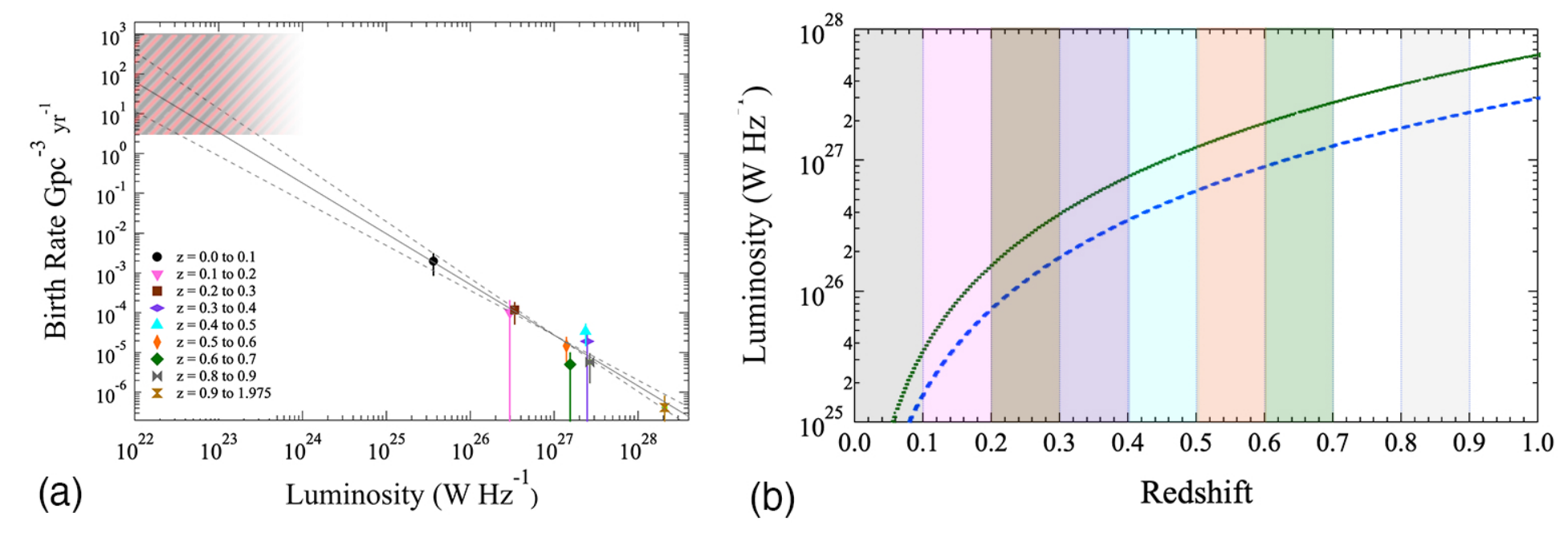}
    \caption{Dependence of birth rates of CSO 2s in the complete PR+CJ1+PW samples on luminosity. (a) The birth rates {\it vs.\/} luminosities of CSO 2s are shown by the plotted points. The solid gray line is the least squares fit to the points and the dashed lines show the $\pm 1\sigma$ ranges of the fit.  The approximate range of birthrates for jetted TDEs is  shown by the red and gray hatched region (see text).   (b) the luminosity cutoffs of the PW  (green line) and PR+CJ1 (blue dashed line) samples. The colors of the panels are faded versions of the colors in the different redshift ranges shown in (a). 
    }
    \label{plt:birthrates} 
\end{figure*}  

\section{The  Birth Rates of CSO~2\lowercase{s} and TDE\lowercase{s}}\label{sec:birthrate}

We have seen in Paper 2 that the redshift distribution of our sample of 54 CSOs is strongly affected by selection bias. By considering only the PR+CJ1+PW complete samples, we can use the CSO~2 ages derived above to estimate the birth rate of CSO~2s.  For these objects, from \cref{plt:prcj1pwages} we see that the oldest CSO~2s are estimated to have ages of $\sim 5000$~yr, assumed here to be the typical CSO 2 lifetime. 
In~\cref{plt:histogramredshift} we show the redshift distribution for the PR+CJ1+PW complete samples as well as that of the CSO~2s in these complete samples.  Apart from the outlier, J1227+3635, at $z=1.975$, we see that the CSO~2s are approximately uniformly distributed in redshift out to a redshift of $z = 0.9$.  It seems, therefore, that CSO~2s began to appear in appreciable numbers at a redshift of $z\sim 0.9$, i.e., about 7 Gyr ago.  

  We now make a very rough estimate of the birth rate of CSO 2s, which we refine in the next section.  Given the lifetime of $\sim 5000$ years, this  means that there is a probability of $\sim 5\times 10^3/7\times 10^9 \sim 7 \times 10^{-7}$ of seeing a particular CSO 2 at the present time, for the moment ignoring that fact that objects at high redshifts cannot be seen over the whole 7 billion years because of the light travel time, which we calculate correctly in the next section. The comoving volume of the universe out to redshift $z=0.9$ is  $\sim 120 \; {\rm Gpc}^3$, so the space density of CSO 2s that is required to observe one CSO 2 at the present time is $\sim 1/(7 \times 10^{-7}\times 120)\;{\rm Gpc}^{-3}=1.2 \times 10^4 \; {\rm Gpc}^{-3}$, and the birth rate, given that we observe 16 CSO 2s out to redshift $z = 0.9$, is $\sim 16 \times 1.2 \times 10^4/ 7 \times 10^9\;{\rm Gpc}^{-3}\,{\rm yr}^{-1} \sim 3\times 10^{-5}\;{\rm Gpc}^{-3}\,{\rm yr}^{-1}$.

  \subsection{The Tip of the Iceberg}\label{sec:tip}
  
  The birth rate of $\sim 3 \times 10^{-5}\;{\rm Gpc}^{-3}\,{\rm yr}^{-1}$  is based on the PR+CJ1+PW samples, which are complete down to 0.7 Jy at 5 GHz and 1.5 Jy at 2.7 GHz, respectively. Because of the steepness of the luminosity function, we are only sampling the most luminous CSO 2s in each redshift range.  The samples are therefore luminosity-limited depending on the redshift of the CSO 2s.   We have divided  16 CSOs into redshift bins of width $\Delta z = 0.1$ out to a redshift of $z=0.9$, as shown in Figs. \ref{plt:birthrates}(a) and (b), and then one bin covering $0.9<z\leq 1.975$ to accommodate the lone CSO 2 at $z=1.975$ (J1227+3635).  The colors of the panels in Fig.~\ref{plt:birthrates}(b) indicate  $\Delta z  =0.1$ bins and are diluted versions of the colors show in Fig.~\ref{plt:birthrates}(a). We calculate the CSO 2 luminosities in Fig.~\ref{plt:birthrates}(b) as the average luminosity in each bin. All bins show luminosity variations of a factor $<3$, so we do not expect this choice to significantly affect our results. In future work, with a larger CSO 2 sample, we will perform a rigorous likelihood analysis to calculate the CSO 2 luminosity function and birth rate; based on preliminary tests applying such a framework to the sample in this work, we do not expect that result to be significantly different from the one found here using this simplified calculation, so we choose to adopt this simpler, more intuitive, methodology.

  The luminosity cutoffs corresponding to 1.5\,Jy (the PW sample) and 0.7\,Jy (the PR+CJ1 sample)  are shown as a function of redshift by the green and blue dashed  curves in Fig.~\ref{plt:birthrates}(b), respectively. We see that the three CSO 2s we have observed in the redshift range $0.0<z<0.1$, which all have luminosities below $4 \times 10^{25} \;{\rm W\, Hz^{-1}}$, would have fallen below the flux density limits in the higher redshift bins. Similarly, the four CSO 2s we have observed in the redshift range $0.1<z<0.3$, which all have luminosities below $4 \times 10^{26} \;{\rm W\, Hz^{-1}}$, would have fallen below the flux density limits in the higher redshift bins.  It is clear, therefore, that in these flux-density-limited complete samples we are seeing only the tip of the iceberg, i.e., only the most luminous CSO 2s in each redshift range.

  The solid line in Fig.~\ref{plt:birthrates}(a)  is the least squares fit to the log(birthrate) {\it vs.} log(luminosity) data of the CSO 2s, and has {a slope of} $-1.27 \pm 0.15$.  The upper and lower $1\sigma$ limits on the slopes are indicated by the dotted lines in Fig.~\ref{plt:birthrates}(a).

The birthrate analysis may offer a prediction of the finite-lifetime scenario for CSO 2s. Multi-epoch wide area radio surveys such as the VLA Sky Survey \citep[VLASS;][]{Lacy2020}, the DSA-2000 \citep{2019BAAS...51g.255H},  and those made with SKA pathfinders \citep[e.g,][]{Murphy2013} should be able to detect the births of CSO 2s as radio transients. For example, VLASS is sensitive to $\sim1$\,mJy transients that rise over the six years between epochs \citep[e.g.,][]{Dong2021}. To a distance of 1\,Gpc, where the VLASS sensitivity limit corresponds to $1.2\times10^{23}$\,W\,Hz$^{-1}$, the survey should observe the birth of $\sim50$ CSO 2s. Likewise, transient searches performed by comparing the FIRST survey to VLASS are sensitive to ${\sim}1-2$ decade timescale transients, and could observe the birth of CSO 2s. D.~Z.~Dong et al. (in prep) have identified all FIRST versus VLASS transients hosted by galaxies within 200 Mpc, and found ${\sim}2$ events that are consistent with being young CSO 2s. This is consistent with expectations: given the selection criteria used in D.~Z.~Dong et al. (in prep), we would expect ${\sim}1$ new CSO 2 to be detected. Significant caveats to these predictions are that the ultimate luminosity of a CSO 2 is likely not attained near birth, and, as, we  see in Fig. \ref{plt:prcj1pwages}(a), there might be a correlation between the (peak) luminosity of a CSO 2 and its age. Nonetheless, time-domain radio surveys provide the opportunity to characterize the processes driving the births of powerful radio AGN.

\subsection{The Birth Rates and Radio Luminosities of Jetted-TDEs}\label{sec:tderate}

The  birth rate of jetted-TDEs has been shown to be weakly constrained to be (0.003--1)$\times$  the total TDE rate by \citet{2020NewAR..8901538D}, which implies a birth rate of 3--1000\,${\rm Gpc^{-3}\,yr^{-1}}$.  This range is indicated by the  red and gray hatched region in Fig.~\ref{plt:birthrates}(a). The luminosity of jetted TDEs can be estimated roughly from the observed maximum radio flux density at multiple wavelengths  of $\approx$10 mJy for AT2022cmc \citep{2022Natur.612..430A}: In a source at rest emitting isotropically,   observed distantly with a flux density $S'$, and in which the jet is continuous, the observed flux density is $S= {\cal D} ^{2-\alpha}S'$ \citep{1979Natur.277..182S}, where  ${\cal D}$ is the Doppler factor:
\begin{equation}
{\cal D} =\frac1{\Gamma(1-{\vec{\beta}}\cdot{\vec{ n}})}
\end{equation}
and $\Gamma=(1-\beta^2)^{-1/2}$ is the Lorentz gamma factor. 

Here we use $\Gamma=12$,  as estimated for AT2022cmc \citep{2022Natur.612..430A}, and we assume the angle between the jet axis and the line of sight is $\theta \sim 1/\Gamma$, at which angle $\Gamma= {\cal D} $ so that the luminosity for the off-axis TDEs is  reduced by the factor $ \Gamma^2$, assuming spectral index $\alpha=0$. We have therefore reduced the peak flux density of AT2022cmc of $\sim 10$ mJy observed in this on axis jetted-TDE to 0.07 mJy, which corresponds to a peak luminosity of $6.5 \times 10^{23}\;{\rm W \, Hz^{-1}}$ for the unbeamed jetted-TDE for comparison with the luminosities of the (unbeamed) CSO 2s.  This upper cutoff is indicated by the fading red and gray hatched bar in Fig. \ref{plt:birthrates}.

\begin{figure*}
    \centering
    \includegraphics[width=0.8\linewidth]{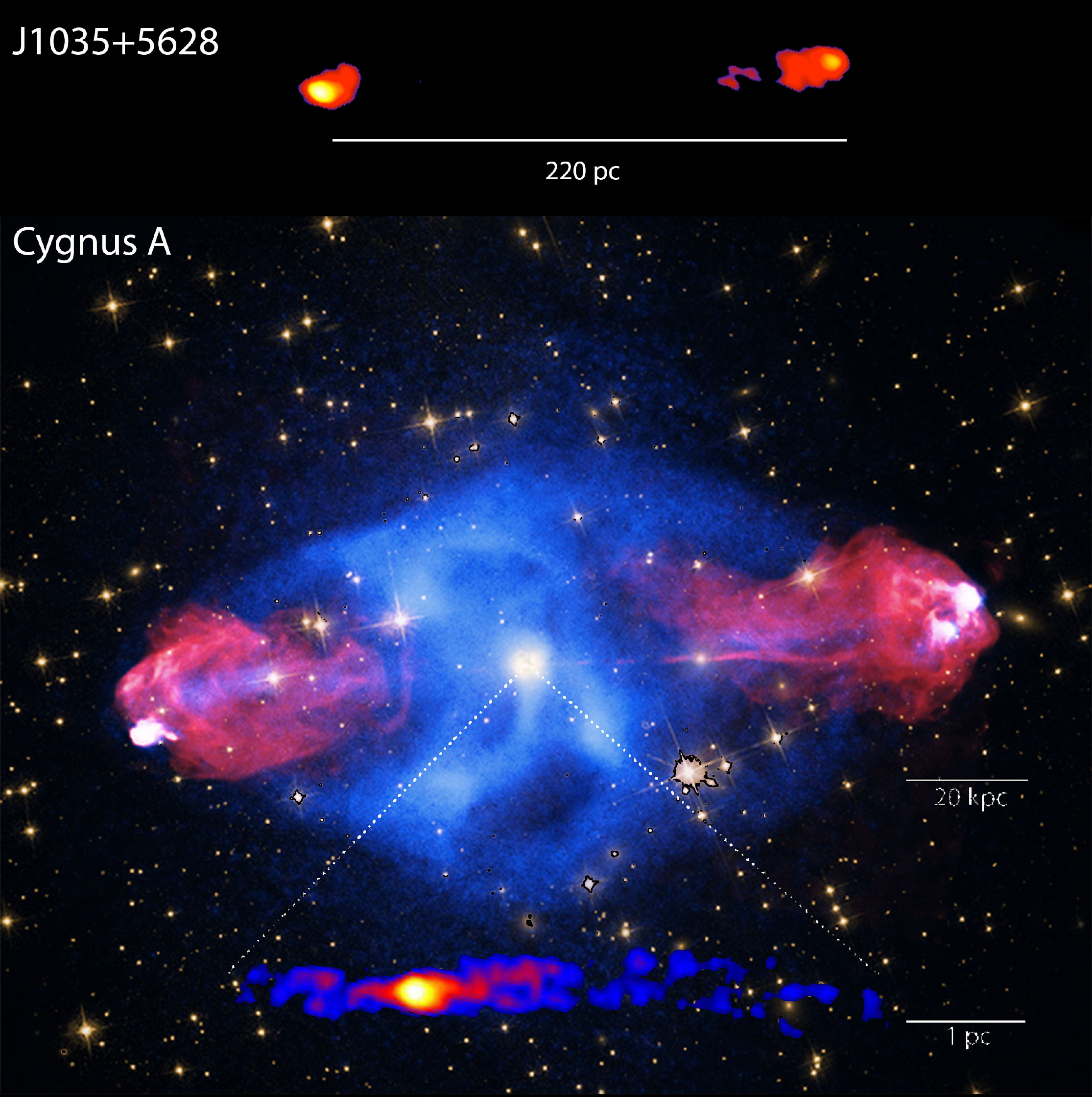}
    \caption{Upper panel: VLBA image of the CSO~2.0 object J1035+5628 \citep{2000ApJ...541..112T}. Lower panel: Composite Cygnus A image from \citet{2019ARAandA..57..467B} in which the radio emission is shown in red and the X-ray emission is shown in blue, all overlaid on the optical image.  VLBI Cygnus A inset, \citet{2017A&ARv..25....4B}.  Although the ranges of luminosities of CSO~2s and FR~IIs are indistinguishable, only $\lesssim 1\%$ of CSO~2s go on to form FR~IIs.  The remaining  $\gtrsim 99 \%$ disappear before reaching a size of $\sim$500 pc and an age of $\sim$5000  yr. }
    \label{plt:cygnusA}
\end{figure*}

\section{An Hypothesis Regarding Fuelling of CSO 2\lowercase{s} and FR II\lowercase{s}}\label{sec:coherent}

The SMBH in the central engines of FR II objects, such as Cygnus A (Fig.~\ref{plt:cygnusA} bottom panel), are continuously fueled for $\sim 10^7$ yr \citep{1991ApJ...383..554C}. We propose the hypothesis that CSO 2s (Fig.~\ref{plt:cygnusA} top panel)  are continuously fueled for timescales of years to $\sim 5 \times 10^3$ yr, and that something truncates the fuelling of CSO 2s after this time, otherwise they would, no doubt, go on to become FR IIs. This fuelling disparity between CSO 2s and FR IIs has no obvious explanation.

\begin{figure*}
    \centering
    \includegraphics[width=0.8\linewidth]{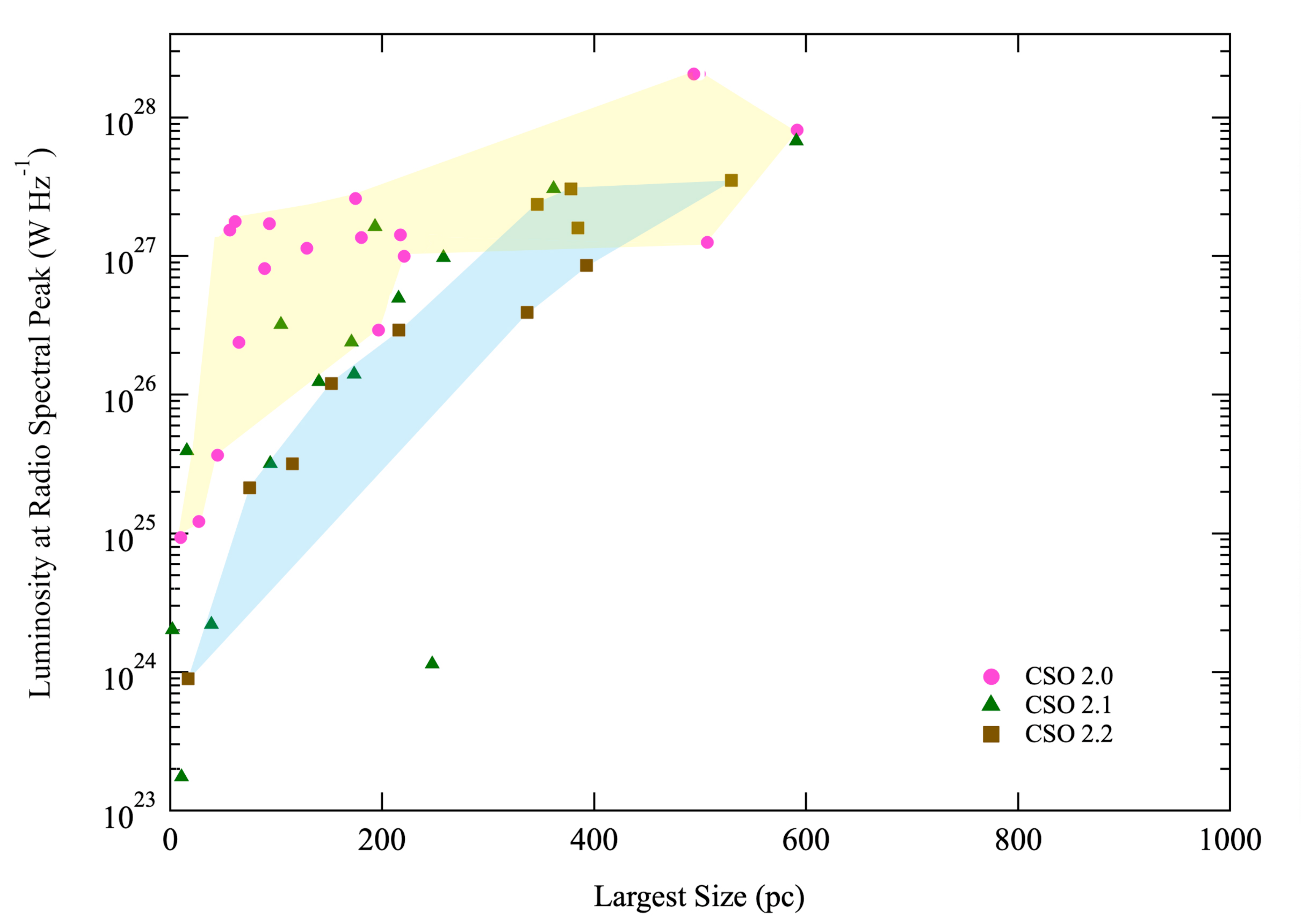}
    \caption{CSO~2s in the $(P,D)$ plane: pink circles - CSO~2.0; green triangles - CSO~2.1; brown squares  - CSO~2.2. The faint yellow and blue regions show the areas of the $(P,D)$ plane occupied by the CSO~2.0 and CSO~2.2 objects, respectively. In this log-linear plot the size cutoff and evolution from CSO~2.0 to CSO~2.2 is more clear than in the log-log plot of Fig.~1. Note the absence of any CSO 2s in the size range 600 pc -- 1000 pc.  This shows the abrupt cutoff in CSO 2 size well before reaching the nominal cutoff imposed by the CSO size selection criterion, as discussed in detail in Paper 2. }
    \label{plt:evolution}
\end{figure*}

 Paper~2 provides compelling evidence for a sharp cutoff in size of CSO~2s at $\approx 500$ pc.  This is consistent with our interpretation of the  evolutionary sequence proposed in this paper, in the sense that the evolutionary sequence, which is based on morphology alone, without reference to the sizes, suggests that an upper size cutoff must exist for CSO~2s. This relationship between evolution and size cutoff can be seen clearly in \cref{plt:evolution}, where the CSO~2.0s are generally smaller than the CSO~2.2s and of greater luminosity in the same size range.

As can be seen from \cref{plt:linsizedist} and \citet{1974MNRAS.167P..31F}, the luminosities of CSO~2s and FR~IIs  \citep{1974MNRAS.167P..31F} cover the same range from $\sim 10^{25}$ W/Hz to $\sim 10^{28}$ W/Hz. Note that these CSO~2 luminosities are at the peak frequency, which, as can be seen in Paper 1, in almost all cases lies between 100\,MHz and 10\,GHz.  However, the total energies of the most luminous CSO~2.0s are approximately $2 \times 10^4$ times less than the total energy of Cygnus A ($\sim 7 \times 10^{58}$ erg), which we have calculated based on the flux densities, angular sizes,  and spectral indices for the different components of Cygnus A given by \citet{1974MNRAS.166..305H}. So it appears that had CSO~2s simply continued at the same luminosity for $\sim 10^4 \times$ longer, they would have evolved into FR~IIs, like Cygnus A, with comparable kpc sizes and total energies.

Clearly, in CSO~2s there is a physical process operating that does not operate in FR~II objects. This unknown physical process terminates CSO~2 development as jetted-AGN after a few tens of years up to 5000\,yr or so.  As pointed out in R94, one simple way to accomplish this might  be through a single isolated fuelling event, such as stellar capture.

The fact that CSO 2s only began to appear in significant numbers about 7 billion years ago, at redshift $z = 0.9$, may support the hypothesis of a single isolated fuelling event, because prior to that time there was significantly more galaxy merging providing fuelling to AGN, which would drown out any effects due to the capture of a single star. On the other hand, galaxy mergers can also enhance the TDE rate \citep{Pfister2019}, and partial stellar disruptions may also play a significant role in fuelling SMBHs of the mass considered herein with order unity duty cycle \citep{Macleod2012}. However, self-consistent modeling of TDE rates with the SMBH mass function and star-formation histories \citep{Kochanek2016} robustly predicts that TDE rates drop precipitously with redshift for all SMBH masses, consistent with the redshift distribution of the CSO 2 population. Further, if CSO 2s of the luminosities exhibited by our sample require SMBHs as massive as those inferred above, evolution in the SMBH mass function also supports the late appearance of the CSO 2 population.

The  highly statistically significant cutoff in the size of CSO 2s discussed in Paper 2 cannot be due to episodic fuelling, because that would not produce the observed sharp cutoff in the numbers of CSO 2s at around 500 pc.  Random fuelling could produce a slow drop-off in the numbers of CSO 2s, but not the abrupt change that is observed.

\section{CSO 2\lowercase{s} and TDE\lowercase{s}}\label{sec:TDEs}

This cutoff  is clearly telling us something important about the origins and nature of this unique class of jetted AGN. It should not, of course,  be assumed that all CSO 2s have  the same origin, but Occam's razor should apply and we should consider multiple origins only when demanded by the phenomenology.  Since we are not able, as yet, to make a compelling case for the origin we here investigate different possibilities.

In the following, we first demonstrate the plausibility of \textit{some} CSO~2s being formed through TDEs, as first suggested by R94. We then compare the CSO~2 birthrate with the expected rate of TDEs for the SMBH masses we derive above, and show that it is possible that \textit{all} CSO~2s represent TDEs. We finish by identifying certain predictions of the TDE hypothesis for CSO~2s. 

\subsection{CSO 2\lowercase{s} powered by TDE\lowercase{s}}

TDEs occur when stars pass within the tidal radius, $R_{T}$, at the pericenter of their orbit around an SMBH. The tidal radius depends primarily on the stellar radius, $R_*$, the stellar mass, $M_*$, the SMBH mass, $M_{\rm BH}$, and on the internal stellar structure. Neglecting the effects of SMBH spin, the ratio of $R_{T}$ to the Schwarzschild radius is \citep[e.g.,][]{Macleod2012}
\begin{equation}
R_{T}/R_{S} \approx 1.6 m_{*,3}^{-1/3} m_{\rm BH,9}^{-2/3} r_{*,10} \, ,
\end{equation}
where $m_{*,3} \equiv M_*/3M_{\odot}$, $r_{*,10} \equiv R_*/10R_{\odot}$, and $m_{BH,9} \equiv M_{\rm BH}/10^{9}M_{\odot}$. Stars that are not massive enough or too compact will be disrupted within the event horizon, and the results will thus be invisible. 

The characteristic timescale on which post-TDE bound material falls back onto the SMBH is the fallback timescale, $t_{\rm fb}$. Defining $\beta$ as the ratio between the stellar-dynamical timescale, $\sqrt{R^{3}/GM_{\rm BH}}$, and the pericenter passage timescale ($R_{p}/v_{p}$, where $R_{p}$ is the pericenter radius and $v_{p}$ is the corresponding stellar velocity), the fallback timescale is given by \citep{1988Natur.333..523R}
\begin{equation}
t_{\rm fb}\approx 37\beta^{-3} m_{\rm BH,9}^{1/2} m_{*,3}^{-1} r_{*,10}^{3/2}\,{\rm yr}\,.
\end{equation}
Encounters with small impact parameters will have $\beta\gg1$. The fallback timescale is defined in terms of accretion of the most bound material. Following this time, the accretion rate is expected to taper according to a power-law decay, with an index of $-5/3$ expected for a flat distribution of disrupted material in orbital-energy space. Hydrodynamical simulations of TDEs find rough agreement with this scenario for a variety of stellar types, although the power-law slopes of the accretion-rate decay depend sensitively on the nature of the disruption (e.g., impact parameter) and the stellar structure \citep[e.g.,][]{Macleod2012,GR2013,Norman2021}. The maximal accretion rate can be estimated as the ratio $M_{*}/t_{\rm fb}$:
\begin{equation}
\dot{M}_{\rm max} \approx 0.08 \beta^{3} m_{\rm BH,9}^{-1/2} m_{*,3}^{2}r_{*,10}^{-3/2}\,{\rm M_{\odot}\,yr^{-1}}\,.
\end{equation}

\begin{figure}
    \centering
    \includegraphics[width=\linewidth]{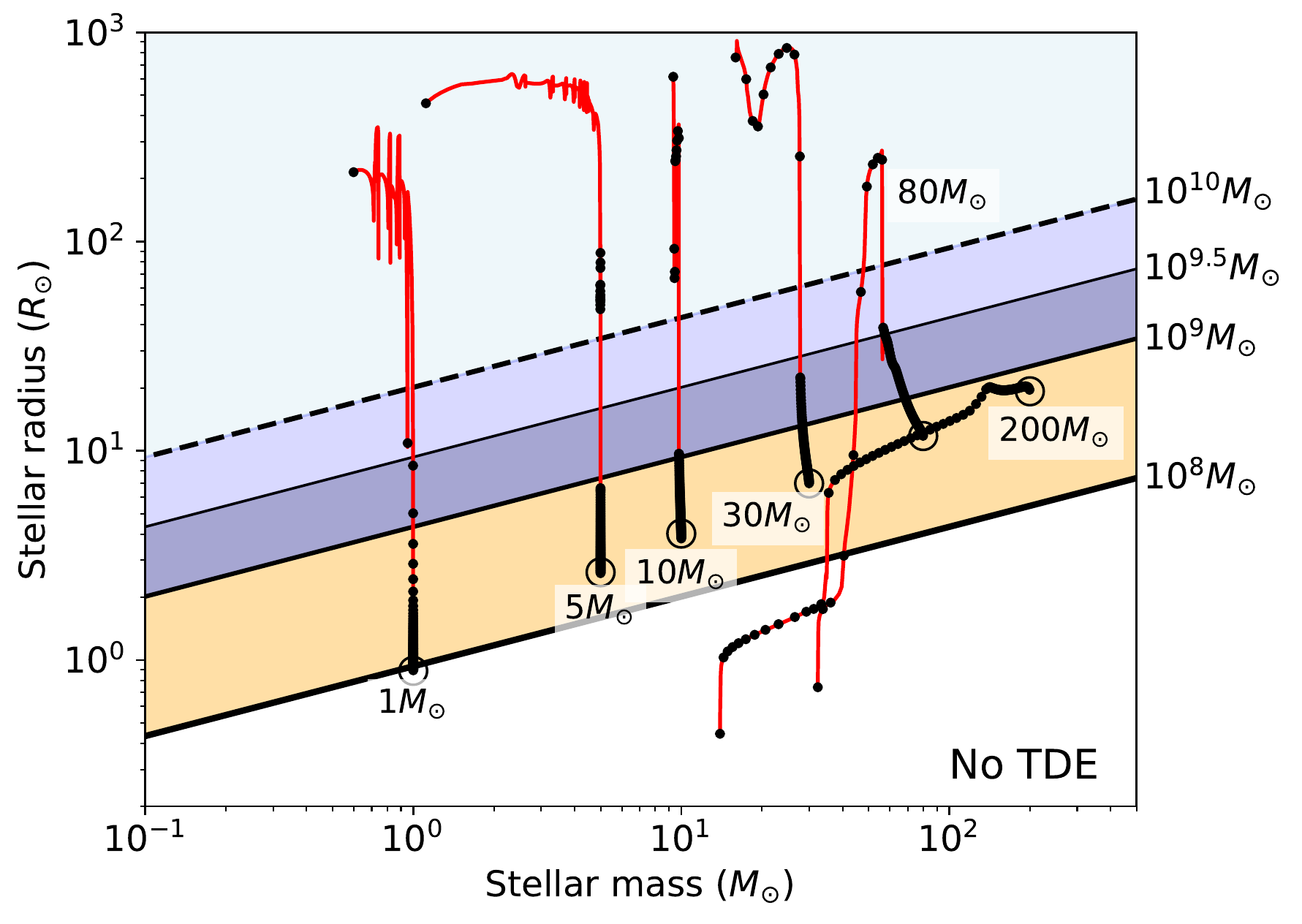}
    \caption{Regions in the stellar radius-mass space where disruption is possible outside the event horizons of SMBHs of different masses. The lower limits, corresponding to $R_{T}/R_{S}=1$ for SMBHs of various masses as labeled, are shown as lines, and the shaded area indicates where disruption is observable. We also show evolutionary tracks for solar-metallicity stars of $1M_{\odot}$, $5M_{\odot}$, $10M_{\odot}$, $30M_{\odot}$, $80M_{\odot}$, and $200M_{\odot}$ interpolated from the MESA Isochrones and Stellar Tracks \citep[MIST;][]{MIST}. The stars are evolved from the zero-age main sequence (indicated as open circles) either to the end of thermal pulsations on the asymptotic giant branch, or to supernova. Each track is garnished with black dots that indicate intervals of $1/100$ of the lifetime of each star. For example, a $1M_{\odot}$ star can be disrupted by a $10^{9}M_{\odot}$ SMBH for $\sim4/100$ of its lifetime.}
    \label{plt:rad_mass}
\end{figure}

Only massive and/or post main sequence stars can be observably disrupted by SMBHs of the $\gtrsim10^{9}M_{\odot}$ masses inferred for the CSO~2 sample herein. Figure~\ref{plt:rad_mass} shows the regions of the stellar radius-mass space where disruption is possible, together with sample stellar evolutionary tracks. The figure indicates that disruption is not possible for the most massive CSO-2 SMBHs for most of the lifetimes of most stars. We note however that the treatment of the extended envelopes of the most massive ($\gtrsim100M_{\odot}$) stars in the Modules for Experiments in Stellar Astrophysics \citep[MESA;][]{Paxton2018} may not be entirely accurate based on observations of WNh stars in the Small Magellanic Cloud \citep{Grafener2021}. The precise amount of mass donated to an SMBH in either a full or partial disruption (e.g., of just an envelope), and whether or not disruption occurs for different $\beta$ and SMBH spins, requires numerical simulations. For example, \citet{Macleod2012} performed detailed simulations of the disruption of stars with masses up to $5M_{\odot}$ and showed that $>10^{9}M_{\odot}$ SMBHs could only disrupt such stars in post main sequence phases. In general, the maximal accretion rate will be much less than the  Eddington  rate. 

The energetics and timescales of CSO~2s are consistent with expectations for TDEs around the most massive SMBHs. Any given TDE around a $\gtrsim10^{9}M_{\odot}$ SMBH is much more likely to be a post main sequence star of one to a few solar masses, rather than a $\gg10M_{\odot}$ star. Although the $80M_{\odot}$ star in Figure~\ref{plt:rad_mass} spends over half its lifetime in a region where a TDE is possible, its lifetime is only $\sim4$\,Myr, and the $5M_{\odot}$ star spends $\sim10$\,Myr in the TDE region of the plot. We now consider the minimum stellar mass for which a TDE can power a CSO~2 jet, which in a fiducial case requires $\sim1M_{\odot}c^{2}$. This can be approximately equated to the total accreted mass (see below) for the sub-Eddington mass-supply rates under consideration, and for a rapidly rotating SMBH. We note that although the relativistic TDE Swift J1644$+$57 converted just $\sim1\%$ \citep{Cendes2021} of the rest-energy of the approximately solar-mass disrupted star \citep{Sasha2014}, such events are entirely different from CSO~2s given the initial super-Eddington accretion phase and prompt jet launching. Then, as approximately a third of the mass of an evolved star will be accreted \citep{Macleod2012}, the fiducial scenario assumed above of a $3M_{\odot}$ horizontal-branch star with a radius $10R_{\odot}$ is energetically reasonable for a CSO~2. The rise and decay timescales of the accretion will be less than about 100 yr.  A somewhat more massive star in an AGB phase (AGB-star disruptions contribute a smaller fraction of the envelope mass to accretion) disrupted by a $10^{9.5}M_{\odot}$ SMBH can easily accommodate the $\sim10^{3}$\,yr durations inferred for the CSO~2s considered here. 

\subsection{Comparing the CSO 2 and TDE rates}

The CSO~2 birthrate of $\sim3\times10^{-5}$\,Gpc$^{-3}$\,yr$^{-1}$ can be compared with the empirical TDE rate of $\sim10^{3}$\,Gpc$^{-3}$\,yr$^{-1}$ found by \citet{svv2018}. A direct comparison is fraught with uncertainty. 
\begin{itemize}
    \item First, the IMF in nuclear regions, and in particular for stars in CSO~2 host galaxies on radial orbits, may be substantially different to the field IMF in the Galaxy. For example, \citet{2005MNRAS.364L..23N} have argued that the IMF in the centers of galaxies is top-heavy by at least a factor 10. \citet{2007ApJ...669.1024M} have presented compelling evidence for a long-standing top-heavy IMF in the central parsec of our own galaxy based on the work of \citet{2006ApJ...643.1011P}.  They conclude that the model that fits the observations best has IMF slope $x=-0.85$, in comparison with the \citet{1955ApJ...121..161S} IMF slope of $x = -2.35$. More recent observations on a larger sample have shown the IMF in the center of our galaxy to be extremely top-heavy with $x=-0.45$ \citep{2010ApJ...708..834B}. There is indeed evidence for the preferential disruption of $\gtrsim2M_{\odot}$ stars in nearby TDE hosts \citep{Mockler2022}. The uncertainty in the IMF needs to be taken into account, as does the prospect of evolved-star disruptions contributing significantly  to the putative CSO~2 TDEs.

    \item Second, the fraction of disruptions that result in radio emission from relativistic jets is largely unconstrained by the observations thus far. Although it is possible to consider the fraction of relativistic TDEs in relation to the overall TDE population \citep[e.g.,][]{2022Natur.612..430A}, these relativistic TDEs are potentially very different to the TDEs that could correspond to CSO~2s. The inferred accretion rates are disparate by more than four orders of magnitude, and the selection of impact parameters, SMBH spins, and pre-existing AGN activity is unknown. Additionally, evidence is mounting for delayed radio emission in a wider sample of TDEs \citep{Horesh2021,2023ApJ...945..142S}, and the full extent of this phenomenon is yet to be observed. A recent review \citep{colle_lu2020} posits that the fraction of TDEs that launch jets is largely unconstrained in the range of $3\times10^{-3}-1$.

    \item Third, evolution in the SMBH mass function, and stellar populations available for disruption, must be accounted for in comparing local TDE rates with the higher-redshift CSO~2 sample. This latter effect may result in a near order-of-magnitude decline in the total TDE rate between $0<z<1$ \citep{Kochanek2016}. 

\end{itemize}
  
We nonetheless argue that the CSO~2 birthrate is consistent with expectations from the TDE rate. For SMBH masses $\lesssim10^{8}M_{\odot}$, consistent with most TDEs observed as transients, giant stars are likely to contribute 10\% of the rate \citep{MT1999,Macleod2012}. Approximately $3\times10^{-4}$ of the SMBHs in the local universe have masses in excess of $10^{9.4}M_{\odot}$ \citep{Shankar2009}, representing the typical SMBH masses of the CSO~2 sample. The remaining factor of $10^{3}$ discrepancy can be explained by the need to disrupt the most massive evolved stars (or even more massive main-sequence stars) likely in the tips of the red giant or AGB sequences, possibly on deeply plunging orbits, and potentially in the presence of previous AGN activity and/or a rapidly spinning SMBH. Further, as indicated in Figure~\ref{plt:birthrates}, an extrapolation of the birthrates of CSO~2s is consistent with the range of relativistic-TDE birthrates inferred by \citet{colle_lu2020}, assuming a maximum luminosity consistent with observed relativistic TDEs \citep{2022Natur.612..430A}. This hints at a continuum of jet powers between the TDEs observed as transients and CSO~2s, where the abscissa is determined by several factors including SMBH properties, and the physical properties and orbits of the disrupted stars. 

\subsection{Predictions of the CSO~2 TDE scenario}

Although it is plausible that CSO~2s can be explained by TDEs around the most massive SMBHs, testable predictions of this scenario are required in order to proceed. First, evidence for significant ongoing accretion, in particular $\gg10^{-4}$ of the Eddington rate, should be absent from CSO~2s. Identifying a sample of slowly evolving nuclear transients that can be linked to evolved-star TDEs around lower-mass SMBHs can determine the corresponding rate, and enable tests of the outcomes of extended slow accretion, in particular with regards to jet launching. 

We are only seeing the highest luminosity CSO 2s in each redshift range. The lowest-luminosity CSO 2s in our complete samples have luminosities $\sim 3 \times 10^{25}\; {\rm W\, Hz^{-1}}$ but the lowest luminosity CSO 2s we have observed have luminosities $\sim 10^{23}\;{\rm W\, Hz^{-1}}$. 

As shown in Table \ref{tab:energies} with the object J1205+2031 (\# 43 in Fig.~\ref{plt:linsizedist}),  the radio energy requirement is  four orders of magnitude lower than for a fiducial CSO 2 (J2355+4950). The energy requirement is only  $\approx 10^{-4}M_\odot {\rm c^2}$, well within the energy range of current TDEs.  So it is clearly possible to observe CSO~2s that have both energy and  luminosity comparable to those of TDEs caused by $\approx 1 M_\odot$ stars.

We suggest, therefore, that these low energy CSO~2s are indeed the results of TDEs of $\approx 1 M_\odot$ stars, and that as we explore smaller CSO~2s at these low energies we will be able to form a connection with TDEs.  A number of the bona fide CSO~2s have estimated ages in the 100--200\,yr range.  It is to be expected that as we push to higher resolution and lower flux densities, and continue to apply the CSO~2 selection, we will find CSO 2s in the 10--20 yr age range. This presents the prospect of studying the initial formation events in archival survey data. 

We can push our limits on complete samples down a factor 10 by observing the steep spectrum CSOs in the sample that complements the incomplete flat spectrum VIPS sample, and we have undertaken a program to do just this.

\section{Relativistic Jet Scenarios}\label{sec:conuncon}

If the expanded classification of CSO~2s into CSO~2.0s, 2.1s, and 2.2s represents a continuum of CSO evolution, as we contend, a coherent physical model should describe their sourcing and morphology. We now return to the topic of CSO~2 ignition and evolution discussed in \S\,\ref{sec:evolution} and \S\,\ref{sec:TDEs}, and interpret it in the context of a uniquely-fueled relativistic jet propagating into an external medium.    We develop a model for CSOs in a separate paper (in preparation). For that reason, although much good work has been done on ``Young CSO'' models, we do not refer to that work here, but discuss it fully in our upcoming paper.  For the present we refer the interested reader to OS21 on this topic. 

\subsection{Fueling by Quasi-stationary Disks}

\subsubsection{Inflow Models}

The traditional way to discuss the energetics of massive black holes in galactic nuclei is to suppose that the black holes are essentially dormant until they are supplied with gas. If this happens through an orbiting disk at a steady rate $\dot M$ under the action of a local (usually magnetic) torque, then the  radiative luminosity is $L=\epsilon\dot Mc^2$, where $\epsilon$ is the radiative efficiency \citep{shakura73, balbus91}. For a thin disk, which is commonly believed to be appropriate when the mass supply rate is modest relative to the Eddington rate, $\epsilon$ is related to the binding energy of the innermost stable circular orbit, ISCO, which depends on the black hole spin, and leads to an estimate $\epsilon\sim0.05-0.4$. For $r \gg r_{\rm ISCO}$, the outward transport of angular momentum and energy by the torque contributes to the dissipation at a rate three times the local release of gravitational binding energy. Energy is conserved because a zero torque boundary condition at the ISCO leads to a deficit in the release of energy by the disk, close to the ISCO. The energy released is presumed to be radiated and outflows are supposed to be dynamically unimportant. This is now called ``Standard and Normal Evolution'' or SANE accretion \citep{curd22a}. 

When the mass supply rate is either much smaller or much larger than this, it is commonly argued that the accretion is radiatively inefficient close to the black hole and $\epsilon \ll 1$. For a low mass supply rate, it is supposed that a thick, ion pressure-supported torus forms. The ions are accompanied by much cooler electrons which have to maintain a near-Maxwellian distribution function at $\sim0.1$ of the ion temperature, despite being collisionless and in the presence of plasma turbulence. When the mass supply rate is large, the gas becomes radiation-dominated and the inefficiency follows from photon trapping. In either case, the torus' funnel supposedly traps magnetic flux in what is now called a ``Magnetically Arrested Disk'' or MAD state \citep[][and references therein]{narayan22}. If the black hole is spinning fast, a significant amount of flux threads the event horizon and two electromagnetically-powered jets are formed. As the gas flow is essentially conservative, it was, traditionally, supposed to pass through a cusp, located in the equatorial plane between the marginally stable and marginally bound orbits, before plunging, invisibly, into the horizon\footnote{More recent, MAD simulations exhibit a strong toroidal current sheet \citep{parfrey19,ripperda22}. The continued avoidance of reconnection, particle acceleration and efficient radiative emission in this current sheet is problematic.}. 

Under these circumstances, jet formation has been seen, implicitly, as a consequence of accretion with jet power efficiency $L_{\rm jet}=k\dot Mc^2$ with $k$ variously estimated to increase from $\sim0.3$ to $\sim2$ --- we adopt $k=1$ --- as the angular frequency of the black hole, $\Omega_{\rm H}$, increases from $\sim0.5\Omega_{\rm max}$ to $\Omega_{\rm max}\equiv c^3/2GM$ \citep{tchekhovskoy11}. A near-maximally rotating black hole will put roughly ten times as much power into the jets as would have been released by the accreted gas in a steady, radiatively efficient disk flow and much more than is released in a radiatively inefficient flow. On the other hand, the jet production efficiency estimated for a small sample of CSOs does not seem to reach the highest efficiency levels \citep{2020ApJ...892..116W}.

This view has been developed with the aid of powerful, general relativistic magnetohydrodynamic and particle-in-cell simulations  \citep{tchekhovskoy11, parfrey19, 2020ARA&A..58..407D}. They usually derive from initial conditions with mass and magnetic flux orbiting the black hole \citep{narayan22}. These simulations frequently exhibit winds, driven by radiation pressure, gas pressure or magnetic field, though most of the mass supplied passes through the torus to the black hole. After settling down to a quasi-steady state the disk can also exhibit local dynamo action, so that the polarity of the magnetic field threading the horizon alternates relatively rapidly \citep{2011ApJ...730...94S}. These field reversals are thought to occur on many radial scales with clear, observational implications for the jets.

\subsubsection{Outflow Models}
An alternative model of the gas flow is suggested by the EHT observations of M87 \citep{EHT19}. In its most extreme form, when the black hole is near-maximally rotating, most of the gas supplied to the outer disk is carried off by a hydromagnetic wind \citep{blandford22}. The power for driving this wind comes primarily from the spin of the black hole, not the release of gravitational energy by infalling gas\footnote{If the black hole rotational energy is not tapped, it is still energetically possible for most of the mass supplied at large radius to escape to freedom in a wind, from intermediate radius \citep{blandford99}. This can be powered, gravitationally, by a small fraction of the supplied mass, that makes it to the ISCO, and which, altruistically, sacrifices itself for this purpose. However, this arrangement is unlikely to form powerful jets and is not of interest for CSOs.}. 

In the outflow model, it is conjectured that the black hole is immersed in an ergomagnetosphere with very little gas present and that electromagnetic energy is transported radially outward from the hole, through a small scale (a ``clutch'')  or large scale (a ``capstan'') electromagnetic field, connecting the horizon to the disk, as well as parallel to the spin axis to form the jets \citep{blandford22}. The magnetic flux passes through the equatorial plane, even close to the horizon, without forming a dissipative current sheet. It is ``hemmed in'', not by the pressure of hot gas orbiting in a thick torus, as in the inflow model, but by the inertia of cold gas orbiting in a thin disk and the vertical magnetic field threading the disk, strong enough to suppress the magnetorotational instability (MRI) \citep{blandford22}. Stresses applied to this vertical field below the Alfv\'en point are communicated to the disk. The transition from magnetic to inertial dominance happens at a radius $r_{\rm ring}$, and the mass of orbiting gas in this region is $m_{\rm ring}$. Presumably, $r_{\rm ring}$ and $m_{\rm ring}$ increase with $\Omega_{\rm H}$. Rotation power is unimportant below some intermediate value of $\Omega_{\rm H}$.

Within the ring, the strength of the poloidal magnetic field must be roughly constant. Outside the ring, the magnetic field strength declines with radius, in such a way as to maintain the magnetocentrifugal wind, although the total flux threading the disk resides mostly at large radius. This wind is then responsible for the jet collimation, extending far along the jet and interacting with its surface.  Under these circumstances, the jet power should be bounded above by $\sim (\Omega_{\rm H}/2\pi\Omega_{\rm max})^2(GM/r_{\rm ring}c^2)^{-4}(m_{\rm ring}/M)$ times the gravitational power $c^5/G\sim3\times10^{52}\,{\rm W}$, with $\Omega_{\rm max}=c^3/2GM$. This allows the jet power to be much larger than ever observed, in principle. 

However, it is unreasonable to expect the gas to remain in the ring forever. Either an interchange-like instability will allow it to accrete onto the black hole, as in the inflow model, or magnetic stress will expel it as the innermost part of a wind, as in the outflow model. If we suppose that the residence time of the gas in the ring is $N$ times the dynamical time at the ring, $t_{\rm  ring}=N(r_{\rm ring}^3/GM)^{1/2}$, then we can introduce the flow rate through the ring $\dot m_{\rm ring}=m_{\rm ring}/t_{\rm ring}$, to obtain $L_{\rm jet}\sim N(\Omega_{\rm H}/2\pi\Omega_{\rm max})^2(GM/r_{\rm ring}c^2)^{-5/2}\dot m_{\rm ring}c^2$, accommodating both models. What is important is that the total jet energy produced, under either idealization, can equal, and may even exceed, the rest mass energy of the gas supplied to the ring, provided the black hole rotates rapidly.

\subsection{Ignition Models}
\label{subsec:fueling}

\subsubsection{Tidal Disruption Events}

We consider two possible explanations for the sources of CSO 2s to account for the most energetic (Table \ref{tab:energies}) and the most luminous (Fig. \ref{plt:comp1vs2}(b)) CSO 2s. The first is that the CSO 2s are associated with  black holes that spin at least modestly fast, $\Omega_{\rm H}\gtrsim0.5\Omega_{\rm max}$ and capture a single star. We introduced this model in section \ref{sec:TDEs} \citep{1988Natur.333..523R,2021AandARv..29....3O}, and elaborate here on the plausibility of this model on physical grounds. The maximum CSO 2 energies measured so far, $\sim 7\,{\rm M}_\odot c^2$, can be accommodated by a single stellar capture under either the inflow or the outflow models. When $\Omega_{\rm H}$ is large, the Lense-Thirring precession of the orbits of the debris will lead to the presence of infalling gas from many directions which, under the outflow model, is conducive to the formation of a collimated jet. Less energetic CSOs are easily explained with lower $\Omega_{\rm H}$ or partial tidal stripping.   

Models of TDEs generally suppose that stars are ripped apart on relativistic parabolic orbits and that the bound debris returns to the black hole and eventually settles into a disk at a rate $\dot M$. As previously discussed, the simplest dynamical models of this give $\dot M\propto (t/t_0)^{-5/3}$ \citep{ 1989IAUS..136..543P}, though simulations show a richer pattern of outcomes \citep{Bonnerot2020, 2021MNRAS.507.3207C}. For the inflow model, we expect the amount of mass that falls in after time $t$, will decrease $\propto t^{-2/3}$. Less than about $10^{-5} M_* $ will be available to power a CSO 2.0 after $\sim 1000\,{\rm yr}$ which seems inadequate. However, for the outflow model, mass may be cycled, through a magnetocentrifugal wind \citep{blandford82}, many times to large radius without becoming unbound so that it will fall back again. This can prolong the timescale over which a jet can be powered.

If TDEs operating under the outflow scenario are responsible for most CSO 2s, then it is reasonable to expect that CSO 2s should be accompanied by disk luminosity corresponding to a net inflow $\dot{M}\sim10^{-3}\,{\rm M}_\odot\,{\rm yr}^{-1}$ or total disk luminosity $\lesssim10^{43}\,{\rm erg\,s}^{-1}$. In addition, the central black holes should be rapidly spinning, as might be made apparent by X-ray spectroscopy of fluorescent iron lines.

More generally, many new observational capabilities are coming online and much should be learned about TDEs, reconciling radio, optical and X-ray perspectives. This will provide a better framework in which to interpret our, no less rapidly, developing understanding of CSOs. This includes the possibility that we will be able to rule out the TDE model.

\subsubsection{Disk Instabilities}

An alternative model for CSOs\footnote{As with supernovae and Gamma Ray Bursts, it would not be a surprise if the full class of CSOs involved quite different physical explanations.} supposes that the disk is thin and accretes gas at a modest rate. The outer disk may evolve to a state of instability, possibly thermal \citep{2009ApJ...698..840C}, similar to what is observed in dwarf novae \citep{1984AcA....34..161S, 1996PASP..108...39O}. Such an instability induces a sudden transition to a high-torque state when the mass inflow increases by several orders of magnitude. For example, if the steady rate of inflow is $\dot{M}\sim10^{-5}\,{\rm M}_\odot\,{\rm yr}^{-1}$, and it increases by a hundred at a radius $\sim10^{17}\,{\rm cm}$, where there is $\sim1\,{\rm M}_\odot$ of gas, then it might be possible to account the higher level of jet power needed to account for a CSO\footnote{If this model were also to explain some TDEs the leading edge of the inflow would have to steepen to form a front and not diffuse.}. 
This model might be challenged to account for the apparent upper cutoff in the CSO 2 energies. Again, disk emission might be expected. 

\subsection{Propagation into External Media}

\subsubsection{Early Life Expansion}\label{sec:propagation}

Once the jets of a CSO~2 have been launched, they propagate into the external medium in much the same manner as that of a typical FR II source. We suppose that in this stage---corresponding with the CSO~2.0 phase---the actively-launched jet collides with the external medium and shocks form at its head.  At these shocked regions, the hot spots, like those apparent in J1035+5628 in Fig.~\ref{plt:cygnusA}, form due to particle heating of the jetted material in the shocked regions, which simultaneously amplifies magnetic fields and synchrotron emission \citep{Begelman1984, 2019ARAandA..57..467B}. At the site of this hot spot, the jetted material becomes redirected upstream and inflates a cocoon bounding the jet. This cocoon will also emit synchrotron radiation and should produce the less bright radio emission surrounding the hot spots \citep{Begelman1984}.

Directing relativistic outflows out to distances $\gtrsim$ 10 pc, as seen in CSO 2s and indeed larger double radio sources, requires a large-scale collimation mechanism. Explanations of this collimation involve pure gas pressure from either the cocoon or the ambient medium \citep[e.g.,][]{2011ApJ...740..100B, 2018MNRAS.477.2128H}, or large scale magnetic fields \citep[e.g.,][]{2009ApJ...698.1570L, 2011PhRvE..83a6302L}. In the former scenario, the ambient pressure directly balances the component of the jet ram pressure projected perpendicular to the jet axis \citep{2011ApJ...738..148M}. A strong collimation shock forms at the site of the pressure balance, and redirects the jetted material along the jet axis. In magnetized jets, the hoop stress of the magnetic field assists in collimation. While these two models do not give significantly different phenomenological descriptions of jets (i.e., that the jet propagates into the external medium, inflates a cocoon to pressure confine the jet), the additional magnetic pressure  can require lower external gas pressure for collimation \citep{2014MNRAS.443.1532B}. Distinguishing between these models should be achievable observationally.

\begin{figure}
    \includegraphics[width=1.0\columnwidth]{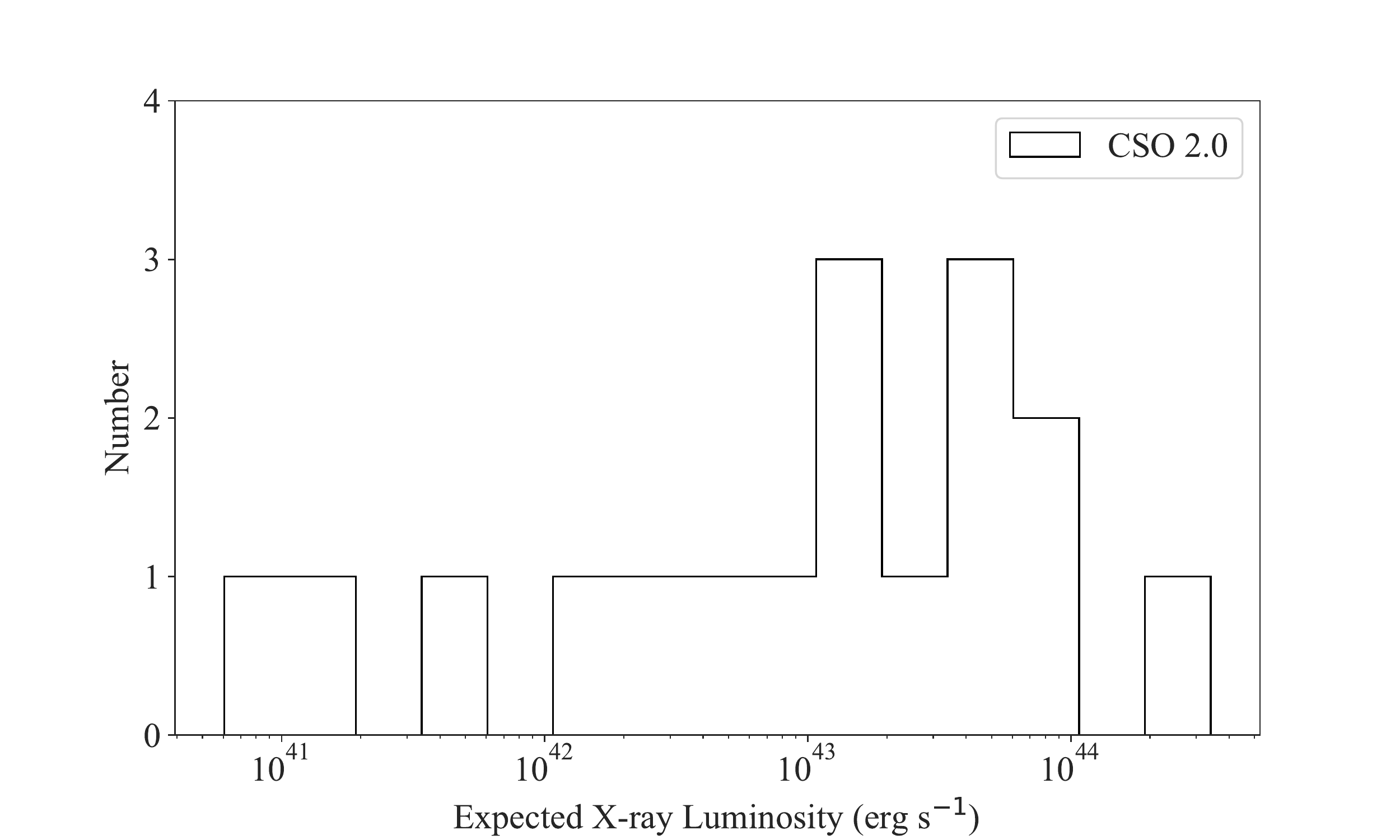}
    \caption{The expected thermal X-ray luminosities of the 17 bona fide CSO~2.0s with spectroscopic redshift if collimated by gas-pressure.}
    \label{plt:CSO2.0expectedx-rays}
\end{figure}

 Luminous thermal X-rays surrounding the jet represent one potential probe. Since in a gas-collimated jet the thermal pressure of the external medium must balance the pressure in the jet $p_j$, we expect
\begin{equation}
\label{eq:pressurebalance}
    \frac{1}{3}\frac{U_{\rm tot,eq}}{V}=n_a k_B T_a,
\end{equation}
where $U_{\rm tot,eq}$ is once again the equipartition energy of the CSOs, $V$ is the volume of the emitting region, $n_a$ is the number density of the ambient gas, $k_B$ is Boltzmann's constant, and $T_a$ is the external temperature. In Eq.~\ref{eq:pressurebalance} for pressure balance, we have assumed that the CSO 2 jet pressure is one third the average equipartition energy density. With an ambient gas temperature of $T=10^7$ K, we estimate the external particle density and consequently the luminosity of its thermal emission \citep[e.g., ][]{Rybicki1985}. We show the anticipated thermal X-ray luminosities for CSO~2.0s in Fig. \ref{plt:CSO2.0expectedx-rays}.
We expect gas-pressure-collimated CSO~2.0s to have X-ray luminosities in excess of $\sim10^{41}$ erg s$^{-1}$.  X-ray emission in excess of this threshold (although with low spatial resolution) has already been observed among some CSO~2s \citep{2007A&A...476..759B, 2016ApJ...823...57S, 2020ApJ...897..164K}, giving credence to the gas-pressure-collimated hypothesis. However, inverse-Compton emission originating in jets \citep{2008ApJ...680..911S,2022ApJ...941...52S} and thermal X-rays from the accretion disk \citep{1999PASP..111....1K,2018MNRAS.480.1247K,2022ApJ...941...52S} may also account for an X-ray excess. Although difficult, the resolution of X-ray emission from nearby CSOs at sub-kpc scales of the core could help distinguish these origins. 

Simultaneously, constraints on magnetized jets may come from additional radio polarization measurements as well as improved radio spectra. The stronger and more orderly magnetic fields in this model should induce greater degrees of magnetic polarization. Another difference between the gas pressure dominated and magnetized models lies in the advance speed of the jet head. The heads of Poynting flux-dominated jets are predicted to propagate at relativistic speeds, notably faster than their totally gas-pressure-collimated counterparts, as the lack of collimation shocks near the base of the jets does not slow the jets down \citep{2014MNRAS.443.1532B}. The subrelativisitic advance speeds observed in CSO~2.0s makes this globally magnetized jet scenario unlikely.  Further measurements of the speeds of advance of CSO 2s can constrain this model considerably. Nevertheless, these jets could still be strongly magnetized at small scales with instabilities dissipating larger scale magnetic fields, giving way to a principally cocoon gas-pressure-collimated jet at those scales \citep{2013ApJ...764..148L}. In such cases, standing transverse magnetosonic oscillations in the jet on pc scales \citep{2009ApJ...698.1570L, 2017MNRAS.465.1608L} or significant rotation measures \citep[e.g.][]{2009ApJ...694.1485K} may give credence to this model.

\subsubsection{Late Life Rise}

Motivated by the morphological structure previously described, we suppose that the transition that separates CSO~2.0s from CSO~2.2s occurs when jet fueling either severely weakens or terminates completely. While the intermediate stage between these two states remains obscure, due to the ambiguous nature of the CSO~2.1 class, we believe it possible that after a period of jet activity, the CSO 2  jet shuts off (possibly due to one of the models considered in \S\,\ref{subsec:fueling}), and the remaining downstream material in the cocoon begins to rise convectively as a turbulent plume. In this model, the rising plume phase represents the CSO~2.2 state.

Before discussing the consequences of the rising plume model, we first briefly consider our motivation for interpreting late-life CSO 2s as lacking active jets. Without active jets, they should emit less radio synchrotron emission as there is no longer a continuously supplied stream of energetic particles flowing from the core. Additionally, we also expect less X-ray emission than early life CSO 2s as the material in the radio emitting lobes will be less energetic, and no longer require the strong collimation needed in the CSO~2.0 case. Objects we classify as CSO~2.0s represent nearly twice as many X-ray sources as those we classify as CSO~2.2s in recent observations \citep{2020ApJ...897..164K}. With both lower radio luminosities and even fewer X-ray detections than CSO~2.0s, we suppose that the radio lobes of late-life CSO~2.2s may rise as plumes rather than through strong jet power.

A turbulent plume rises buoyantly as long as the density inside the plume remains less than the density of the ambient medium. This occurs subsonically as the plume both collides with the external medium at its head and entrains ambient material without forming a strong shock. If the ambient medium is convectively stable, i.e., the ambient density decreases with distance from the central engine, the plume will reach a terminal distance, at which point it begins to spread horizontally \citep{1956RSPSA.234....1M} and produces the characteristic mushroom cloud shape observed in volcanic plumes \citep{2010AnRFM..42..391W}. In fact, \cite{1956RSPSA.234....1M} showed that a plume rising in any convectively stable ambient medium will reach a maximum distance from the central engine and begin to spread horizontally. 

This picture can explain the observed behavior of CSO~2.2s, whose outward-advancing speeds are negligible and which exhibit sizable features orthogonal to the jet direction. Consider an ambient density profile of $\rho_a=\rho_0\left(\zeta_0/\zeta\right)^a$, where $\zeta$ is the distance from the center of the AGN. We may estimate the change in distance of the buoyant plume $\Delta \zeta$ from the central engine by the scale height of the ambient medium 
\begin{equation}
    \Delta \zeta\sim \left[\rho_a\left\vert\frac{d \rho_a}{d\zeta}\right\vert^{-1}\right]_{\zeta_0}=\frac{\zeta_0}{a},  
\end{equation}
where $\zeta_0$ is the distance at which the plume begins to rise. $\zeta_0$ may represent the distance from the central engine reached by the active CSO~2.0, so the maximum distance achieved by the buoyant inactive CSO~2.2 will be $\zeta_\textrm{max}=\zeta_0+\Delta \zeta$. Therefore, the maximum size of a CSO 2 should be 
\begin{equation}
D_{\textrm{max}}=\left(1+\frac{1}{a}\right)D_0,
\end{equation}
where $D_\textrm{max}=2\zeta_\textrm{max}$ and $D_0=2\zeta_0$ (since total CSO 2 size contains both lobes). Setting $D_0$ as the median CSO~2.0 size $129$ pc and using a fiducial value for $a$ of 2 \citep{2022ApJ...928....7W}, we obtain an expected size of CSO~2.2s of 200 pc, about a 40\% difference from the observed median size of 340 pc. A less steep profile of $a=0.75$ gives an expected size of CSO~2.2s of 300 pc, nearly an exact match to the observed median size. (On much longer timescales, it is possible that these plumes might evolve to very faint bubbles, analogous to those seen in rich clusters of galaxies.) Consequently, this simple pictures comports with observed CSO 2 properties.

\section{Future Observations}\label{sec:future}

If the size statistics discussed in Paper 2 hold once the steep spectrum counterpart of the VIPS sample has been observed, then a highly significant new avenue for the study of jetted-AGN has been discovered. 
  The follow-up of these results with VLBI observations of larger, deeper, samples of CSO 2s, including at higher frequencies so as to include smaller CSO 2s, can address, in both the time domain and the structure dimensions, questions on the fuelling, launching, and propagation, of relativistic jets, in ways that were previously inaccessible.

From the statistics of CSOs in the VIPS and PR+CJ1+PW samples (Paper 1) we estimate that over the whole sky there are $\sim 300$—$1000$ CSOs with $S_{\rm 5\,GHz} \gtrsim 85$/mJy that will be amenable to study with current and planned instruments, greatly improving the power of statistical tests. It will also be possible to look for even fainter CSOs in selected regions to probe the population of low-luminosity CSOs. 

Current surveys include the Very Large Array Sky Survey (VLASS) \citep{Lacy2020} which covers declination $> -40^\circ$  at 2--4 GHz with a resolution of 2.5 arcsec, and detects sources brighter than 1\,mJy. With its three epochs of observation, it will provide a list of compact, low-variability sources that can be followed up with VLBI. In addition the MEERKAT \citep{2016mks..confE...1J} and ASKAP \citep{2008ExA....22..151J} instruments provide powerful CSO search capabilities in the southern hemisphere.

Future instruments  will also find more CSOs. The proposed Deep Synoptic Array (DSA-2000) \citep{2019BAAS...51g.255H}, covering frequencies 0.7--2.0 GHz and 4 month cadence over 16 epochs will provide an ideal complement to the VLASS. The sensitivity of 2 $\mu$Jy/beam at each epoch would enable sensitive monitoring of variability of CSOs down to 1 mJy in total flux density, or deeper if necessary. 

To improve our understanding of the astrophysics of CSOs, deeper and higher-resolution observations will be needed. We can increase the resolution and sensitivity by going to higher frequencies (45 GHz and 90 GHz), using global VLBI arrays that incorporate large antennas, such as the Green Bank Telescope, the Effelsberg Telescope, Deep Space Network antennas, and the Large Millimeter Telescope. In the longer term, the proposed Next Generation Very Large Array (ngVLA) will be much more sensitive than the VLBA, and its large bandwidth will allow, for example, all of the 3.5--12.3 GHz band to be covered simultaneously. The rms noise in a 1-hour observation is expected to be about 3.2 $\mu$Jy/beam at 8 GHz, and an order of magnitude lower at higher frequencies, for a 1 mas beamwidth.\footnote{\url{https://ngvla.nrao.edu/page/performance}}

To exploit the full potential the class of CSO 2s offers for understanding the origins and astrophysics of jetted-AGN, we will need several orders of magnitude improvement in dynamic range, and a factor $\sim4$ improvement in resolution. This will require extending the VLBI arrays (such as the ngVLA) to near earth orbit (NEO). A single 10m antenna in NEO, operating up to 90\,GHz, would allow observations of  300 sources in 18 months down to a flux density limit of $<85$\, mJy at 5 GHz.With a 6 year mission lifetime, i.e., four eighteen-month cycles, then, separation speeds down to $0.03c$ could be measured with the $(u,v)$ coverage shown in  Appendix D. 

\section{Conclusion}\label{sec:conclusion}

We have examined a carefully-selected sample of 54~bona fide CSOs for which we have spectroscopic redshifts. We have defined three new morphological classes of CSOs: CSO~2.0, CSO~2.1, and CSO~2.2.  We have classified the 54~CSOs in our sample into these three CSO~2 classes and the CSO~1 class, using blind tests.  

We find it remarkable that, 
 in spite of the very wide variety of morphologies displayed by CSOs, it is possible to classify them into just four morphological classes. We also find the agreement in our blind tests noteworthy.

 The resulting classifications of CSO~2s, and their different positions in the $(P,D)$ plane can be explained on the simple hypothesis that there is an evolutionary sequence ${\rm CSO} \;2.0 \rightarrow {\rm CSO}\; 2.1 \rightarrow {\rm CSO} \;2.2$, with CSO~2.0s being ``early-life'', CSO~2.1s ``mid-life'', and CSO~2.2s ``late-life'' CSOs.  
 
 Such an evolutionary sequence provides an explanation for the size cutoff of CSO 2s that was found in Paper~2, and is consistent with all that is known about the phenomenology of CSO 2s. The theory could be tested by carrying out a systematic study of the speeds of advance of CSO~2.2s compared to those of CSO~2.0s and also by looking for systematic differences in ages between these two classes.
 
 The origins of CSO 2s and the reasons for their relatively short  lives as jetted-AGN will be the subjects of continuing studies. At this stage, none of the possibilities we have considered for ignition, including tidal disruption events and MHD instabilities in the accretion disk, and  the extraction of the spin energy of the SMBH can be definitively ruled out.  It is entirely possible that multiple mechanisms play a role in CSOs  since it is clear that CSOs constitute not so much a ``class'' but more of a ``family'' of AGN that is, as yet, relatively unexplored, and has the potential for studying AGN and relativistic jets that will lead well beyond what we can imagine at this stage.

 All of these considerations make it clear  that CSOs provide a unique and  vastly under-unexploited window into  the study of SMBHs, their accretion disks, and the birth of relativistic jets. This amounts to a new, and compelling, science driver for both the DSA-2000 and the ngVLA. The multi-epoch DSA-2000 surveys will have the sensitivity to dig down to very low luminosity to provide samples of weak CSOs.  The ngVLA, operating up to 90 GHz in conjunction with the existing global VLBI network, where needed, will provide studies of CSOs that could revolutionize the study of AGN and, by extension, other sources known to form relativistic jets such as X-ray binaries, Gamma Ray Bursts and pulsars. These include the creation, launching and fuelling of relativistic jets, and, most importantly,  whether the existing paradigm for the fuelling of relativistic jets, via accretion, is always correct, or whether their energy can derive from the spin of the black hole.

\begin{acknowledgments}

 We thank Martin Rees for discussions that led to a significant improvement in our interpretation of these results.
We thank Christopher O'Dea for providing the data on the large-scale structure of 0108+388.  We thank the reviewer of this paper for many helpful suggestions that have clarified several important aspects of this work.
This research has made use of data from the MOJAVE database that is maintained by the MOJAVE team \citep{MOJAVE_XV}. The MOJAVE program is supported by NASA-Fermi grant 80NSSC19K1579.
S.K. and K.T. acknowledge support from the European Research Council (ERC) under the European Unions Horizon 2020 research and innovation programme under grant agreement No.~771282.
KT acknowledges support from the Foundation of Research and Technology - Hellas Synergy Grants Program through project POLAR, jointly implemented by the Institute of Astrophysics and the Institute of Computer Science.
A.S. acknowledges support from the NASA contract NAS8-03060 (Chandra X-ray Center).  GT and ES acknowledge support from NSF Grant AST-1835400.
N.G.’s research is supported by the Simons Foundation, the Chancellor Fellowship at UCSC and the Vera Rubin Presidential Chair.

This paper depended on a very large amount of VLBI data, almost all of which was taken with the Very Long Baseline Array.  The National Radio Astronomy Observatory is a facility of the National Science Foundation operated under cooperative agreement by Associated Universities, Inc.

 In recognition of his many important contributions to astrophysics and cosmology, and in particular of his work on jetted-AGN, this paper is dedicated to Mark Birkinshaw. Mark was working hard on this paper right up until his death from cancer on 23 July 2023. Mark is sorely missed world-wide by his colleagues and friends.
\end{acknowledgments}
\newpage
\appendix

\section{Table of Bona Fide CSOs in Right Ascension Order}
This Table has the same information as that in Table \ref{tab:bonafides}, but 
 with the sources arranged in order of right ascension.

\begin{deluxetable*}{ccDl@{\hskip5mm}|@{\hskip5mm}ccDl}[h]
\tablecaption{Key to the CSOs in \cref{plt:linsizedist} in Right Ascension order \label{tab:bonafides2}}.
\decimals
\tablehead{Source Name 	&	ID \# 	&	\multicolumn2c{$z$}&	Class	&Source Name	 & ID \#	&	\multicolumn2c{$z$}&	Class}
\startdata
J0029+3456	&	8	&	0.517	&	CSO~2.0	&	J1311+1658	&	38	&	0.081408	&	CSO~1	\\
J0111+3906	&	3	&	0.668	&	CSO~2.0	&	J1313+5458	&	20	&	0.613	&	CSO~2.2	\\
J0119+3210	&	35	&	0.0602	&	CSO~2.2	&	J1326+3154	&	13	&	0.37	&	CSO~2.2	\\
J0131+5545	&	41	&	0.03649	&	CSO~2.2	&	J1347+1217	&	28	&	0.121	&	CSO~2.2	\\
J0405+3803	&	53	&	0.05505	&	CSO~2.0	&	J1400+6210	&	15	&	0.431	&	CSO~2.2	\\
J0713+4349	&	11	&	0.518	&	CSO~2.0	&	J1407+2827	&	52	&	0.077	&	CSO~2.1	\\
J0741+2706	&	9	&	0.772137	&	CSO~2.1	&	J1414+4554	&	36	&	0.186	&	CSO~2.1	\\
J0825+3919	&	18	&	1.21	&	CSO~2.1	&	J1434+4236	&	22	&	0.452	&	CSO~2.2	\\
J0832+1832	&	54	&	0.154	&	CSO~1	&	J1440+6108	&	30	&	0.445365	&	CSO~2.1	\\
J0855+5751	&	39	&	0.025998	&	CSO~2.1	&	J1508+3423	&	33	&	0.045565	&	CSO~2.1	\\
J0906+4124	&	46	&	0.027	&	CSO~1	&	J1511+0518	&	51	&	0.084	&	CSO~2.0	\\
J0909+1928	&	44	&	0.027843	&	CSO~1	&	J1559+5924	&	47	&	0.0602	&	CSO~1	\\
J0943+1702	&	10	&	1.601115	&	CSO~2.0	&	J1602+5243	&	26	&	0.105689	&	CSO~1	\\
J1025+1022	&	40	&	0.045805	&	CSO~1	&	J1609+2641	&	14	&	0.473	&	CSO~2.1	\\
J1035+5628	&	12	&	0.46	&	CSO~2.0	&	J1644+2536	&	23	&	0.588	&	CSO~2.1	\\
J1111+1955	&	1	&	0.299	&	CSO~2.0	&	J1723-6500	&	49	&	0.01443	&	CSO~2.1	\\
J1120+1420	&	21	&	0.362	&	CSO~2.0	&	J1734+0926	&	6	&	0.735	&	CSO~2.0	\\
J1148+5924	&	42	&	0.01075	&	CSO~1	&	J1735+5049	&	4	&	0.835	&	CSO~2.0	\\
J1158+2450	&	32	&	0.203	&	CSO~2.2	&	J1816+3457	&	31	&	0.245	&	CSO~2.1	\\
J1159+5820	&	17	&	1.27997	&	CSO~2.0	&	J1915+6548	&	25	&	0.486	&	CSO~2.1	\\
J1205+2031	&	43	&	0.024037	&	CSO~2.1	&	J1939-6342	&	7	&	0.1813	&	CSO~2.0	\\
J1220+2916	&	45	&	0.002	&	CSO~1	&	J1944+5448	&	29	&	0.263	&	CSO~2.0	\\
J1227+3635	&	16	&	1.975	&	CSO~2.0	&	J1945+7055	&	37	&	0.101	&	CSO~2.2	\\
J1234+4753	&	34	&	0.373082	&	CSO~2.1	&	J2022+6136	&	2	&	0.227	&	CSO~2.1	\\
J1244+4048	&	19	&	0.813586	&	CSO~2.2	&	J2203+1007	&	5	&	1.005	&	CSO~2.0	\\
J1247+6723	&	50	&	0.107219	&	CSO~2.0	&	J2327+0846	&	27	&	0.02892	&	CSO~1	\\
J1254+1856	&	48	&	0.1145	&	CSO~1	&	J2355+4950	&	24	&	0.238	&	CSO~2.2	\\							
\enddata
\tablecomments{The ID \# is the reference number  in \cref{plt:linsizedist}.    References for the redshifts are given in Paper 1. }
\end{deluxetable*}
\clearpage

\section{Individual Classification Notes \\
for the 54~bona fide CSOs that have spectroscopic redshifts}\label{sec:notes}

Numbers in [square] brackets are those assigned to the objects in \cref{plt:linsizedist},  \cref{tab:bonafides}, and \cref{tab:bonafides2}.
Dagger$^\dagger$ indicates the reference to the map displayed in \cref{plt:linsizedist}.

\vskip 6pt
\noindent
J0029+3456, CSO~2.0 [8]: Maps at 5\,GHz \citep{2000ApJS..131...95F}$^\dagger$ and 15\,GHz (MOJAVE stacked epoch) show clear hot spots at the opposite outer edges of the lobes.
\vskip 6pt
\noindent
J0111+3906,  CSO~2.0 [3]:  \citet{1996ApJ...463...95T}$^\dagger$, using multi-frequency, multi-epoch observations, pinpointed the flat-spectrum center of activity - marked by the red cross in \cref{plt:linsizedist}[3], and provided compelling evidence of steep-spectrum hot spots situated at the outer edges of the lobes straddling the nucleus. This was the first CSO to be shown to have some faint large-scale structure, and thus to have had previous activity, which established the existence of multiple epochs of activity in some CSOs \citep{1990AA...232...19B}.
\vskip 6pt
\noindent
J0119+3210,  CSO~2.2 [35]: The 5\,GHz map by \citet{2003AandA...399..889G}$^\dagger$ shows weak hot spots in a highly resolved lobes perpendicular to the source axis.   \vskip 6pt
\noindent
J0131+5545, CSO~2.2 [41]: The stacked epoch MOJAVE$^\dagger$ image shows two highly resolved, edge-brightened lobes with no hot spots,
\citep{2020ApJ...899..141L}, making this is a CSO~2.2 object.  \citet{2020ApJ...899..141L} interpret this and the emission gap between the lobes and bright inner jet as evidence of two separate epochs of activity in this object, which may be correct, but since compact components are seen along the jets of many CSO 2s, and since we interpret the structure as showing two such compact components (see \S \ref{sec:J0131}, and Fig.~\ref{plt:sixmaps} (e)), we assume here that this is a normal CSO and not an object displaying two separate epochs of acitvity.
\vskip 6pt
\noindent
J0405+3803, CSO~2.0 [53]: The multifrequency  observations of \citet{2004ApJ...602..123M}$^\dagger$ show clearly that there are hot spots at the outer extremities of the inner lobe structure.   In addition, as these authors show, there are two active nuclei in this object, so this is also a rare example of a binary supermassive black hole (SMBH). This  CSO  exhibits more than one epoch of activity \citep{2004ApJ...602..123M}.
\vskip 6pt
\noindent
J0713+4349, CSO~2.0 [11]: \citet{1984IAUS..110..131R} showed that this object has a flat spectrum core, a steep spectrum jet, and two steep spectrum lobes. \citet{1996ApJ...463...95T} pinpointed the center of activity and showed that there are hot spots at the extremities of the lobes. The image shown is a MOJAVE$^\dagger$ stacked image.
\vskip 6pt
\noindent
J0741+2706, CSO~2.1 [9]: \citet{2016MNRAS.459..820T}$^\dagger$ observed this object at 5\,GHz and 8\,GHz, and identified the flat spectrum core and two steep spectrum lobes.  The object is edge-dimmed - i.e., the regions of highest surface brightness outside of the core are closer to the core than to the outer edges of the lobes.  There are no clear hot spots. The blind tests were split equally between a 2.1 and a 2.2.  This shows that the classification is indeterminate, and hence this is a CSO~2.1 object.
\vskip 6pt
\noindent
J0825+3919, CSO~2.1 [18]:  In view of the absence of dominant hot spots at the outer edges of the lobes \citep{2004AandA...426..463O}$^\dagger$, we classify this as a CSO~2.1. 
\vskip 6pt
\noindent
J0832+1832, CSO~1.0 [54]: Like J0741+2706, \citet{2016MNRAS.459..820T}$^\dagger$ observed this object at 5\,GHz and 8\,GHz, and identified the flat spectrum core and two steep spectrum lobes.  The object is edge-dimmed - i.e., the regions of highest surface brightness outside of the core are closer to the core than to the outer edges of the lobes.  There are no clear hot spots. The edge-dimmed morphology makes this a CSO~1  object.
\vskip 6pt
\noindent
J0855+5751, CSO~2.1 [39]: \citet{2005ApJS..159...27T}, in pilot VIPS 5\,GHz observations, identified this object as a CSO based on the morphology of the two lobes and the hot spots in the lobes. \citet{2016MNRAS.462.2819B}$^\dagger$ re-reduced the data and their map makes clear that the northern hot-spot is at the outer extremity of the lobe, whereas  the southern hot-spot is not at the outer extremity of the lobe.  This discrepancy in the morphologies of the two lobes leads to the classification as ``intermediate'', or CSO~2.1.
\vskip 6pt
\noindent
J0906+4124, CSO~1 [46]: The EVN images at 5\,GHz and 8\,GHz by \citet{2021MNRAS.506.1609C}$^\dagger$ show a flat-spectrum core flanked  by two oppositely directed edge-dimmed jets in this low-luminosity CSO~1 object.
\vskip 6pt
\noindent
J0909+1928, CSO~1 [44]: The VLBA image at 8.4\,GHz by \citet{2018ApJ...863..155C}$^\dagger$ shows a strong core flanked  by two oppositely directed edge-dimmed jets in this low-luminosity CSO~1 object.
\vskip 6pt
\noindent
J0943+1702, CSO~2.0 [10]: \citet{2016MNRAS.459..820T} identified this as a CSO based on 5\,GHz and 8\,GHz observations, which showed a clear flat-spectrum core and two steep spectrum lobes. The lobes are barely resolved in  the \citet{2016MNRAS.459..820T} maps, but the MOJAVE$^\dagger$ 15\,GHz observations show clear hot spots at the outer edges of the lobes, making this a clear CSO~2.0 object. This is one of only three CSOs whose classification was revised after the blind tests\ddag. The CSO~2.0 classification was agreed unanimously in a group discussion of how the MOJAVE image was embedded in the 5\,GHz image and the positions of the 15\,GHz hot spots at the outer edges of the 5\,GHz lobes.

\vskip 6pt
\noindent
J1025+1022, CSO~1 [40]: The VLBA image at 8.4\,GHz by \citet{2018ApJ...863..155C}$^\dagger$ shows a strong core flanked  by two oppositely directed edge-dimmed jets in this low-luminosity CSO~1 object.
\vskip 6pt
\noindent
J1035+5628, CSO~2.0 [12]: \citet{1996ApJ...463...95T} made 8.4\,GHz and 15\,GHz observations of this Pearson-Readhead \citep{1988ApJ...328..114P} CSO  and identified its flat spectrum core and two leading-edge brightened lobes, which are seen clearly in the MOJAVE$^\dagger$ 15\,GHz image.
\vskip 6pt
\noindent
J1111+1955, CSO~2.0 [1]: The 8.4\,GHz image of \citep{2000ApJ...534...90P}$^\dagger$ shows clear hot spots at the outer edges of the two lobes. \citet{2005ApJ...622..136G} confirmed this as a CSO based on the steep spectra of the edge-brightened outer lobes.
\vskip 6pt
\noindent
J1120+1420, CSO~2.0 [21]: \citet{1998MNRAS.297..559B}$^\dagger$ identified this as a CSO based on 23\,GHz MERLIN and 1.6\,GHz VLBA observations.  In the Radio Fundamental Catalog\footnote{http://astrogeo.org/rfc/} (RFC) maps, this object has clear hot spots at the outer edges of the lobes, making this a CSO~2.0 object.
\vskip 6pt
\noindent
J1148+5924, CSO~1 [42]: The \citet{2016MNRAS.459..820T} 5\,GHz and 8\,GHz maps show this to be an edge-dimmed object, as is also seen in the MOJAVE$^\dagger$ image.  The morphology and low luminosity indicate this is a CSO~1 object. This object shows evidence of a previous epoch of activity.
\vskip 6pt
\noindent
J1158+2450, CSO~2.1 [32]: A very unusual object, as revealed in the 5\,GHz, 8\,GHz and 15\,GHz maps of \citet{2008ApJ...684..153T}$^\dagger$.  The 5\,GHz map shows an apparent east-west double source, but the 15\,GHz map shows that the eastern component has a north-south CSO embedded in the slightly larger east-west structure. The 8--15\,GHz spectral index map of \citet{2008ApJ...684..153T} clearly identifies the flat spectrum core and steep spectrum lobes in the north-south structure in the eastern component, making this a clear CSO~2.2.  This is one of only two CSOs whose classification was revised after the blind tests had been carried out\ddag.  The final CSO~2.2 classification was unanimously agreed in a group discussion of the \citet{2008ApJ...684..153T} 8\,GHz and 15\,GHz maps.
\vskip 6pt
\noindent
J1159+5820, CSO~2.0 [17]: The extended steep spectrum, edge-brightened by hot spots outer lobes revealed in the 5\,GHz and 8\,GHz images of \citet{2016MNRAS.459..820T}$^\dagger$ mark this objects as a CSO~2.0.
\vskip 6pt
\noindent
J1205+2031, CSO~2.1 [43]: The EVN images at 5\,GHz and 8\,GHz by \citet{2021MNRAS.506.1609C}$^\dagger$ show two resolved lobes in this low-luminosity CSO~2.1 object.
\vskip 6pt
\noindent
J1220+2916, CSO~1 [45]: This object is observed in the 5\,GHz and 8\,GHz observations of \citet{2016MNRAS.459..820T}$^\dagger$ to have a flat spectrum core, indicated by the red cross, and two edge-dimmed jets extending in opposite directions from the core.  The morphology and low luminosity mark this as a CSO~1 object.
\vskip 6pt
\noindent
J1227+3635, CSO~2.0 [16]: This object is observed in the 1.7\,GHz and 5\,GHz observations of \citet{2013MNRAS.433..147D}, as well as the 5\,GHz and 8\,GHz observations of \citet{2016MNRAS.459..820T}$^\dagger$ to have a flat spectrum core straddled by two steep spectrum lobes, one of which is at the end of a long, narrow jet and is strongly edge brightened. 
The optical identification is a quasar at redshift 1.973, which suggests that the jet axis cannot be close to the plane of the sky.  The most likely interpretation is that this is a CSO where some effects of relativistic beaming are being seen, which might also explain the fact that this is the highest luminosity CSO in this sample of 55.  The side opposite the jet on this interpretation is de-boosted and the hot-spot radiation on that side is likely also de-boosted.   The morphology of the very narrow jet and strong hot-spot on one side mark this as a CSO~2.0 object. 
\vskip 6pt
\noindent
J1234+4753, CSO~2.1 [34]: \citet{2016MNRAS.459..820T}$^\dagger$ show that this object has a flat ($\alpha \sim -0.5$ from  5 to 8\,GHz) to steep ($\alpha \sim -1.0$ from 8 to 15\,GHz) spectrum core and edge-dimmed CSO~1 morphology.  
\vskip 6pt
\noindent
J1244+4048, CSO~2.2 [19]: The \citet{2004AandA...426..463O}$^\dagger$ map shows a nucleus and two-sided edge-dimmed jet. The 5\,GHz and 8\,GHz images of  \citet{2016MNRAS.459..820T} show a flat spectrum core and two-sided steep spectrum jet with highly-resolved hot spots, indicative of a CSO~2.2.
\vskip 6pt
\noindent
J1247+6723, CSO~2.0 [50]: The 15\,GHz map of \citet{2003PASA...20...16M}$^\dagger$ shows hot spots towards the outer edges of the lobes, making this a CSO~2.0 object. This object is a clear double on larger scales, showing evidence of a previous epoch of activity.
\vskip 6pt
\noindent
J1254+1856, CSO~1 [48]: The 5\,GHz and 8\,GHz maps of \citet{2016MNRAS.459..820T} show a flat spectrum core with two fading, edge-dimmed steep spectrum jets.  The morphology and low luminosity clearly indicate a CSO~1. The MOJAVE$^\dagger$ image shows the same features.
\vskip 6pt
\noindent
J1311+1658, CSO~1 [38]: 5\,GHz and 8\,GHz observations of \citet{2016MNRAS.459..820T}$^\dagger$  show this object to have a flat spectrum core and two edge-dimmed steep spectrum jets. It  is therefore a CSO~1 object. 
\vskip 6pt
\noindent
J1313+5458, CSO~2.2 [20]: The 5\,GHz map of \citet{1994ApJS...95..345T}$^\dagger$ shows two heavily-resolved lobes. 5\,GHz and 8\,GHz observations of \citet{2016MNRAS.459..820T} shows these lobes to have steep spectra.  The  highest surface brightness region, possibly a hot-spot, in one lobe is not at the extremity of the source, and the other lobe has an amorphous structure with no hot-spot.
\vskip 6pt
\noindent
J1326+3154, CSO~2.2 [13]: The 5\,GHz map of \citet{2007ApJ...658..203H}$^\dagger$ shows two highly resolved lobes with hot spots at extremities of the lobes in some directions, but not the overall extremities of the source, making this a CSO~2.2 object.
\vskip 6pt
\noindent
J1347+1217, CSO~2.2 [28]: The 5\,GHz map of \citet{1997A&A...325..943S}$^\dagger$ hows a well-resolved lobe on one side, and a weak lobe on the other side of the presumed nucleus. This object shows evidence of a previous epoch of activity.
\vskip 6pt
\noindent
J1400+6210, CSO~2.2 [15]: \citet{1996ApJ...463...95T} pinpointed the center of activity in this object, which made clear that it is a CSO.   They also pointed out that, while there is a lot of structure in the jets, there are no visible hot spots. So this is a CSO~2.2 object.
\vskip 6pt
\noindent
J1407+2827, CSO~2.1 [52]: The strange morphology of this source, as shown in the MOJAVE$^\dagger$ image, makes this object difficult to classify - hence ``indeterminate'', i.e., CSO~2.1.
\vskip 6pt
\noindent
J1414+4554, CSO~2.1 [36]: The 5\,GHz map of \citet{2005ApJ...622..136G}$^\dagger$ shows an edge-brightened southern lobe with a resolved hot-spot, and an edge-dimmed northern lobe, making this an ``indeterminate'' CSO~2.1 object.
\vskip 6pt
\noindent
J1434+4236, CSO~2.2 [22]: 5\,GHz and 8\,GHz observations of \citet{2016MNRAS.459..820T}$^\dagger$  show this object to have two resolved lobes, each with well-resolved extensions perpendicular to the source axis.  Hence this is a CSO~2.2 object,
\vskip 6pt
\noindent
J1440+6108, CSO~2.2 [30]: 5\,GHz and 8\,GHz observations of \citet{2016MNRAS.459..820T}$^\dagger$  show this object to have two resolved lobes, each with well-resolved extensions perpendicular to the source axis.  Hence this is a CSO~2.2 object.
\vskip 6pt
\noindent
J1508+3423, CSO~2.1 [33]: The  5\,GHz MERLIN observations by  \citet{2010MNRAS.408.2261K}$^\dagger$ show a complex structure indicative of a CSO~2.1 object.
\vskip 6pt
\noindent
J1511+0518, CSO~2.0 [51]: The 22\,GHz image of \citet{2006AA...450..959O}$^\dagger$ shows slight edge-dimming in both lobes in this very compact source.  We have provisionally classified it as a CSO~2.0 because of its very compact, unresolved, structure at all frequencies, from 2.3\,GHz to 22\,GHz, including in the MOJAVE stacked image,  perpendicular to the axis, but further observations at higher frequencies are needed to confirm this.
\vskip 6pt
\noindent
J1559+5924, CSO~1 [47]: The MOJAVE$^\dagger$ and the 5\,GHz and 8\,GHz observations of \citet{2016MNRAS.459..820T} show this to be an edge-dimmed, low-luminosity CSO~1 object.
\vskip 6pt
\noindent
J1602+5243, CSO~1 [26]: The image of \citet{2009A&A...498..641D}$^\dagger$ shows two edge-dimmed jets and is clearly a CSO~1 object. 
\vskip 6pt
\noindent
J1609+2641, CSO~2.1 [14]: 5\,GHz and 8\,GHz observations of \citet{2016MNRAS.459..820T}$^\dagger$  show this object to have a hot-spot at the end of one lobe and the other lobe to be well-resolved with no hot-spot, making this an ``indeterminate'' CSO~2.1 object.
\vskip 6pt
\noindent
J1644+2536, CSO~2.1 [23]: 5\,GHz and 8\,GHz observations of \citet{2016MNRAS.459..820T}$^\dagger$  show this object to have a flat spectrum core flanked by two edge-dimmed jets, making this a CSO~2.1 object.
\vskip 6pt
\noindent
J1723-6500, CSO~2.1 [49]: The 8.4\,GHz image of \citet{2019AA...627A.148A}$^\dagger$ of this low-luminosity object shows a very curious. edge-dimmed structure, making this a CSO~2.1 object.
\vskip 6pt
\noindent
J1734+0926, CSO~2.0 [6]: As shown in the MOJAVE$^\dagger$ map, this is a double-lobed object with hot spots at the outer edges of the lobes, and hence this is a CSO~2.0 object.
\vskip 6pt
\noindent
J1735+5049, CSO~2.0 [4]: \citet{2014MNRAS.438..463O}$^\dagger$ carried out 6-frequency VLBA observations of this object, which revealed its flat-spectrum core, and steep-spectrum lobes with hot spots at the outer edges being self-absorbed below 6\,GHz. So this is clearly a CSO~2.0 object.
\vskip 6pt
\noindent
J1816+3457, CSO~2.1 [31]: The \citet{2005ApJ...622..136G}$^\dagger$ 8.4\,GHz map shows resolved lobes, edge-dimming in the southern lobe, and no hot-spot in the northern lobe. Hence the CSO~2.1 classification. Note that the image has been rotated to fit into the available space in Fig.~1, so that the southern lobe is the lobe at the upper right.
\vskip 6pt
\noindent
J1915+6548, CSO~2.1 [25]: As described by \citet{2007ApJ...661...78G}$^\dagger$, this is a very curious object, and is therefore of indeterminate class: CSO~2.1.
\vskip 6pt
\noindent
J1939-6342, CSO~2.0 [7]: As shown in this 8.4\,GHz image by \citet{2002A&A...392..841T}$^\dagger$, one of the lobes is unresolved at the outer edge, and the other is unresolved near the outer edge. Both lobes are unresolved perpendicular to the jet axis, making this a CSO~2.0 object.
\vskip 6pt
\noindent
J1944+5448, CSO~2.0 [29]: The 8.4\,GHz map of \citet{2016AN....337...42R}, and this MOJAVE$^\dagger$ image, both show the core and two outer lobes with hot spots at the outer edges, which make this a CSO~2.0 object.
\vskip 6pt
\noindent
J1945+7055, CSO~2.2 [37]: In the 8\,GHz map of this much-studied CSO \citet{1999ApJ...521..103P}, we see the core and two linear curving jets, the southern one culminating in a resolved component and the northern one having a hot-spot coincident with a sharp bend in the jet transitioning to a fainter component further out. Since this object shows edge-dimming at the end of the two oppositely-directed jets, as can be seen in this MOJAVE$^\dagger$ image, we classify it as a CSO~2.2 object.
\vskip 6pt
\noindent
J2022+6136, CSO~2.1 [2]: In this object one lobe is edge-brightened and the other is edge-dimmed -- hence the CSO~2.1 classification. This image is that of
\citet{2000ApJ...541...66L}$^\dagger$.
\vskip 6pt
\noindent
J2203+1007, CSO~2.0 [5]: The 15\,GHz image of \citet{2014MNRAS.438..463O}$^\dagger$ shows clear hot spots at the outer edges of the two lobes, making this a CSO~2.0 object.
\vskip 6pt
\noindent
J2327+0846, CSO~1 [27]: This highly unusual CSO was mapped at 1.4\,GHz by \citet{2003ApJ...597..809M}$^\dagger$, who showed that its age as a jetted-AGN is a few Myr. Note that the center of activity has been identified by \citet{2003ApJ...597..809M} with the highest surface brightness component,  This source shows strong edge-dimming and is therefore a CSO~1 object. 
\vskip 6pt
\noindent
J2355+4950, CSO~2.2 [24]: This object has been extensively studied as the archetypical CSO by R96, who showed that it has two extended lobes - the southern one with  a hot-spot at the outer edge of he lobe, and the northern one having no hot-spot. Although the southern lobe has a hot-spot, material is spilling out of this hot-spot towards the east.  In addition, at lower frequencies the structure extends outside of the jets and hot spots to form a common envelope.  The position of the flat-spectrum nucleus was pinpointed by \citet{1996ApJ...463...95T}.
R96 showed that there is a bright jet component in the  northern jet, and they also showed that the northern jet is an order of magnitude brighter than the southern jet.  This strongly suggests that the northern side is the approaching side and that we are seeing slightly beamed radiation from the two jets.  On this interpretation, the northern lobe is seen at a considerably later epoch than the southern lobe, which could explain the lack of a hot-spot in the northern lobe if the jet is no longer powering the hot-spot in the northern lobe. There is, therefore, a strong hint of evolution in the jets and lobes. These considerations plus that extended envelope and significant extension of the source normal to the jet axis lead us to classify this as a CSO~2.2 object.
The image shown is that of
\citet{1999NewAR..43..669O}$^\dagger$.

\section{The Blind Tests}\label{sec:blind1}

To conduct the blind tests, we formed four independent teams each consisting of two of the co-authors of this paper. These tests were ``blind'' in the following two senses.  First, the four teams worked completely independently, with no knowledge of the classifications of the other three teams.  Second, the classifications were based on morphology alone, with no cognizance of redshift, physical size, and luminosity, thereby precluding any influence from location in the $(P,D)$ plane.   We re-classified three objects because not all of the relevant maps had been taken into account in the initial classifications (J0943+1702, J1158+2450 and J2327+0846) after we had finished the blind tests, but these re-classifications were done after completing the blind tests and carrying out the statistical tests on the blind tests, and so do not affect the statistical results of the binomial tests quoted below. 

The two members of each team individually classified the 54~CSOs, then compared their results and agreed on a team classification.  The results of the four teams' classifications are shown in Table \ref{tab:blindtests}.  The results show remarkable consistency, given the variety of morphologies observed, and the small number of classes. In 47\% of the cases all four teams agreed on the classification, in 85\% of the cases three or four teams agreed, in 5\% of the cases the classifications were split 2:2, and in 9\% they were split 2:1:1. The final classifications are also given in Table \ref{tab:blindtests}.

The probabilities shown in \cref{tab:binomial} are derived as follows.  There are four possible ways all four teams can choose the same classification ($p={4\over256}=0.0156$), and 4!~possible ways ($p = {24 \over 256} = 0.0938$) all four teams could choose different classifications. In the case of exactly three teams agreeing, the number of possibilities is (4 ways to pick the agreeing classification) $\times$ (3 ways to pick the non-agreeing classification) $\times$ (4 ways to arrange these) = 48 ($p = {48 \over 256} = 0.1875$). That leaves 180 ways that exactly two teams (or two pairs of teams) could choose the same classification ($p = {180 \over 256} = 0.7031$).

The probabilities, under the random assumption,  of agreement by all four teams, three or four teams, two teams, and by no teams, are given  in the third column of \cref{tab:binomial}. The differences between the actual numbers (fifth column) and the expected number (fourth column) are striking. 

In the cases of the three objects marked with a double dagger (\ddag), which includes one of the 2:2 splits, the objects were discussed and the final decision on the correct classification was unanimous.

\begin{deluxetable*}{cccccc@{\hskip5mm}|@{\hskip5mm}cccccc}
\tablecaption{Results of the Blind Classifications of CSOs in \cref{plt:linsizedist} }.
\tablehead{CSO&Team&Team&Team&Team&Final&CSO&Team&Team&Team&Team&Final\\
Name&	\#1 	&	  \#2&\#3	&\#4&Class&Name&	\#1 	&	  \#2&\#3	&\#4&Class}
\startdata
J0029+3456	&	2.0	&	2.0	&	2.0	&	2.0	&	2.0&	J1311+1658	&	1	&	1	&	1	&	1	&	1\\	
J0111+3906	&	2.0	&	2.0	&	2.0	&	2.0	&	2.0&	J1313+5458 & 2.2&2.2&2.0&2.2&2.2	\\
J0119+3210	&	2.1	&	2.2	&	1.0	&	2.2	&	2.2&	J1326+3154	&	2.2	&	2.2	&	2.0	&	2.2	&	2.2	\\
J0131+5545	&	2.2	&	2.2	&	1	&	2.2	&	2.2&	J1347+1217	&	2.2	&	2.2	&	2.2	&	1	&	2.2	\\
J0405+3803	&	2.0	&	2.0	&	2.1	&	2.2	&	2.0&	J1400+6210	&	2.2	&	2.2	&	2.1	&	1	&	2.2	\\
J0713+4349	&	2.0	&	2.2	&	2.0	&	2.0	&	2.0&	J1407+2827	&	2.1	&	2.1	&	2.1	&	2.1	&	2.1	\\
J0741+2706	&	2.2	&	2.1	&	2.1	&	2.2	&	2.1&	J1414+4554	&	2.1	&	2.1	&	2.0	&	2.1	&	2.1	\\
J0825+3919	&	2.2	&	2.1	&	2.2	&	2.1	&	2.1&	J1434+4236	&	2.2	&	2.2	&	2.2	&	2.2	&	2.2	\\
J0832+1832	&	1	&	1	&	2.1	&	1.0	&	1&	J1440+6108	&	2.1	&	2.1	&	2.0	&	2.1	&	2.1	\\
J0855+5751	&	2.1	&	2.1	&	2.1	&	2.1	&	2.1&	J1508+3423	&	2.1	&	2.1	&	1.0	&	2.1	&	2.1	\\
J0906+4124	&	1	&	1	&	1	&	2.0	&	1&	J1511+0518	&	2.0	&	2.0	&	2.0	&	2.2	&	2.0	\\
J0909+1928	&	1	&	1	&	1	&	1	&	1&	J1559+5924	&	1	&	1	&	2.1	&	1	&	1	\\
J0943+1702$^\ddag$	&	2.0	&	2.1	&	2.1	&	2.1	&	2.0&	J1602+5243	&	1	&	1	&	1	&	1	&	1	\\
J1025+1022	&	1	&	1	&	1	&	2.2	&	1&	J1609+2641	&	2.1	&	2.0	&	2.2	&	2.1	&	2.1	\\
J1035+5628	&	2.0	&	2.0	&	2.0	&	2.0	&	2.0&	J1644+2536	&	2.1	&	2.1	&	2.1	&	1	&	2.1	\\
J1111+1955	&	2.0	&	2.0	&	2.0	&	2.0	&	2.0&	J1723-6500	&	2.1	&	2.1	&	2.0	&	2.2	&	2.1	\\
J1120+1420	&	2.0	&	2.0	&	2.0	&	2.0	&	2.0&	J1734+0926	&	2.0	&	2.0	&	2.0	&	2.0	&	2.0	\\
J1148+5924	&	1	&	1	&	1	&	1	&	1&	J1735+5049	&	2.0	&	2.0	&	2.0	&	2.0	&	2.0	\\
J1158+2450$^\ddag$	&	2.2	&	2.2	&	2.1	&	2.1	&	2.2&	J1816+3457	&	2.1	&	2.1	&	2.2	&	2.1	&	2.1	\\
J1159+5820	&	2.0	&	2.0	&	2.0	&	2.0	&	2.0&	J1915+6548	&	2.1	&	2.1	&	2.1	&	2.1	&	2.1	\\
J1205+2031	&	2.1	&	2.1	&	2.0	&	2.1	&	2.1&	J1939$-$6342	&	2.0	&	2.0	&	2.0	&	2.1	&	2.0	\\
J1220+2916	&	1	&	1	&	1	&	1	&	1&	J1944+5448	&	2.0	&	2.0	&	2.0	&	2.1	&	2.0	\\
J1227+3635	&	2.0	&	2.0	&	2.0	&	2.0	&	2.0&	J1945+7055	&	2.2	&	2.2	&	2.0	&	2.2	&	2.2	\\
J1234+4753	&	2.1	&	2.1	&	2.1	&	2.1	&	2.1&	J2022+6136	&	2.1	&	2.1	&	2.1	&	2.1	&	2.1	\\
J1244+4048	&	2.2	&	2.2	&	2.2	&	2.2	&	2.2&	J2203+1007	&	2.0	&	2.0	&	2.0	&	2.0	&	2.0	\\
J1247+6723	&	2.0	&	2.0	&	2.0	&	2.0	&	2.0&	J2327+0846$^\ddag$	&	2.2	&	2.2	&	1.0	&	2.2	&	1	\\
J1254+1856	&	1	&	1	&	1	&	1	&	1&	J2355+4950	&	2.2	&	2.2	&	2.2	&	2.2	&	2.2	\\								\enddata
\tablecomments{Four teams of two carried out the classification independently.  At first each member of a team made their own independent classification, and then they conferred to agree on a final classification. The results are listed for each team.  The final classification was chosen based on the majority vote.  There were only three cases in which there was an even 2:2 split, and in the cases of J0741+2706 and J0825+3919 the indeterminate (2.1) class was chosen.  In the cases of the three objects marked with a double dagger ($^\ddag$), which includes one of the 2:2 splits, the objects were discussed and the final decision on the correct classification was unanimous.  This was done without reference to the redshifts, luminosities or physical sizes of these three objects.}
\label{tab:blindtests} 
\end{deluxetable*}

\clearpage

\section{$(u,v)$ Coverage with a Near Earth Orbiter}\label{sec:spaceuv}

\begin{figure*}
        \includegraphics[width=1.0\linewidth]{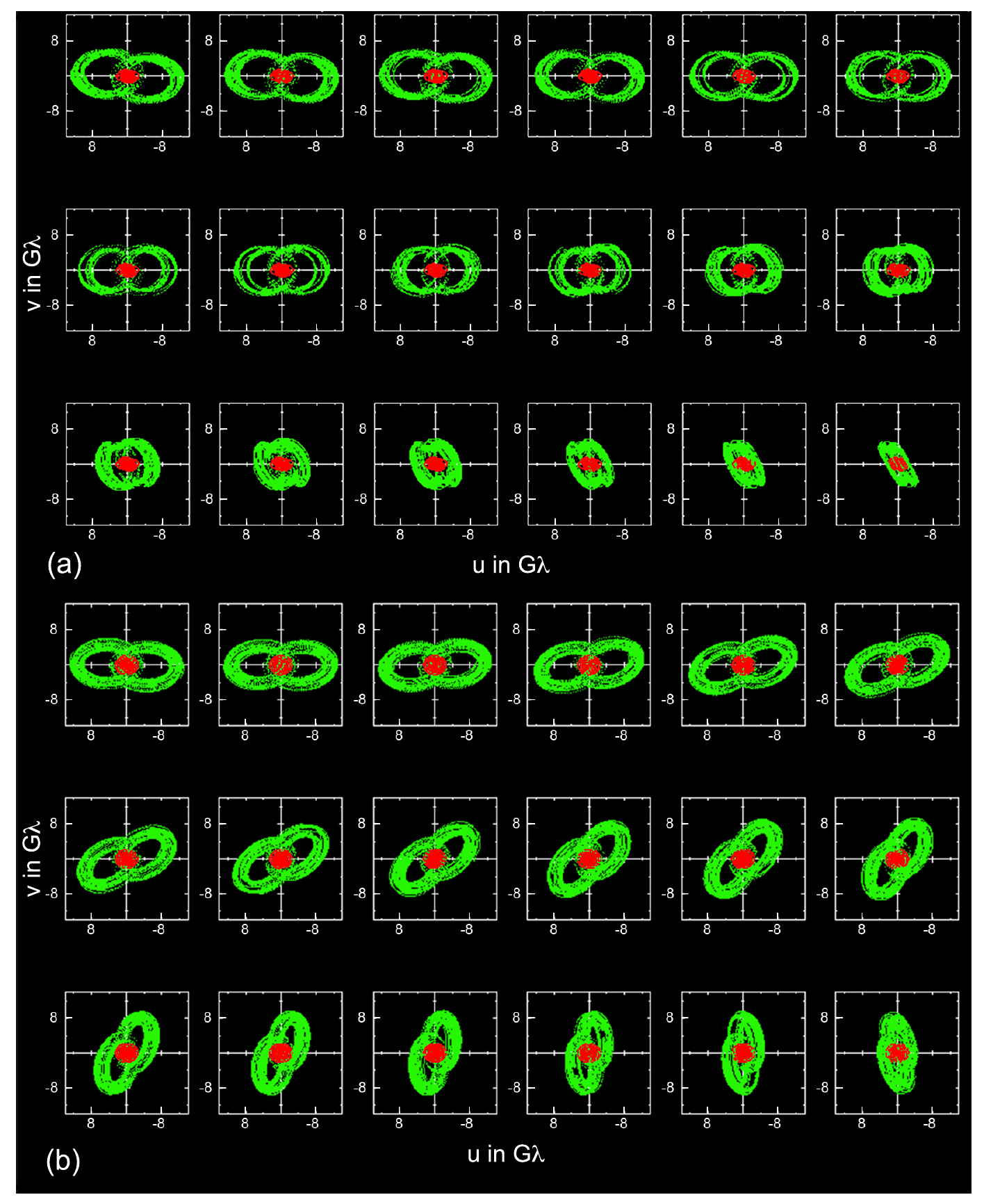}
        \caption{$(u,v)$ coverage at 86 GHz with a NEO of apogee 2.5 earth diameters at inclination $60^\circ$ operating in conjunction with a ground array. Red tracks show earth-earth baselines and green tracks show space-earth baselines. The successive frames show observations of duration 1 day spaced 1 month apart over an 18 month period. (a) For a source at declination $\delta=30^\circ$; (b) For a source at declination $\delta=60^\circ$. Over an 18 month period holes in the $(u,v)$ coverage are well filled. In order to fill the  $(u,v)$ gaps adequately, the observations over 1 day would not need to be continuous, and many objects could be observed each day. Such a NEO could observe many hundreds of CSOs (see text).}
        \label{plt:space}.
    \end{figure*}

In Fig. \ref{plt:space} we show the $(u,v)$ coverage that could be obtained on a typical CSO with observations spread over 1 day per month over 18 months. For the chosen apogee and perigee heights of 32,000 km and 1,000 km the precession rates of the right ascension of the ascending node ($\Omega$) and the the argument of perigee  ($\omega$) are $-71.4^{\circ}$/year and $+17.8^{\circ}$/year respectively for an orbit inclination of $60^{\circ}$. This allows the large $(u,v)$ holes obtained when observing the source closest to the orbit normal direction (as occurs in the first $(u,v)$ coverage for each sequence shown) to be filled in with subsequent shorter-baseline observations when the source is observed further away from this direction due to the $\Omega$ precession.

    \clearpage

\bibliographystyle{aasjournal}





\end{document}